\documentclass[11pt,a4paper]{article}
\pdfoutput=1
\usepackage{amsfonts}
\usepackage{amssymb}
\usepackage[centertags]{amsmath}
\usepackage{graphicx}
\usepackage{hyperref}
\usepackage{booktabs}
\usepackage{theorem}
\usepackage{array}
\usepackage{epsfig}
\usepackage{colordvi}
\usepackage{color}
\usepackage{ulem}
\usepackage[small,bf]{caption}
\usepackage{cite}
\usepackage{color}
\usepackage{subfigure}

\def\1{\mathbf{1}}
\def\3{\mathbf{3}}
\def\2{\mathbf{2}}

\newcommand{\betabeta}{\mbox{$(\beta \beta)_{0 \nu}  $}}
\def\ltap{\ \raisebox{-.4ex}{\rlap{$\sim$}} \raisebox{.4ex}{$<$}\ }

\allowdisplaybreaks[1]

\usepackage{a4wide}

\newcommand{\bec}{\begin{cases}}
\newcommand{\eec}{\end{cases}}
\newcommand{\beq}{\begin{equation*}}
\newcommand{\eeq}{\end{equation*}}
\newcommand{\be}{\begin{equation}}
\newcommand{\ee}{\end{equation}}
\newcommand{\ba}{\begin{eqnarray}}
\newcommand{\ea}{\end{eqnarray}}

\makeatletter

\newcommand{\Rmnum}[1]{\expandafter\@slowromancap\romannumeral #1@}
\makeatother

\begin{document}

\begin{titlepage}

\vspace*{-15mm}
\begin{flushright}
SISSA 58/2014/FISI\\
IPMU14-0331\\
arXiv:1410.8056
\end{flushright}
\vspace*{0.7cm}

\begin{center}
{\bf\LARGE {Determining the Dirac CP Violation Phase in the}}\\
[4mm]
{\bf\LARGE {Neutrino Mixing Matrix from Sum Rules}}\\
[8mm]
\vspace{0.4cm} I. Girardi$\mbox{}^{a)}$, S. T. Petcov$\mbox{}^{a,b)}$
\footnote{Also at: Institute of Nuclear Research and Nuclear Energy,
Bulgarian Academy of Sciences, 1784 Sofia, Bulgaria.}
and A. V. Titov$\mbox{}^{a)}$
\\[1mm]
\end{center}
\vspace*{0.50cm}
\centerline{$^{a}$ \it SISSA/INFN, Via Bonomea 265, 34136 Trieste, Italy }
\vspace*{0.2cm}
\centerline{$^{b}$ \it Kavli IPMU (WPI), University of Tokyo,\\
5-1-5 Kashiwanoha, 277-8583 Kashiwa, Japan}
\vspace*{1.20cm}

\begin{abstract}
\noindent
Using the fact that the neutrino mixing matrix
$U = U^\dagger_{e}U_{\nu}$, where $U_{e}$ and $U_{\nu}$
result from the diagonalisation of the charged lepton
and neutrino mass matrices, we analyse the sum rules which
the Dirac phase $\delta$ present in $U$
satisfies when $U_{\nu}$ has a form dictated by, 
or associated with, discrete symmetries and
$U_e$ has a ``minimal'' form (in
terms of angles and phases it contains)
that can provide the requisite
corrections to $U_{\nu}$, so that
reactor, atmospheric and solar neutrino mixing angles
$\theta_{13}$, $\theta_{23}$ and  $\theta_{12}$
have values compatible with the current data.
The following symmetry forms are considered:
i) tri-bimaximal (TBM), ii) bimaximal (BM)
(or corresponding to the conservation of the
lepton charge $L' = L_e - L_\mu - L_{\tau}$ (LC)),
iii) golden ratio type A (GRA),
iv) golden ratio type B (GRB),
and v) hexagonal (HG).
We investigate the predictions
for $\delta$ in the cases of
TBM, BM (LC), GRA, GRB and HG forms
using the exact and the leading order sum rules
for $\cos\delta$ proposed in the literature,
taking into account also the uncertainties
in the measured values of
$\sin^2\theta_{12}$, $\sin^2\theta_{23}$ and $\sin^2\theta_{13}$.
This allows us, in particular, to assess
the accuracy of the predictions for $\cos\delta$
based on the leading order sum rules and
its dependence on the values
of the indicated neutrino mixing parameters
when the latter are varied in their respective
3$\sigma$ experimentally allowed ranges.
\end{abstract}

\vspace{0.5cm}

\end{titlepage}
\setcounter{footnote}{0}

\newpage

\section{Introduction}

One of the major goals of the future
experimental studies in neutrino physics
is the searches for CP violation (CPV) effects
in neutrino oscillations
(see, e.g., \cite{PDG2014,LBLFuture13}).
It is part of a more general and ambitious program
of research aiming to determine the status of the
CP symmetry in the lepton sector.

 In the case of the reference 3-neutrino mixing scheme
\footnote{All compelling data on
neutrino masses, mixing and oscillations are
compatible with the existence of mixing of three
light neutrinos $\nu_i$, $i=1,2,3$,
with masses $m_i \ltap 1$ eV in the weak
charged lepton current (see, e.g., \cite{PDG2014}).},
CPV effects in the flavour
neutrino oscillations,
i.e., a difference between the probabilities of
$\nu_l \rightarrow \nu_{l'}$ and
$\bar{\nu}_l \rightarrow \bar{\nu}_{l'}$
oscillations in vacuum \cite{Cabibbo:1977nk,BHP80},
$P(\nu_l \rightarrow \nu_{l'})$ and
$P(\bar{\nu}_l \rightarrow \bar{\nu}_{l'})$,
$l\neq l' =e,\mu,\tau$,
can be caused, as is well known, by the
Dirac phase present in the Pontecorvo,
Maki, Nakagawa and Sakata (PMNS)
neutrino mixing matrix $U_{\rm PMNS} \equiv U$.
If the neutrinos with definite masses
$\nu_i$, $i=1,2,3$, are Majorana particles,
the 3-neutrino mixing matrix contains
two additional Majorana CPV phases
\cite{BHP80}. However, the flavour neutrino
oscillation probabilities
$P(\nu_l \rightarrow \nu_{l'})$ and
$P(\bar{\nu}_l \rightarrow \bar{\nu}_{l'})$,
$l,l' =e,\mu,\tau$, do not depend on
the Majorana phases
\footnote{The Majorana phases can play
important role,
e.g., in $|\Delta L| = 2$
processes like neutrinoless double beta
($\betabeta$-) decay $(A,Z) \rightarrow (A,Z+2) + e^- + e^-$,
$L$ being the total lepton charge,
in which the Majorana nature of
massive neutrinos $\nu_i$, if any, manifests itself
(see, e.g., \cite{BiPet87,BPP1,WRodej10}).}
\cite{BHP80,Lang87}.
Our interest in the CPV
phases present in the neutrino mixing matrix
is stimulated also by the intriguing possibility
that the Dirac phase and/or the Majorana phases in
$U_{\rm PMNS}$ can provide the CP violation
necessary for the generation of the observed
baryon asymmetry of the Universe
\cite{Pascoli:2006ie,Pascoli:2006ci}.

  In the standard parametrisation \cite{PDG2014}
of the PMNS matrix we are going to employ
in our further discussion,
$U_{\rm PMNS}$ is expressed in terms of the solar,
atmospheric and reactor neutrino mixing
angles $\theta_{12}$,  $\theta_{23}$ and
$\theta_{13}$, respectively, and the Dirac and
Majorana CPV phases, as follows:
\begin{equation}
U= VQ\,,~~~
Q = {\rm diag}\left(1, e^{i \frac{\alpha_{21}}{2}}, e^{i \frac{\alpha_{31}}{2}}\right)\,,
\label{eq:VQ}
\end{equation}
%
where $\alpha_{21,31}$
are the two Majorana CPV
phases and $V$ is a CKM-like matrix,
\begin{equation}
\begin{array}{c}
\label{eq:Vpara}
V = \left(\begin{array}{ccc}
 c_{12} c_{13} & s_{12} c_{13} & s_{13} e^{-i \delta}  \\[0.2cm]
 -s_{12} c_{23} - c_{12} s_{23} s_{13} e^{i \delta}
 & c_{12} c_{23} - s_{12} s_{23} s_{13} e^{i \delta}
 & s_{23} c_{13} 
\\[0.2cm]
 s_{12} s_{23} - c_{12} c_{23} s_{13} e^{i \delta} &
 - c_{12} s_{23} - s_{12} c_{23} s_{13} e^{i \delta}
 & c_{23} c_{13} 
\\
  \end{array}
\right)\,.
\end{array}
\end{equation}
%
\noindent
In eq.~(\ref{eq:Vpara}),
$\delta$ is the Dirac CPV phase,
$0 \leq \delta \leq 2\pi$,
we have used the standard notation
$c_{ij} = \cos\theta_{ij}$,
$s_{ij} = \sin\theta_{ij}$, and
$0 \leq  \theta_{ij} \leq \pi/2$.
If CP invariance holds, we have
$\delta =0,\pi,2\pi$,
the values 0 and $2\pi$ being physically
indistinguishable.

  The existing neutrino oscillation data
 allow us to determine
the neutrino mixing parameters
$\sin^2\theta_{12}$, $\sin^2\theta_{23}$ and $\sin^2\theta_{13}$,
which are relevant for our further analysis,
with a relatively good precision
\cite{Capozzi:2013csa,Gonzalez-Garcia:2014bfa}.
The best fit values and the 3$\sigma$ allowed ranges of
$\sin^2\theta_{12}$, $\sin^2\theta_{23}$ and $\sin^2\theta_{13}$,
found in the global analysis
in ref. \cite{Capozzi:2013csa} read:
\begin{eqnarray}
\label{th12values}
(\sin^2 \theta_{12})_{\rm BF} = 0.308\,,~~~~
 0.259 \leq \sin^2 \theta_{12} \leq 0.359\,,\\ [0.30cm]
\label{th23values}
(\sin^2\theta_{23})_{\rm BF} = 0.437~(0.455)\,,~~~~
 0.374~(0.380) \leq \sin^2\theta_{23} \leq 0.626~(0.641)\,,\\[0.30cm]
\label{th13values}
(\sin^2\theta_{13})_{\rm BF} = 0.0234~(0.0240)\,,~~~~
0.0176~(0.0178) \leq \sin^2\theta_{13} \leq 0.0295~(0.0298)\,,
\end{eqnarray}
%
where the values (values in brackets)
correspond to
neutrino mass spectrum with normal ordering
(inverted ordering) (see, e.g., \cite{PDG2014}),
denoted further as NO (IO) spectrum.

 In the present article we will be concerned
with the predictions for the Dirac phase
$\delta$ and will not discuss the
Majorana phases in what follows.
More specifically, we will be interested
in the predictions for the Dirac CPV phase
$\delta$  which are based on the
so-called ``sum rules'' for $\cos\delta$
\cite{Antusch:2005kw,Marzocca:2013cr,Petcov:2014laa}
(see also, e.g.,
\cite{King:2005bj,King:2014nza,King:2013eh}).
The sum rules of interest appear in an approach
aiming at quantitative understanding of the pattern
of neutrino mixing
on the basis of symmetry considerations. In
this approach one exploits the fact that,
up to perturbative corrections,
the PMNS matrix has an approximate form, $U_{\nu}$,
which can be dictated by symmetries.
The matrix $U_{\nu}$ is assumed to originate
from the diagonalisation of the
neutrino Majorana mass term.
The angles in $U_{\nu}$ have specific
symmetry values which differ, in general, from
the experimentally determined values of the PMNS
angles $\theta_{12}$, $\theta_{13}$ and
$\theta_{23}$, and thus need to be corrected.
The requisite perturbative corrections,
which modify the values of the angles in $U_{\nu}$
to coincide with the measured values
of $\theta_{12}$, $\theta_{13}$ and
$\theta_{23}$, are provided by the matrix
$U_e$ arising from the diagonalisation of
the charged lepton mass matrix,
$U = U_e^{\dagger}\,U_{\nu}$.
In the sum rules we will analyse in detail in the present article
the Dirac phase $\delta$, more precisely, 
$\cos\delta$, is expressed, in general,
in terms of the mixing angles $\theta_{12}$,
$\theta_{13}$ and $\theta_{23}$ of the PMNS matrix $U$
and the angles present in $U_{\nu}$, whose
values are fixed, being dictated by an underlying
approximate discrete symmetry of the lepton sector
(see, e.g., \cite{King:2014nza}).

%
\section{The Sum Rules}
%
%
In the framework of the reference 3 flavour neutrino mixing
we will consider,
the PMNS neutrino mixing matrix
is always given by
\be
U = U_e^{\dagger} U_{\nu} \,,
\label{UeUnu}
\ee
%
where $U_e$ and $U_{\nu}$ are $3 \times 3$ unitary matrices
originating from the diagonalisation of the charged lepton and
the neutrino (Majorana) mass terms.
As we have already indicated,
we will suppose in what follows that
$U_{\nu}$ has a form which is dictated
by symmetries. More specifically,
we will assume that
\be
U_\nu  = \Psi_1\,\tilde{U}_{\nu}\,Q_0 =
\Psi_1\,R_{23} \left( \theta^\nu_{23} \right)
R_{12} \left( \theta^\nu_{12} \right) Q_0\,,
\label{UnuQ0}
\ee
%
where $R_{23}(\theta^\nu_{23})$ and $R_{12}(\theta^\nu_{12})$ are
orthogonal matrices describing rotations in the 2-3 and 1-2 planes,
respectively, and $\Psi_1$ and $Q_0$ are diagonal phase matrices
each containing two phases.
 Obviously, the phases in the matrix $Q_0$
give contribution to the Majorana phases in the PMNS matrix.
In the present article we will consider
the following symmetry forms of the matrix $\tilde{U}_{\nu}$:
i) tri-bimaximal (TBM) \cite{TBM},
ii) bimaximal (BM), or due to a symmetry corresponding
to the conservation of the lepton charge
$L' = L_e - L_{\mu} - L_{\tau}$ (LC) \cite{SPPD82,BM},
iii) golden ratio type A (GRA) form \cite{Everett:2008et,GRAM},
iv) golden ratio type B (GRB) form \cite{GRBM}, and
v) hexagonal (HG) form\cite{HGM,Kim:2010zub}.
The TBM, BM, GRA, GRB and HG forms can be 
obtained respectively from, e.g., $T'$/$A_4$, $S_4$, $A_5$, $D_{10}$ and $D_{12}$ 
discrete (lepton) flavour symmetries (see, e.g., 
\cite{King:2014nza,Chen:2009gf,Girardi:2013sza,
Everett:2008et,GRAM,GRBM,Kim:2010zub}).
In all these cases we have $\theta^\nu_{23} = -\pi/4$, and the
matrix  $\tilde{U}_{\nu}$ is given by
\begin{equation}
\tilde{U}_{\nu} = \begin{pmatrix}
\cos \theta^{\nu}_{12} & \sin \theta^{\nu}_{12} & 0 \vspace{0.2cm} \\
- \dfrac{\sin \theta^{\nu}_{12}}{\sqrt{2}} &
\dfrac{\cos \theta^{\nu}_{12}}{\sqrt{2}} &
- \dfrac{1}{\sqrt{2}} \vspace{0.2cm} \\
- \dfrac{\sin \theta^{\nu}_{12}}{\sqrt{2}}  &
\dfrac{\cos \theta^{\nu}_{12}}{\sqrt{2}} &
\dfrac{1}{\sqrt{2}}
\end{pmatrix} \;.
\label{Unu1}
\end{equation}
%
The TBM, BM (LC), GRA, GRB and HG forms of $\tilde{U}_{\nu}$
correspond to different fixed values of $\theta^{\nu}_{12}$
and thus of $\sin^2\theta^{\nu}_{12}$, namely, to
i)  $\sin^2\theta^{\nu}_{12} = 1/3$,
ii)  $\sin^2\theta^{\nu}_{12} = 1/2$,
iii)  $\sin^2\theta^{\nu}_{12} =  (2 + r)^{-1} \cong 0.276$,
$r$ being the golden ratio, $r = (1 +\sqrt{5})/2$,
iv) $\sin^2\theta^{\nu}_{12} = (3 - r)/4 \cong 0.345$, and
v) $\sin^2\theta^{\nu}_{12} = 1/4$.
Thus, the matrix  $U_e$ in eq.~(\ref{UeUnu})
should provide corrections which not only generate
nonzero value of $\theta_{13}$, but also
lead to reactor, atmospheric and solar neutrino
mixing angles $\theta_{13}$, $\theta_{23}$ and  $\theta_{12}$
which have values compatible with the current data,
including a possible sizeable deviation of $\theta_{23}$
from $\pi/4$. As was shown in \cite{Marzocca:2013cr},
the ``minimal'' form of $U_e$, in terms of angles and
phases it contains, that can provide the requisite
corrections to $U_{\nu}$
includes a product of two orthogonal matrices describing
rotations in the 2-3 and 1-2 planes,
$R_{23}(\theta^e_{23})$ and $R_{12}(\theta^e_{12})$,
$\theta^e_{23}$ and $\theta^e_{12}$
being two (real) angles. In what follows we will
adopt this minimal form of  $U_e$.
It proves convenient to cast it
in the form \cite{Marzocca:2013cr}:
\begin{equation}
U_e  = \Psi_2^{\dagger} \, \tilde U_e = 
\Psi_2^{\dagger} \, R^{-1}_{23} \left( \theta^e_{23} \right)
R^{-1}_{12} \left( \theta^e_{12} \right)\,,
\label{eq:nuU}
\end{equation}
%
where  $\Psi_2$ is a diagonal phase matrix
including two phases, and
\begin{equation}
R_{12}\left( \theta^e_{12} \right) = \begin{pmatrix}
\cos \theta^e_{12} & \sin \theta^e_{12} & 0\\
- \sin \theta^e_{12} & \cos \theta^e_{12} & 0\\
0 & 0 & 1 \end{pmatrix} \;,
\quad
R_{23}\left( \theta^e_{23} \right) = \begin{pmatrix}
1 & 0 & 0\\
0 & \cos \theta^e_{23} & \sin \theta^e_{23} \\
0 & - \sin \theta^e_{23}  & \cos \theta^e_{23} \\
\end{pmatrix} \;.
\label{R1223}
\end{equation}
%
Thus, the PMNS matrix in the approach we are following
is given by
\begin{equation}
U = U_e^{\dagger}\,U_\nu =
R_{12} \left( \theta^e_{12} \right)\,
R_{23} \left( \theta^e_{23} \right)\,
\Psi\,
R_{23} \left( \theta^\nu_{23} \right)
R_{12} \left( \theta^\nu_{12} \right) Q_0\,,~\Psi =\Psi_2\Psi_1\,,~
\theta^\nu_{23} = -\,\frac{\pi}{4}\,.
\label{UUedagUnu}
\end{equation}
%
The matrices $\Psi$ and $Q_0$
are diagonal phase matrices each containing, in general,
two physical CPV phases
\footnote{The diagonal phase matrix $\Psi$, as we see, can
originate from the charged lepton or the neutrino sector, or else
can receive contributions from both sectors \cite{FPR04}.}
\cite{FPR04}:
\begin{equation}
\Psi =
{\rm diag} \left(1,e^{-i\psi}, e^{-i\omega} \right)\,,~~
Q_0 = {\rm diag} \left(1,e^{i\frac{\xi_{21}}{2}},
e^{i\frac{\xi_{31}}{2}} \right)\,.
\label{PsieQ0}
\end{equation}
%

As was explained earlier, 
the requirement that $U_e$ has a
 ``minimal'' form in terms of angles
and phases it contains, needed  to provide
the requisite corrections to $U_{\nu}$, 
makes not necessary the inclusion 
in $\tilde{U}_{e}$ of the
orthogonal matrix describing the rotation
in the 1-3 plane, $R_{13}(\theta^e_{13})$. 
Effectively, this is equivalent to the assumption that 
the angle $\theta^e_{13}$, if nonzero, 
is sufficiently small and thus is either negligible, 
or leads to sub-dominant effects in the observable 
of interest in the present analysis, $\cos\delta$.
We will use $\theta^e_{13} \cong 0$ 
to denote values of $\theta^e_{13}$ 
which satisfy the indicated condition.

  We note that  $\theta^e_{13} \cong 0$
is a feature of many  theories 
of charged lepton and neutrino mass generation
(see, e.g.,\cite{Girardi:2013sza,Marzocca:2011dh,Antusch:2011qg,
Chen:2009gf,Everett:2008et,CarlMCC}). 
The assumption that  $\theta^e_{13} \cong 0$ 
was also used in a large number of
studies dedicated to the problem
of understanding the origins of the observed pattern
of lepton mixing
(see, e.g., \cite{Shimizu:2014ria,GTani02,FPR04,Romanino:2004ww,
Antusch:2005kw,Chao:2011sp,Hall:2013yha}).
In large class of GUT inspired
models of flavour,  
the matrix $U_e$ is directly related to
the quark mixing matrix
(see, e.g., \cite{Marzocca:2011dh,Antusch:2011qg,Chen:2009gf,Gehrlein:2014wda}).
As a consequence, in this class of
models we have $\theta^e_{13}\cong 0$. 
We will comment later on the possible effects of 
$\theta^e_{13}\neq 0$, $|\sin\theta^e_{13}| \ll 1$, 
on the predictions for $\cos\delta$, which are of 
principal interest of the present study.

More generally, the approach to understanding the observed 
pattern of neutrino mixing on the basis of 
discrete symmetries employed in the present article, 
which leads to the sum rule of interest for $\cos\delta$, is 
by no means unique --- it is one of the several 
possible approaches discussed in the literature 
on the subject (see, e.g.,  \cite{King:2013eh}).
It is employed in a large number of phenomenological studies 
(see, e.g., \cite{Shimizu:2014ria,GTani02,FPR04,Romanino:2004ww,Antusch:2005kw,Chao:2011sp,Hall:2013yha}) 
as well as in a class of models 
(see, e.g., \cite{Girardi:2013sza,Marzocca:2011dh,Antusch:2011qg,Chen:2009gf,Gehrlein:2014wda}) 
of neutrino mixing based on discrete symmetries.
However, it should be clear that the conditions  
which define the approach used in the present article
are not fulfilled in all models with discrete flavour symmetries. 
For example, they are not fulfilled in the models with 
discrete flavour symmetry  $\Delta (6n^2)$ 
studied in \cite{King:2013vna,Hagedorn:2014wha}, 
with the $S_4$ flavour symmetry constructed in \cite{Luhn:2013lkn} 
and in the models discussed in \cite{Altarelli:2012bn}.

 Following \cite{Marzocca:2013cr},
we will use the following rearrangement of the 
product of matrices $R_{23}(\theta^e_{23})  \Psi\\
 R_{23}(\theta^{\nu}_{23} = -\pi/4)$ in the expression
eq.~(\ref{UUedagUnu}) for $U_{\text{PMNS}}$:
\begin{equation}
R_{23}( \theta^e_{23})\, \Psi \, R_{23}(\theta^{\nu}_{23} = -\pi/4) =
P_1\, \Phi\, R_{23}(\hat\theta_{23})\,Q_1\,,
\label{Phi}
\end{equation}
%
where the angle $\hat\theta_{23}$ is determined by
\begin{equation}
\sin^2\hat\theta_{23} =
\frac{1}{2}\,\left (1 - 2\sin\theta^e_{23}\cos \theta^e_{23}\cos(\omega -\psi)\right )\,,
\label{th23hat}
\end{equation}
%
and 
\begin{equation}
P_1={\rm diag} \left(1,1, e^{-\, i \alpha} \right)\,,~
\Phi = {\rm diag} \left(1,e^{i \phi},1 \right)\,,~
Q_1 = {\rm diag} \left(1,1, e^{i \beta} \right)\,.
\label{PPhitQ}
\end{equation}
%
In eq. (\ref{PPhitQ})
\begin{equation}
\alpha = \gamma + \psi + \omega  \,,~~~~
\beta = \gamma - \phi\,,
\label{alphabeta}
\end{equation}
%
and
\begin{equation}
\gamma = \arg \left (\,-e^{ -i \psi}\cos \theta^e_{23}
+ e^{-i \omega}\sin\theta^e_{23}\right)\,,~~
\phi= \arg \left (e^{ -i \psi}\cos \theta^e_{23}
+ e^{-i \omega}\sin\theta^e_{23}\right)\,.
\label{gammaphi}
\end{equation}
%
The phase $\alpha$ in the matrix $P_1$ 
can be absorbed in the $\tau$ lepton field and, thus, 
is unphysical. The phase $\beta$ gives a 
contribution to the matrix $\hat{Q} = Q_1\,Q_0$; 
the diagonal phase matrix $\hat{Q}$
 contributes to the matrix
of physical Majorana phases.
In the setting considered 
the PMNS matrix takes the form:
\begin{equation}
U_{\text{PMNS}}=
R_{12}(\theta^e_{12})\,\Phi(\phi)\, R_{23}(\hat\theta_{23})\,
R_{12}(\theta^{\nu}_{12})\,\hat{Q}\,,
\label{UPMNSthhat1}
\end{equation}
%
where $\theta^{\nu}_{12}$ has a fixed value
which depends on the symmetry form of $\tilde{U}_\nu$ used.
For the angles $\theta_{13}$, $\theta_{23}$ and $\theta_{12}$
of the standard parametrisation of the PMNS matrix $U$
we get in terms of the parameters in the expression 
eq.~(\ref{UPMNSthhat1}) for $U$ \cite{Marzocca:2013cr}:
\begin{align}
\label{s2th13}
    \sin \theta_{13} &= \left| U_{e3} \right| =
\sin \theta^e_{12} \sin \hat{\theta}_{23} \,, \\[0.30cm]
\label{s2th23}
\sin^2 \theta_{23} &=
\frac{\left| U_{\mu3} \right|^2}{1- \left| U_{e 3} \right|^2 }
= \sin^2 \hat{\theta}_{23}\,
\frac{\cos^2 \theta_{12}^e}{1 - \sin^2 \theta^e_{12} \sin^2 \hat{\theta}_{23} }
= \frac{\sin^2\hat\theta_{23} - \sin^2 \theta_{13}}
{1 - \sin^2\theta_{13}}\,, \\[0.30cm]
\nonumber
\sin^2 \theta_{12}
\nonumber
&= \frac{\left| U_{e2} \right|^2}{1- \left| U_{e3} \right|^2 } =
\left (1 - \cos^2\theta_{23}\cos^2\theta_{13}\right )^{-1}\,
\Big [ \sin^2\theta^{\nu}_{12}\sin^2\theta_{23}   \\[0.3cm]
&+ \cos^2\theta^{\nu}_{12}\cos^2\theta_{23}\sin^2\theta_{13}
+ \frac{1}{2}\,
\sin2\theta^{\nu}_{12}\sin2\theta_{23}\sin\theta_{13}\cos\phi \Big ]\,,
\label{s2th12}
\end{align}
%
where eq.~(\ref{s2th13})
was used in order to obtain the expression for
$\sin^2 \theta_{23}$ in terms of $\hat{\theta}_{23}$ and $\theta_{13}$,
and eqs. (\ref{s2th13}) and  (\ref{s2th23}) were used
to get the last expression for $\sin^2 \theta_{12}$.
Within the approach employed,
the expressions in eqs. (\ref{s2th13})~--~(\ref{s2th12}) are exact.

 It follows from  eqs.~(\ref{eq:VQ}), (\ref{eq:Vpara}) and (\ref{UPMNSthhat1})
that the four observables $\theta_{12}$, $\theta_{23}$, $\theta_{13}$
and  $\delta$ are functions of three parameters
$\theta^e_{12}$, $\hat\theta_{23}$ and $\phi$.
As a consequence, the Dirac phase $\delta$ can be expressed
as a function of the three PMNS angles
$\theta_{12}$, $\theta_{23}$ and $\theta_{13}$ \cite{Marzocca:2013cr},
leading to a new ``sum rule''
relating $\delta$ and 
$\theta_{12}$, $\theta_{23}$ and $\theta_{13}$. 
For an arbitrary fixed value of the angle
$\theta^{\nu}_{12}$ the sum rule 
for $\cos\delta$ reads \cite{Petcov:2014laa}:
\begin{equation}
\cos\delta =  \frac{\tan\theta_{23}}{\sin2\theta_{12}\sin\theta_{13}}\,
\left [\cos2\theta^{\nu}_{12} +
\left (\sin^2\theta_{12} - \cos^2\theta^{\nu}_{12} \right )\,
 \left (1 - \cot^2\theta_{23}\,\sin^2\theta_{13}\right )\right ]\,.
\label{cosdthnu}
\end{equation}
%
For $\theta^{\nu}_{12} = \pi/4$ and $\theta^{\nu}_{12}= 
\sin^{-1}(1/\sqrt{3})$ 
the expression eq.~(\ref{cosdthnu}) for $\cos\delta$  
reduces to those found in \cite{Marzocca:2013cr} 
in the BM (LC) and TBM cases, respectively.
 A similar sum rule for an arbitrary $\theta^{\nu}_{12}$
can be derived for the phase $\phi$
\cite{Marzocca:2013cr,Petcov:2014laa}.
It proves convenient for our further discussion
to cast the sum rules for $\cos\delta$ and $\cos\phi$ 
of interest in the form:
\begin{equation}
\sin^2\theta_{12} = \cos^2\theta^{\nu}_{12} +
\frac{\sin2\theta_{12}\sin\theta_{13}\cos\delta
 -  \tan\theta_{23}\cos2\theta^{\nu}_{12}}
{\tan\theta_{23} (1 - \cot^2\theta_{23}\,\sin^2\theta_{13})}\,,
\label{s2th12cosdthnu}
\end{equation}
%
\begin{equation}
\sin^2\theta_{12} = \cos^2\theta^{\nu}_{12} +
\dfrac{1}{2} \sin 2 \theta_{23}\, \frac{\sin2\theta^{\nu}_{12}\,\sin\theta_{13}\,\cos\phi
 -  \tan\theta_{23}\cos2\theta^{\nu}_{12}}
{(1 - \cos^2\theta_{23}\,\cos^2\theta_{13})}\,.
\label{s2th12cosphithnu}
\end{equation}
%
The phases $\delta$ and $\phi$ are related by \cite{Petcov:2014laa}:
\begin{align}
\label{sindsinphi}
\sin\delta =
&\; -\, \frac{\sin2\theta^{\nu}_{12}}{\sin2\theta_{12}}\,\sin\phi\,,
\\[0.30cm]
\cos\delta =&\; \frac{\sin2\theta^{\nu}_{12}}{\sin2\theta_{12}}\,\cos\phi\,
\left (-1 + \frac{2\sin^2\theta_{23}}
{\sin^2\theta_{23}\cos^2\theta_{13} + \sin^2\theta_{13}}\,\right )
\nonumber \\[0.30cm]
+ &\; \frac{\cos2\theta^{\nu}_{12}}{\sin2\theta_{12}}\,
\frac{\sin2\theta_{23}\, \sin\theta_{13}}
{\sin^2\theta_{23}\cos^2\theta_{13} + \sin^2\theta_{13}}\,.
\label{cosdcosphi}
\end{align}
%
Within the scheme considered 
the sum rules eqs.~(\ref{cosdthnu})~--~(\ref{s2th12cosphithnu})
and the relations eqs.~(\ref{sindsinphi}) and (\ref{cosdcosphi}) are exact.
In a complete self-consistent theory of (lepton) flavour based 
on discrete flavour symmetry, the indicated sum rules and relations 
are expected to get corrections due to, e.g., $\theta^e_{13} \neq 0$, 
renormalisation group (RG) effects, etc. 
Analytic expression for the correction 
in the expression for $\cos\delta$, eq. (\ref{cosdthnu}), 
due to $|\sin\theta^e_{13}| \ll 1$
was derived in \cite{Petcov:2014laa}. 
As was shown in \cite{Petcov:2014laa}, 
for the  best fit values of the lepton mixing 
angles  $\theta_{12}$, $\theta_{13}$  and $\theta_{23}$,
a nonzero $\theta^e_{13} \ltap 10^{-3}$  
produces a correction to the value of $\cos\delta$ 
obtained from the ``exact'' sum rule eq.~(\ref{cosdthnu}),  
which does not exceed 
11\% (4.9\%) in the TBM (GRB) cases and is even smaller 
in the other three cases of symmetry forms 
of $\tilde{U}_{\nu}$ analysed in the 
present article. A value of $\theta^e_{13} \ltap 10^{-3}$ 
is a feature of many  theories and models 
of charged lepton and neutrino mass generation 
(see, e.g., 
\cite{Everett:2008et,Girardi:2013sza,Marzocca:2011dh,Antusch:2011qg,Chen:2009gf,
CarlMCC}).
The RG effects on the lepton mixing angles and the 
CPV phases are known to be negligible for 
hierarchical neutrino mass spectrum 
(see, e.g., \cite{RGE,Petcov:2005yh} and the references quoted therein); 
these effects are relatively small for values of 
the lightest neutrino mass 
not exceeding approximately 0.05 eV 
\footnote{In supersymmetric theories 
this result is valid for moderate values 
of the parameter $\tan\beta \ltap 10$ (see \cite{RGE,Petcov:2005yh}); 
for $\tan\beta = 50$ the same statement is true 
for values of the lightest neutrino mass 
smaller than approximately 0.01 eV.}.
We will call the sum rules and the relations given in 
eqs. (\ref{cosdthnu})~--~(\ref{s2th12cosphithnu}),
 (\ref{sindsinphi}) and (\ref{cosdcosphi}) 
``exact'',  keeping in mind that they can be subject 
to corrections, which, however, in a number of  
physically interesting cases, if not absent, can only be  
sub-dominant.

 A parametrisation of the PMNS matrix, similar to that given in
eq.~(\ref{UUedagUnu}), has been effectively employed
in ref. \cite{Antusch:2005kw}:
the hierarchy of values of the angles
in the matrices $U_e$ and $U_{\nu}$
assumed in \cite{Antusch:2005kw}
leads the authors to consider the angles $\theta^e_{13}$ and
$\theta^{\nu}_{13}$ of the 1-3 rotations in
$U_e$ and $U_{\nu}$ as negligibly small.
As a consequence, the PMNS matrix is effectively
parametrised in \cite{Antusch:2005kw}
with four angles $\theta^e_{12}$,
$\theta^e_{23}$, $\theta^\nu_{12}$, $\theta^\nu_{23}$ and
\footnote{In contrast to $\theta^{\nu}_{23} = \pi / 4$
employed in \cite{Antusch:2005kw}, we use $\theta^{\nu}_{23} = -\pi / 4$.
The effect of the difference in the signs of $\sin\theta^e_{12}$
and $\sin\theta^e_{23}$ utilised by us and in \cite{Antusch:2005kw}
is discussed in  Appendix \ref{App:A}.}
four phases
$\delta^e_{12}$, $\delta^e_{23}$, $\delta^\nu_{12}$, $\delta^\nu_{23}$.
As is shown in Appendix \ref{App:A}
(see also ref. \cite{Petcov:2014laa}),
these phases are related to the phases
$\psi$, $\omega$, $\xi_{21}$ and $\xi_{31}$  present
in the parametrisation in eq.~(\ref{UUedagUnu}) as follows:
\begin{align}
\label{omegapsidelt}
& \psi = \delta^e_{12} - \delta^{\nu}_{12}  +\pi \,,
\quad
\omega = \delta^e_{23} + \delta^e_{12}
- \delta^{\nu}_{23} - \delta^{\nu}_{12} \,,\\
& \xi_{21} = -2 \delta^{\nu}_{12} \,,
\quad \xi_{31} = -2 (\delta^{\nu}_{12} + \delta^{\nu}_{23}) \,.
\label{xidelt}
\end{align}
%
Treating $\sin\theta^e_{12}$
and  $\sin\theta^e_{23}$ as small parameters,
$|\sin\theta^e_{12}| \ll 1$, $|\sin\theta^e_{23}| \ll 1$,
neglecting terms of order of, or smaller than,
$O((\theta^e_{12})^2)$, $O((\theta^e_{23})^2)$ and
$O(\theta^e_{12} \theta^e_{23})$, and taking into account that
in this approximation we have
$\sin\theta^e_{12}= \sqrt{2}\sin\theta_{13}$,
the following ``leading order'' sum rule
was obtained in \cite{Antusch:2005kw}:
\be
\theta_{12} \cong \theta^{\nu}_{12} + \theta_{13} \cos \delta \,.
\label{AKth12th12nucosd}
\ee
%
This sum rule can be derived from the sum rule
\be
\sin\theta_{12} \cong
\sin\theta^{\nu}_{12}
 + \frac{\sin2\theta^{\nu}_{12}}{2 \sin \theta^{\nu}_{12}}
 \sin\theta_{13} \cos \delta\,,
\label{sinth12cosd}
\ee
%
by treating
$\sin2\theta^{\nu}_{12}\sin\theta_{13}\cos\delta\cong
\sin2\theta^{\nu}_{12}\theta_{13}\cos\delta$
as a small parameter and using the Taylor expansion
$\sin^{-1}(a + b \, x) \cong \sin^{-1} (a) + b \, x /\sqrt{1-a^2}$,
valid for $|bx| \ll 1$.

  From eqs.~(\ref{s2th12cosdthnu}) and (\ref{s2th12cosphithnu}),
employing the approximations
used in ref. \cite{Antusch:2005kw},  we get:
\begin{equation}
\sin^2\theta_{12} \cong \sin^2\theta^{\nu}_{12} +
\sin2\theta_{12}\sin\theta_{13}\cos\delta\,,
\label{s2th12cosdthnu2}
\end{equation}
%
\begin{equation}
\sin^2\theta_{12} \cong \sin^2\theta^{\nu}_{12} +
\sin2\theta^{\nu}_{12}\,\sin\theta_{13}\,\cos\phi\,.
\label{s2th12cosphithnu2}
\end{equation}
%
The first equation leads (in the leading order approximation
used to derive it and using $\sin2\theta_{12}\cong \sin2\theta^{\nu}_{12}$)
to eq.~(\ref{AKth12th12nucosd}),
while from the second equation we find:
\be
\sin \theta_{12} \cong \sin \theta^{\nu}_{12}
 + \frac{\sin2\theta^{\nu}_{12}}{2 \sin \theta^{\nu}_{12}}
 \sin\theta_{13} \cos \phi\,,
\label{sth12cosphi10}
\ee
%
and correspondingly,
\be
\theta_{12} \cong \theta^{\nu}_{12} + \theta_{13} \cos \phi \,.
\label{th12th12nucosphi}
\ee
%
This implies that in the leading order approximation adopted
in ref.  \cite{Antusch:2005kw} we have
\cite{Petcov:2014laa} $\cos\delta = \cos\phi$.
Note, however, that the sum rules for $\cos\delta$ and $\cos\phi$
given in eqs.~(\ref{s2th12cosdthnu2}) and (\ref{s2th12cosphithnu2}),
differ somewhat by the factors multiplying
the terms $\sim \sin\theta_{13}$.

 As was shown in \cite{Petcov:2014laa},
the leading order sum rule (\ref{AKth12th12nucosd})
leads in the cases of the TBM, GRA, GRB and HG
forms of $\tilde{U}_{\nu}$ to largely imprecise
predictions for the value of $\cos\delta$:
for the best fit values of
$\sin^2\theta_{12} = 0.308$, $\sin^2\theta_{13} = 0.0234$ and
$\sin^2\theta_{23} = 0.425$ used in \cite{Petcov:2014laa},
they differ approximately by factors (1.4~--~1.9)
from the values found from the exact sum rule.
The same result holds for  $\cos\phi$. Moreover,
the predicted values of $\cos\delta$ and
$\cos\phi$ differ approximately by factors of (1.5~--~2.0),
in contrast to the prediction $\cos\delta \cong \cos\phi$
following from the leading order sum rules.
The large differences between the results for
$\cos\delta$ and  $\cos\phi$, obtained using
the leading order and the exact sum rules,
are a consequence \cite{Petcov:2014laa}
of the quantitative importance of the
next-to-leading order terms which are
neglected in the leading order sum rules
(\ref{AKth12th12nucosd})~--~(\ref{th12th12nucosphi}).
The next-to-leading order terms are significant
for the TBM, GRA, GRB and
HG forms of $\tilde{U}_{\nu}$ because
in all these cases the ``dominant''
terms $|\theta_{12} - \theta^{\nu}_{12}|\sim \sin^2\theta_{13}$,
or equivalently
\footnote{Note that \cite{Petcov:2014laa}
since $\cos\delta$ and $\cos\phi$
in eqs.~(\ref{AKth12th12nucosd})~--~(\ref{th12th12nucosphi}) are multiplied by
$\sin\theta_{13}$, the ``dominant'' terms
$|\theta_{12} - \theta^{\nu}_{12}|$
and the next-to-leading order
terms $\sim \sin^2\theta_{13}$ give contributions to
$\cos\delta$ and $\cos\phi$, which are
both of the same order and are $\sim \sin\theta_{13}$.}
$|\sin^2\theta_{12} - \sin^2\theta^{\nu}_{12}|\sim \sin^2\theta_{13}$.
It was shown also in \cite{Petcov:2014laa} that
in the case of the BM (LC) form of  $\tilde{U}_{\nu}$ we have
$|\theta_{12} - \theta^{\nu}_{12}|\sim \sin\theta_{13}$
and the leading order sum rules provide rather
precise predictions for $\cos\delta$ and $\cos\phi$.

 The results quoted above were obtained
in \cite{Petcov:2014laa} for the
best fit values of the neutrino mixing parameters
$\sin^2\theta_{12}$, $\sin^2\theta_{23}$ and $\sin^2\theta_{13}$.
In the present article we
investigate in  detail the predictions
for $\cos\delta$ and $\cos\phi$ in the cases of the
TBM, BM (LC), GRA, GRB and HG forms of $\tilde{U}_{\nu}$ using the
exact sum rules given in eqs.~(\ref{s2th12cosdthnu})
(or (\ref{cosdthnu})) and (\ref{s2th12cosphithnu})
 and the leading order
sum rules in eqs.~(\ref{s2th12cosdthnu2}) and
(\ref{s2th12cosphithnu2}), taking into account
also the uncertainties in the measured values of
$\sin^2\theta_{12}$, $\sin^2\theta_{23}$ and $\sin^2\theta_{13}$.
This allows us to better assess
the accuracy of the predictions
for $\cos\delta$ and $\cos\phi$
based on the leading order sum rules and
its dependence  on the values
of the neutrino mixing angles. We investigate
also how the predictions for $\cos\delta$ and $\cos\phi$,
obtained using the exact and the leading order sum rules,
vary when the PMNS neutrino mixing parameters
$\sin^2\theta_{12}$, $\sin^2\theta_{23}$ and
$\sin^2\theta_{13}$ are varied in their respective
experimentally allowed $3\sigma$ ranges.

 In what follows we will present numerical results
using the values of
$\sin^2\theta_{12}$, $\sin^2\theta_{23}$ and $\sin^2\theta_{13}$
quoted in eqs. (\ref{th12values})~--~(\ref{th13values})
and corresponding to NO spectrum of neutrino masses,
unless another choice is explicitly specified.
The results we obtain in the case of IO spectrum
differ insignificantly from those found for NO spectrum.

%
\section{\texorpdfstring{The Case of Negligible $\theta^e_{23}$}{The Case of Negligible theta-e23}}
%
%

 The case of negligible $\theta^e_{23} \cong 0$ was investigated
by many authors
(see, e.g., \cite{
Shimizu:2014ria,GTani02,Romanino:2004ww,HPR07,
Marzocca:2011dh,Alta,Antusch:2005kw}).
It corresponds to a large number of theories and models of
charged lepton and neutrino mass generation
(see, e.g., \cite{
Marzocca:2011dh,Alta,Antusch:2011qg,
Chen:2009gf,Chao:2011sp}). For $\theta^e_{23}\cong 0$, the sum
rules of interest given in
eqs.~(\ref{s2th12cosdthnu})
(or (\ref{cosdthnu})), (\ref{s2th12cosphithnu})
and in eqs.~(\ref{s2th12cosdthnu2}),  (\ref{s2th12cosphithnu2})
were analysed in detail in ref. \cite{Petcov:2014laa}.

  In the limit of negligibly small $\theta^e_{23}$ we find from
eqs.~(\ref{th23hat}), (\ref{alphabeta}) and (\ref{gammaphi}):
\begin{equation}
\sin^2\hat\theta_{23} =\frac{1}{2}\,,~\gamma= -\psi + \pi\,,~
\phi = - \psi = \delta^{\nu}_{12} - \delta^{e}_{12}  - \pi\,,
~\beta = \gamma - \phi = \pi\,.
\label{th23hatphases}
\end{equation}
%
The phase $\omega$ is unphysical.

 In the limiting case of negligible $\theta^e_{23}$
the exact sum rules for $\cos\delta$ and $\cos\phi$
take the following form \cite{Petcov:2014laa}:
\begin{align}
\label{cosdthnua}
\cos\delta & = \; \frac{(1 - 2\sin^2\theta_{13})^{\frac{1}{2}}}
{\sin2\theta_{12}\sin\theta_{13}}\,
\left [\cos2\theta^{\nu}_{12} +
\left (\sin^2\theta_{12} - \cos^2\theta^{\nu}_{12} \right )\,
\frac{1-3\sin^2\theta_{13}}{1 - 2\sin^2\theta_{13}}\right ]\,,
\end{align}
\begin{align}
\label{cphia}
\cos\phi \,
 & =  \frac{1 - \sin^2\theta_{13}}{\sin 2\theta^\nu_{12}\,  \sin\theta_{13}\,
(1 - 2\sin^2\theta_{13})^{\frac{1}{2}}}
\left[\sin^2\theta_{12} - \sin^2\theta^{\nu}_{12} -
\cos2\theta_{12}^\nu \,\frac{\sin^2\theta_{13}}{1-\sin^2\theta_{13}}\right] \,.
\end{align}
%
From the above equations, to leading order in $\sin \theta_{13}$
we get:
\be
\cos\delta = \frac{1}{\sin2\theta_{12}\sin\theta_{13}}\,
\left (\sin^2\theta_{12} - \sin^2\theta^{\nu}_{12} \right )
+ O(\sin \theta_{13})\,,
\label{cosdthnuLOa}
\ee
\be
\cos\phi = \frac{1}{\sin2\theta^{\nu}_{12}\sin\theta_{13}}\,
\left (\sin^2\theta_{12} - \sin^2\theta^{\nu}_{12} \right )
+ O(\sin \theta_{13})\,,
\label{cphiLOa}
\ee
%
or equivalently,
\be
\sin^2 \theta_{12} = \sin^2 \theta^{\nu}_{12}
+ \sin2\theta_{12} \sin\theta_{13} \cos \delta + O(\sin^2 \theta_{13})\,,
\ee
\be
\sin^2 \theta_{12} = \sin^2 \theta^{\nu}_{12}
+ \sin2\theta^{\nu}_{12} \sin\theta_{13} \cos \phi + O(\sin^2 \theta_{13}) \,.
\ee
%
The last two equations coincide with eqs.~(\ref{s2th12cosdthnu2})
and (\ref{s2th12cosphithnu2}) which were derived
from the exact sum rules keeping the leading order corrections
in both $\sin\theta_{13}$ and $\sin\theta^e_{23}$.
This implies, in particular, that the correction due
to  $|\sin\theta^e_{23}| \ll 1$ appears in the sum rules
of interest only in the next-to-leading order terms.
Casting the results obtained in a form we are going to use
in our numerical analysis, we obtain:
\begin{align}
\label{sth12cosd1}
\sin \theta_{12} & = \sin \theta^{\nu}_{12}
 + \frac{\sin2\theta_{12}}{2 \sin \theta^{\nu}_{12}}
 \sin\theta_{13} \cos \delta + O(\sin^2 \theta_{13}) \\
 & = \sin \theta^{\nu}_{12}
 + \frac{\sin2\theta^{\nu}_{12}}{2 \sin \theta^{\nu}_{12}}
 \sin\theta_{13} \cos \delta + O(\sin^2 \theta_{13}) \,,
 \label{sth12cosd2}
\end{align}
 \be
\sin \theta_{12} = \sin \theta^{\nu}_{12}
 + \frac{\sin2\theta^{\nu}_{12}}{2 \sin \theta^{\nu}_{12}}
 \sin\theta_{13} \cos \phi + O(\sin^2 \theta_{13}) \,.
 \label{sth12cosphi1}
\ee

%
We have replaced $\sin2\theta_{12}$ with
$\sin2\theta^{\nu}_{12}$ in eq. (\ref{sth12cosd2}), so that
it corresponds to eqs.~(\ref{AKth12th12nucosd})
and (\ref{sinth12cosd}). In the cases of
the TBM, GRA, GRB and HG
symmetry forms of $\tilde{U}_{\nu}$
we are considering and for the best fit value
of $\sin^2\theta_{12} = 0.308$ we indeed have
$|\sin\theta_{12} - \sin\theta^{\nu}_{12}|\sim \sin^2\theta_{13}$.
Thus, if one applies consistently the approximations
employed in \cite{Antusch:2005kw}, which
lead to eqs.~(\ref{AKth12th12nucosd})~--~(\ref{th12th12nucosphi})
(or to eqs.~(\ref{cosdthnuLOa}) and (\ref{cphiLOa})),
one should neglect also the difference between
$\theta_{12}$ and $\theta^{\nu}_{12}$.
This leads to $\cos\delta = \cos\phi = 0$.

\begin{figure}[h!]
  \begin{center}
     \hspace{-1.6cm}
   \subfigure
 {\includegraphics[width=12cm]{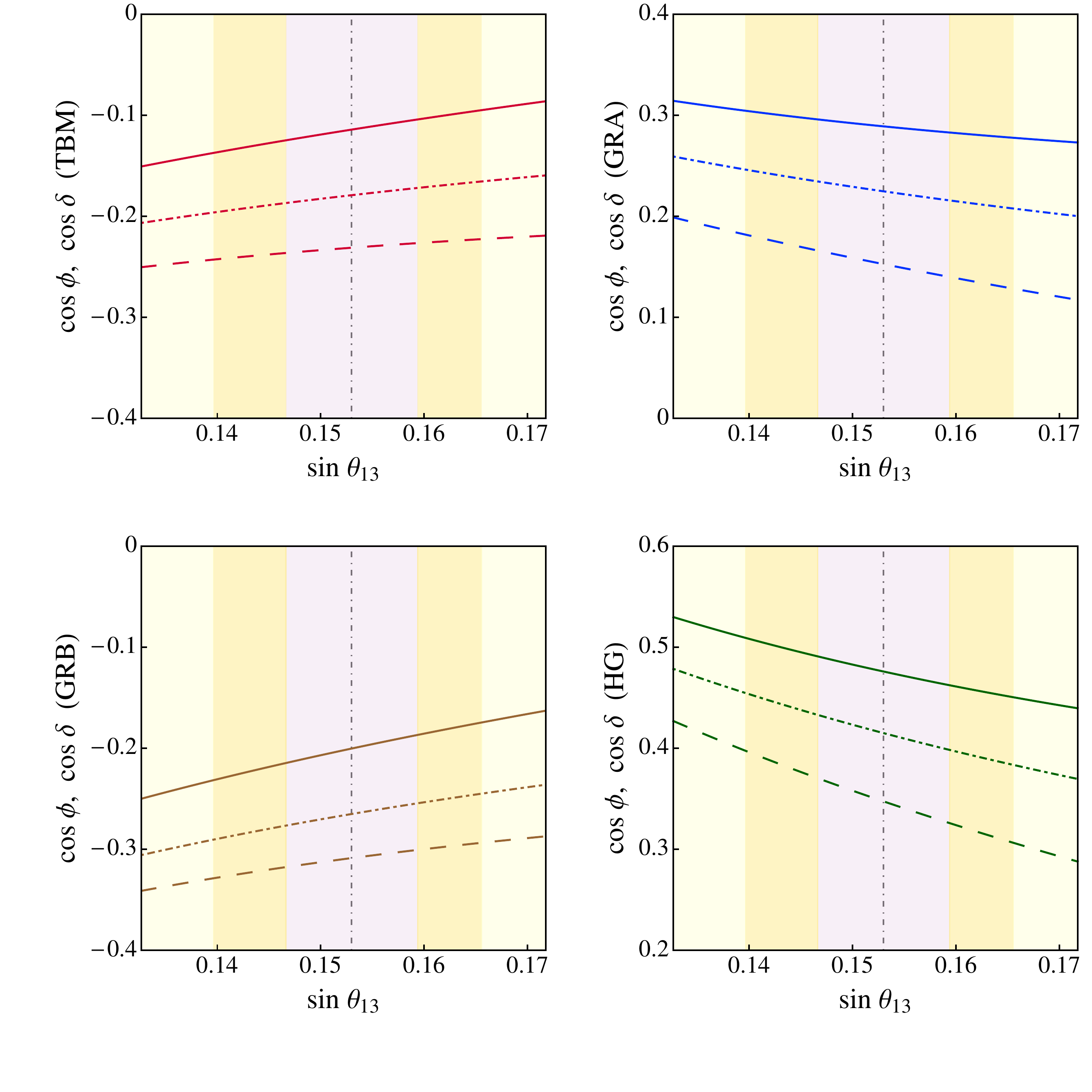}}
  \vspace{5mm}
     \end{center}
\vspace{-1.0cm} \it \caption{\label{Fig:a}
Predictions for $\cos\delta$ and $\cos \phi$
in the cases of the TBM (upper left panel),
GRA (upper right panel), GRB (lower left panel)
and HG (lower right panel)
forms of the matrix  $\tilde{U}_{\nu}$,
as functions of $\sin\theta_{13}$
and for the best fit value of
$\sin^2 \theta_{12} = 0.308$.
The solid lines (dashed lines) correspond
to $\cos\delta$ ($\cos \phi$)
determined from the exact sum rule
given in eq. (\ref{cosdthnua}) (eq.~(\ref{cphia})).
The dash-dotted line in each of the 4 panels
represents $(\cos \delta)_{\rm LO} = (\cos\phi)_{\rm LO}$
obtained from the leading order sum rule in
eq.~(\ref{sth12cosd2}).
The vertical dash-dotted line corresponds to the
best fit value of $\sin^2 \theta_{13} = 0.0234$;
the three coloured vertical bands indicate
the 1$\sigma$, 2$\sigma$ and 3$\sigma$ experimentally
allowed ranges of  $\sin\theta_{13}$
(see text for further details).
}
\end{figure}

 In  Fig.~\ref{Fig:a} we show predictions for
$\cos \delta$ and $\cos \phi$ in the cases
of the TBM, GRA, GRB and HG forms of the matrix  $\tilde{U}_{\nu}$,
as functions of $\sin\theta_{13}$
which is varied in the 3$\sigma$ interval given in
eq.~(\ref{th13values}) and corresponding to
NO neutrino mass spectrum.
The predictions are obtained for the best fit value
of $\sin^2\theta_{12} = 0.308$ using the exact
sum rules eqs.~(\ref{cosdthnua}) and (\ref{cphia})
for $\cos \delta$ (solid lines) and $\cos \phi$ (dashed lines)
and the leading order sum rules eqs.~(\ref{sth12cosd2})
and (\ref{sth12cosphi1}) (dash-dotted lines).
As we see in Fig. \ref{Fig:a}, the predictions for
$\cos\delta$ vary in magnitude and sign
when one varies the symmetry
form of $\tilde{U}_{\nu}$.
More specifically, from the exact sum rule
in eq.~(\ref{cosdthnua}),
using the best fit value  of $\sin^2\theta_{13} = 0.0234$
we get for $\cos\delta$
in the cases of the TBM, BM (LC), GRA, GRB and HG forms
of $\tilde{U}_{\nu}$,  respectively:
$\cos\delta = (-0.114);~(-1.29);~0.289;~(-0.200);~0.476$.

The unphysical value of $\cos\delta$
in the case of the  BM (LC) form of $\tilde{U}_{\nu}$
is a reflection of the fact that
the scheme under discussion with the BM (LC)
form of the matrix $\tilde{U}_{\nu}$ does not provide a good
description of the current data on
$\theta_{12}$, $\theta_{23}$ and $\theta_{13}$
\cite{Marzocca:2013cr}.
One gets a physical result for $\cos\delta$,
$\cos\delta = -0.973$,
for, e.g., values of $\sin^2\theta_{12} = 0.32$, and
$\sin\theta_{13} = 0.16$, lying in the $2\sigma$
experimentally allowed intervals of these
neutrino mixing parameters.
We have checked that
for the best fit value of $\sin^2 \theta_{13}$,
physical values of $(\cos \delta)_{\rm E}$, $(\cos \delta)_{\rm LO}$ and
$(\cos \phi)_{\rm E}$ in the BM (LC) case can be obtained
for relatively large values of $\sin^2 \theta_{12}$.
For, e.g., $\sin^2 \theta_{12} = 0.359$
and $\sin^2 \theta_{13} = 0.0234$ we find
$(\cos \delta)_{\rm E} = -0.915$,
$(\cos \delta)_{\rm LO} = -0.998$ and
$(\cos \phi)_{\rm E} = -0.922$.
In this case the differences between the
exact and leading order sum rule results for
$\cos\delta$ and $\cos\phi$ are
relatively small.

\begin{table}[h!]
\centering
\begin{tabular}{ccccc}
 \toprule
  $\sin^2 \theta_{12} = 0.308$ & TBM & GRA & GRB & HG \\
 \midrule
$(\cos\delta)_{\rm E}$ & $ -0.114$ & $0.289$ & $-0.200$ & $0.476$ \\
$(\cos\delta)_{\rm LO}$ & $ -0.179$ & $0.225$ & $-0.265$ & $0.415$ \\
$(\cos\delta)_{\rm E}/(\cos\delta)_{\rm LO}$  & $ 0.638$ & $1.29$ & $0.756$ & $1.15$ \\
$(\cos\phi)_{\rm E}$ & $ -0.231$ & $0.153$ & $-0.309$ & $0.347$ \\
$(\cos\delta)_{\rm E}/(\cos\phi)_{\rm E}$  & $0.494$ & $1.89$ & $0.649$ & $1.37$ \\
$(\cos\phi)_{\rm E}/(\cos\phi)_{\rm LO}$  & $ 1.29$ & $0.680$ & $1.16$ & $0.837$ \\
 \bottomrule
  \end{tabular}
\caption{The predicted values of
$\cos\delta$ and $\cos \phi$,
obtained from the exact sum rules
in eqs.~(\ref{cosdthnua}) and (\ref{cphia}),
$(\cos \delta)_{\rm E}$ and $(\cos \phi)_{\rm E}$,
and from the leading order sum rule in eq.~(\ref{sth12cosd2}),
$(\cos \delta)_{\rm LO} = (\cos \phi)_{\rm LO}$,
using the best fit values of
$\sin^2 \theta_{13} = 0.0234$ and
$\sin^2 \theta_{12} = 0.308$,
for the  TBM, GRA,  GRB  and HG forms of the matrix
$\tilde{U}_{\nu}$. The values of the ratios
$(\cos\delta)_{\rm E}/(\cos\delta)_{\rm LO}$,
$(\cos\delta)_{\rm E}/(\cos\phi)_{\rm E}$ and
$(\cos\phi)_{\rm E}/(\cos\phi)_{\rm LO}$ are also shown.
\label{tab:1}}
\end{table}

 The above results imply that it would be possible to
distinguish between the different symmetry forms of
 $\tilde{U}_{\nu}$ considered by measuring $\cos\delta$
\cite{Petcov:2014laa}, provided
$\sin^2\theta_{12}$ is known with sufficiently
high precision. Even determining the sign of
$\cos\delta$ will be sufficient to eliminate some of
the possible symmetry forms of $\tilde{U}_{\nu}$.

The leading order sum rules eqs.~(\ref{sth12cosd2})
and (\ref{sth12cosphi1}) lead to values of
$\cos \delta$ and $\cos \phi$, $(\cos \delta)_{\rm LO}$ and
$(\cos \phi)_{\rm LO}$, which coincide:
$(\cos \delta)_{\rm LO} = (\cos \phi)_{\rm LO}$.
These values differ, however, from the values
obtained employing the exact sum rules:
$(\cos \delta)_{\rm E} \neq (\cos \delta)_{\rm LO}$,
$(\cos \phi)_{\rm E} \neq (\cos\phi)_{\rm LO}$.
The exact sum rule values of
$\cos \delta$ and $\cos \phi$ also differ:
$(\cos \delta)_{\rm E} \neq (\cos \phi)_{\rm E}$.
We are interested both in the predictions for the
values of $(\cos \delta)_{\rm E}$, $(\cos \delta)_{\rm LO}$,
$(\cos \phi)_{\rm E}$ and $(\cos \phi)_{\rm LO}$,
and in the differences between
the exact and the leading order
sum rule predictions. In Table~\ref{tab:1}
we give the values of  $(\cos \delta)_{\rm E}$,
$(\cos \phi)_{\rm E}$, $(\cos \delta)_{\rm LO} = (\cos \phi)_{\rm LO}$,
and of the ratios $(\cos \delta)_{\rm E}/(\cos \phi)_{\rm E}$,
$(\cos \delta)_{\rm E}/(\cos \delta)_{\rm LO}$ and
$(\cos \phi)_{\rm E}/(\cos \phi)_{\rm LO}$,
calculated for the best fit values of
$\sin^2 \theta_{13} = 0.0234$ and $\sin^2 \theta_{12} = 0.308$.

 As Fig.~\ref{Fig:a} indicates, the
differences $|(\cos \delta)_{\rm E} - (\cos \delta)_{\rm LO}|$
and $|(\cos \phi)_{\rm E} - (\cos \phi)_{\rm LO}|$
exhibit weak dependence on the value of
$\sin\theta_{13}$ when it is varied
in the 3$\sigma$ interval quoted in
eq.~(\ref{th13values}).
The values of $\cos \delta$,
obtained using the exact sum rule eq.~(\ref{cosdthnua})
in the TBM, GRA, GRB and HG cases, differ from
those calculated using the approximate
sum rule eq.~(\ref{sth12cosd2})
by the factors $0.638$, $1.29$, $0.756$ and $1.15$, respectively.
The largest difference is found to hold in the TBM case.
As was shown in \cite{Petcov:2014laa},
the correction to $(\cos \delta)_{\rm LO}$~---~the
leading order sum rule result
for $\cos\delta$~---~is given approximately by
$\cos2\theta^{\nu}_{12} \sin\theta_{13}/(\sin2\theta_{12})$.
For given $\theta^{\nu}_{12}$,
the relative magnitude of the correction depends
on the magnitude of the ratio
$|\sin^2 \theta_{12} - \sin^2 \theta^{\nu}_{12}|/\sin\theta_{13}$.
The largest correction occurs for the symmetry form of
$\tilde{U}_{\nu}$, for which this ratio has the smallest value.
For the best fit value of $\sin^2 \theta_{12}$,
the smallest value of the ratio of interest
corresponds to the TBM form of $\tilde{U}_{\nu}$
and is equal approximately to $0.166$.

The absolute values of the difference 
$|(\cos \delta)_{\rm E} - (\cos \delta)_{\rm LO}|$ for the 
TBM, GRB, GRA and HG symmetry forms, as it follows 
from Table~\ref{tab:1}, lie in the narrow interval 
(0.061~--~0.065). These differences seem to be rather 
small. However, they are sufficiently large to lead to 
misleading results. Indeed, suppose $\cos\delta$ is measured 
and the value determined experimentally reads:
$\cos\delta = -0.18 \pm 0.025$. If one compares 
this value with the value of $\cos \delta$ 
predicted using the leading order sum rule, 
$(\cos \delta)_{\rm LO}$, one would conclude that 
data are compatible  
with the TBM form of $\tilde{U}_{\nu}$ and that 
all the other forms considered by us are ruled 
out. Using the prediction based on the exact sum rule, 
i.e., $(\cos \delta)_{\rm E}$, would lead to a completely 
different conclusion, namely, that 
the data are compatible only with the GRB form of 
$\tilde{U}_{\nu}$~\footnote{The same hypothetical example
can be used to illustrate the significance of the difference between
the exact and the leading order sum rule predictions for $\cos \delta$
also in the case of $\theta_{23}^e \neq 0$ (see Table~\ref{tab:4}).}. 
In this hypothetical example, which 
is included to illustrate the significance 
of the difference between the predictions for 
$\cos \delta$ obtained using 
the exact and the leading order sum rules,
we have assumed that the prospective uncertainties 
in the predicted values of $(\cos \delta)_{\rm LO}$
and $(\cos \delta)_{\rm E}$ due to the uncertainties 
in the measured values of $\sin^2\theta_{12}$, 
$\sin^2\theta_{13}$ and $\sin^2\theta_{23}$
are sufficiently small. 
These uncertainties will be discussed in Section \ref{sec:5} 
(see Fig.~\ref{Fig:cosdeltaNO_fut}).
The relative difference between 
$(\cos \delta)_{\rm E}$ and $(\cos \delta)_{\rm LO}$, i.e., the ratio 
$|(\cos \delta)_{\rm E} - (\cos \delta)_{\rm LO}|/|(\cos \delta)_{\rm E}|$,
is also significant. For the TBM, GRA, GRB and HG symmetry forms 
it reads: 57.0\%, 22.1\%, 32.5\% and 12.8\%, 
respectively.

 The behaviour of $\cos \delta$ and $\cos \phi$
when $\sin\theta_{13}$  increases is determined by the
sign of $(\sin^2 \theta_{12} - \sin^2 \theta^{\nu}_{12})$:
 $\cos \delta$ and $\cos \phi$ increase (decrease) when this
difference is negative (positive).
For the best fit value of $\sin^2 \theta_{12} = 0.308$,
this difference is negative in the TBM and GRB cases,
while it is positive in the GRA and HG ones.
For the four symmetry forms of $\tilde{U}_{\nu}$,
TBM, GRB, GRA and HG, and the best fit values of
$\sin^2 \theta_{13} = 0.0234$ and
$\sin^2 \theta_{12} = 0.308$, the ratio
$(\sin^2 \theta_{12} - \sin^2 \theta^{\nu}_{12})/\sin\theta_{13}$
reads, respectively: $(-0.166)$, $(-0.245)$,
$0.207$ and $0.379$.

\begin{figure}[h!]
  \begin{center}
     \hspace{-1.6cm}
   \subfigure
 {\includegraphics[width=12cm]{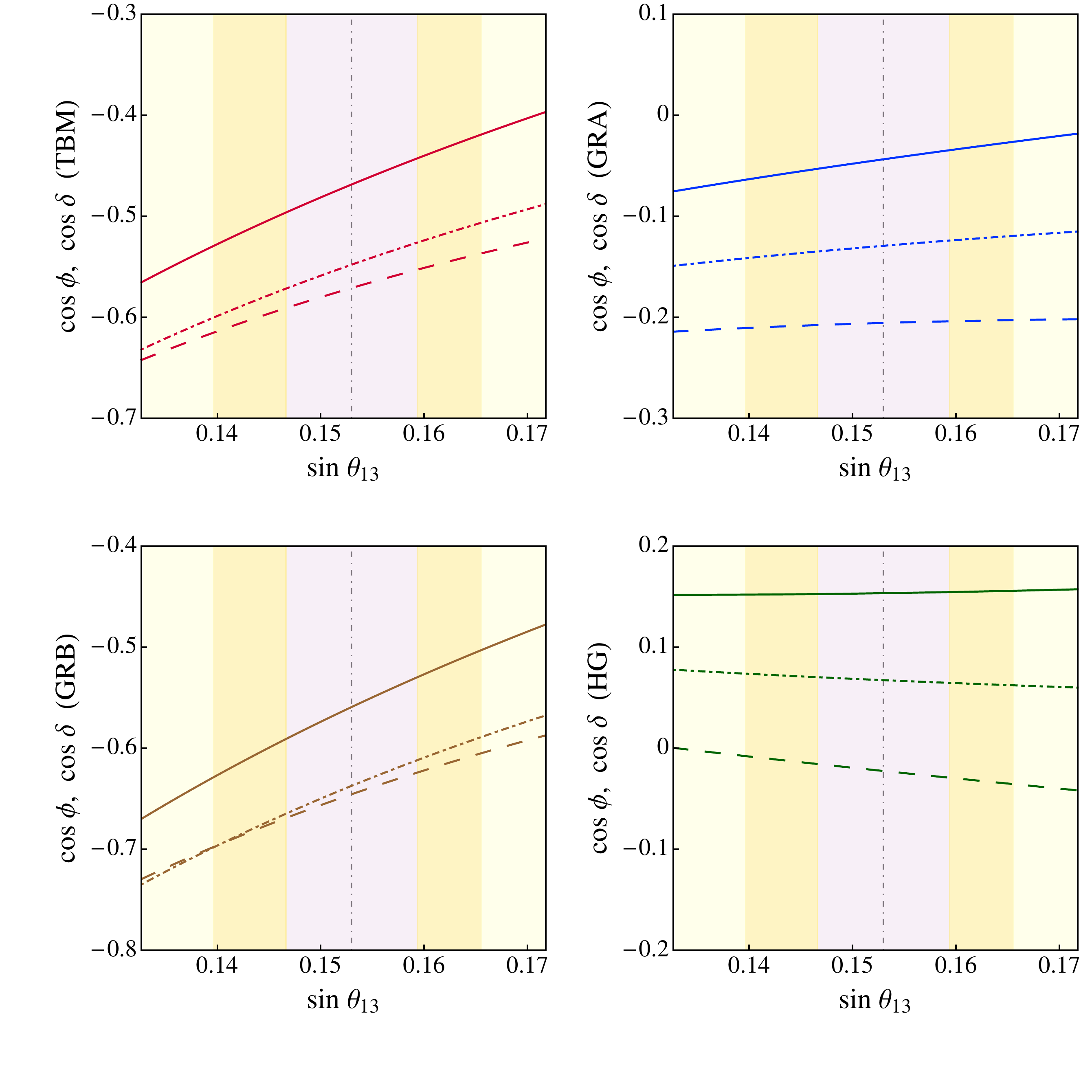}}
  \vspace{5mm}
     \end{center}
\vspace{-1.0cm} \it \caption{\label{Fig:b}
The same as in Fig.~\ref{Fig:a}, but for $\sin^2 \theta_{12} = 0.259$
(see text for further details).
}
\end{figure}

\begin{figure}[h!]
  \begin{center}
     \hspace{-1.6cm}
   \subfigure
 {\includegraphics[width=12cm]{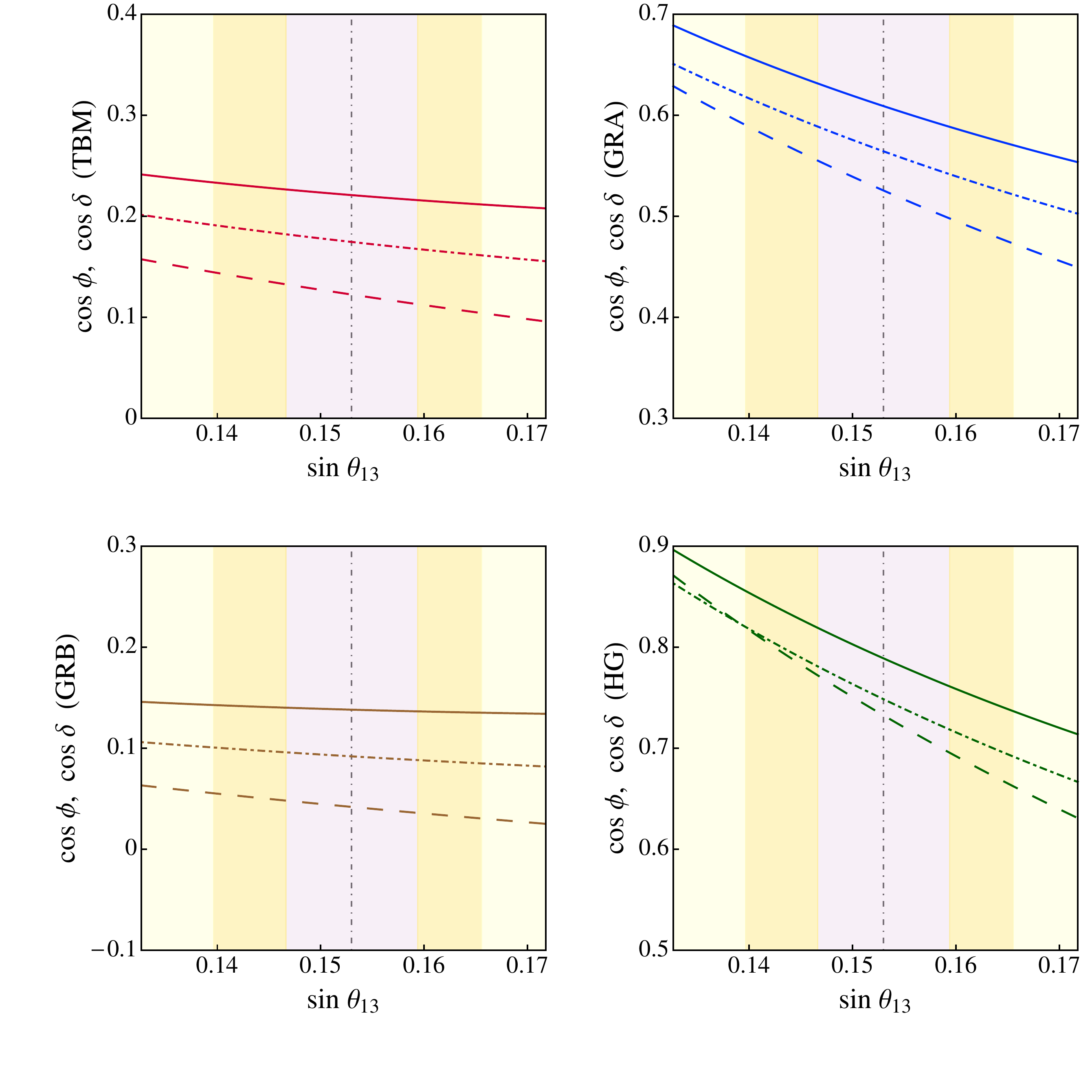}}
  \vspace{5mm}
     \end{center}
\vspace{-1.0cm} \it \caption{\label{Fig:c}
The same as in Fig.~\ref{Fig:a}, but for $\sin^2 \theta_{12} = 0.359$
(see text for further details).
}
\end{figure}

 Given the fact that the magnitude of the ratio
$(\sin^2 \theta_{12} - \sin^2 \theta^{\nu}_{12})/\sin\theta_{13}$
determines the factor by which
$(\cos \delta)_{\rm E}$ and $(\cos \delta)_{\rm LO}$
(and $(\cos \phi)_{\rm E}$ and $(\cos\phi)_{\rm LO}$)
differ, we have checked how the results
described above change when
$\sin^2\theta_{12}$ is varied
in its 3$\sigma$ allowed region, eq.~(\ref{th12values}).
In Figs.~\ref{Fig:b} and \ref{Fig:c}
we show the dependence of the predicted values of
$(\cos \delta)_{\rm E}$, $(\cos \phi)_{\rm E}$
and $(\cos \delta)_{\rm LO} = (\cos\phi)_{\rm LO}$
on $\sin\theta_{13}$ for the minimal and maximal $3\sigma$
allowed values of $\sin^2\theta_{12}$,
$\sin^2\theta_{12} = 0.259$ and 0.359.
The results shown correspond to the
TBM, GRA, GRB, HG forms of  $\tilde{U}_{\nu}$.
For $\sin^2\theta_{12} = 0.259$
($\sin^2\theta_{12} = 0.359$)
and  $\sin^2 \theta_{13} = 0.0234$,
the ratio $(\sin^2 \theta_{12} - \sin^2 \theta^{\nu}_{12})/\sin\theta_{13}$
in the TBM, GRA, GRB and HG cases takes respectively
the values: $(-0.486)$, $(-0.114)$, $(-0.565)$ and $0.059$
($0.168$, $0.540$, $0.088$ and $0.713$).
As in the preceding case,
we give the predicted values of
$(\cos \delta)_{\rm E}$, $(\cos \phi)_{\rm E}$,
$(\cos \delta)_{\rm LO} = (\cos\phi)_{\rm LO}$,
and the ratios between them,
for $\sin^2 \theta_{12} = 0.259$ ($\sin^2 \theta_{12} = 0.359$)
and $\sin^2 \theta_{13} = 0.0234$
in Table~\ref{tab:2}  (Table~\ref{tab:3}).

\begin{figure}[h!]
  \begin{center}
     \hspace{-1.6cm}
   \subfigure
 {\includegraphics[width=12cm]{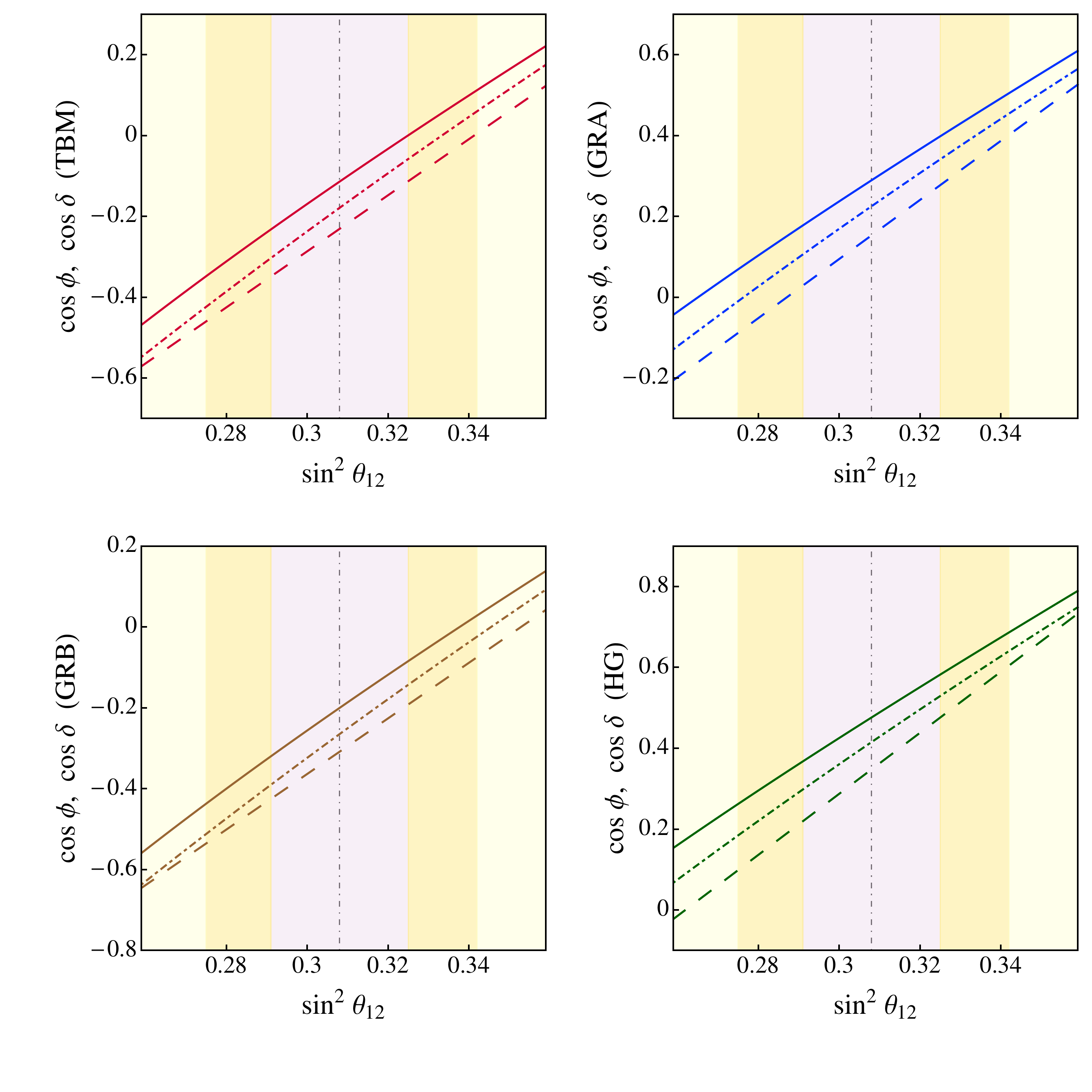}}
  \vspace{5mm}
     \end{center}
\vspace{-1.0cm} \it \caption{\label{Fig:d}
The same as in Fig.~\ref{Fig:a}, but for $\sin^2 \theta_{13} = 0.0234$
and varying $\sin^2 \theta_{12}$ in the $3\sigma$ range.
The vertical dash-dotted line corresponds to
the best fit value of $\sin^2 \theta_{12} = 0.308$
(see text for further details).
}
\end{figure}

 It follows from the results presented in Tables
~\ref{tab:1}~--~\ref{tab:3} that the exact sum rule
predictions of $\cos \delta$, $(\cos \delta)_{\rm E}$,
for the three values of $\sin^2 \theta_{12} = 0.308$, 0.259 and 0.359,
differ drastically. For the TBM form of $\tilde{U}_{\nu}$,
for instance, we get, respectively, the values:
$(\cos \delta)_{\rm E} = (-0.114)$, $(-0.469)$ and $0.221$.
For the GRA and GRB forms of $\tilde{U}_{\nu}$ we have, respectively,
$(\cos \delta)_{\rm E} = 0.289$, $(-0.044)$, $0.609$, and
$(\cos \delta)_{\rm E} = (-0.200)$, $(-0.559)$, $0.138$.
Similarly, for the HG form we find
for the three values of $\sin^2 \theta_{12}$:
$(\cos \delta)_{\rm E} = 0.476$, 0.153, 0.789.
Thus, in the cases of the symmetry forms of
$\tilde{U}_{\nu}$ considered, the exact sum rule
predictions for $\cos\delta$ not only change significantly
in magnitude when $\sin^2 \theta_{12}$ is
varied in its $3\sigma$ allowed
range, but also the sign of $\cos\delta$ changes
in the TBM, GRA and GRB cases
(see Fig.~\ref{Fig:d}).

%
%
%
\begin{table}[t!]
\centering
\begin{tabular}{ccccc}
 \toprule
  $\sin^2 \theta_{12} = 0.259$ & TBM & GRA & GRB & HG \\
 \midrule
$(\cos\delta)_{\rm E}$ & $ -0.469$ & $-0.0436$ & $-0.559$ & $0.153$ \\
$(\cos\delta)_{\rm LO}$ & $ -0.548$ & $-0.129$ & $-0.637$ & $0.0673$ \\
$(\cos\delta)_{\rm E}/(\cos\delta)_{\rm LO}$  & $ 0.855$ & $0.338$ & $0.878$ & $2.28$ \\
$(\cos\phi)_{\rm E}$ & $ -0.571$ & $-0.206$ & $-0.646$ & $-0.0225$ \\
$(\cos\delta)_{\rm E}/(\cos\phi)_{\rm E}$ & $0.821$ & $0.212$ & $0.866$ & $-6.82$ \\

$(\cos \phi)_{\rm E} /(\cos\phi)_{\rm LO}$  & $ 1.04$ & $1.59$ & $1.01$ & $-0.334$ \\
 \bottomrule
  \end{tabular}
\caption{The same as in Table \ref{tab:1}, but for $\sin^2 \theta_{12} = 0.259$.
\label{tab:2}
}
\end{table}

 We observe also that for $\sin^2 \theta_{12} = 0.259$,
the values of $\cos \delta$, obtained
using the exact sum rule eq.~(\ref{cosdthnua})
in the TBM, GRA, GRB and HG cases differ from those
calculated using the leading order
sum rule in eq.~(\ref{sth12cosd2}) by the factors
$0.855$, $0.338$, $0.878$ and $2.28$, respectively;
in the case of $\sin^2 \theta_{12} = 0.359$
the same  factors read:
$1.27$, $1.08$, $1.50$ and $1.05$.

\begin{table}[h!]
\centering
\begin{tabular}{ccccc}
 \toprule
  $\sin^2 \theta_{12} = 0.359$ & TBM & GRA & GRB & HG \\
 \midrule
$(\cos\delta)_{\rm E}$ & $ 0.221$ & $0.609$ & $0.138$ & $0.789$ \\
$(\cos\delta)_{\rm LO}$ & $ 0.175$ & $0.564$ & $0.092$ & $0.749$ \\
$(\cos\delta)_{\rm E}/(\cos\delta)_{\rm LO}$  & $ 1.27$ & $1.08$ & $1.50$ & $1.05$ \\
$(\cos\phi)_{\rm E}$ & $ 0.123$ & $0.526$ & $0.042$ & $0.733$ \\
$(\cos\delta)_{\rm E}/(\cos\phi)_{\rm E}$  & $1.80$ & $1.16$ & $3.29$ & $1.08$ \\
$(\cos\phi)_{\rm E}/(\cos\phi)_{\rm LO}$  & $ 0.702$ & $0.931$ & $0.456$ & $0.979$ \\
 \bottomrule
  \end{tabular}
\caption{The same as in Table \ref{tab:1}, but for $\sin^2 \theta_{12} = 0.359$.
\label{tab:3}
}
\end{table}

For $\sin^2 \theta_{12} = 0.259$,
the largest difference
between the exact and leading order
sum rule results for $\cos\delta$
occurs for the GRA and HG forms of $\tilde{U}_{\nu}$,
while if  $\sin^2 \theta_{12} = 0.359$,
the largest difference holds
for the TBM and GRB forms.

 As Figs. \ref{Fig:a}~--~\ref{Fig:c}
and Tables \ref{tab:1}~--~\ref{tab:3} show,
similar results are valid for $\cos\phi$
obtained from the exact and the leading order sum rules.

 It is worth noting also that the values
of $\cos \phi$ and $\cos \delta$, derived from
the respective exact sum rules
differ significantly for the TBM, GRA, GRB and HG
forms of $\tilde{U}_{\nu}$ considered.
As pointed out in \cite{Petcov:2014laa},
for the best fit values of $\sin^2 \theta_{13}$ and
$\sin^2 \theta_{12}$ they differ
by factors (1.4~--~2.0), as can be seen also
from Table~\ref{tab:1}. This
difference can be much larger for
$\sin^2 \theta_{12} = 0.259$ and 0.359:
for these two values of
$\sin^2 \theta_{12}$,
$\cos\delta$ and $\cos\phi$ differ
in the cases of the different symmetry forms
of interest approximately by factors
(1.2~--~6.8) and (1.1~--~3.3), respectively.

%
\section{\texorpdfstring{The Case of Nonzero $\theta^e_{23}$}{The Case of Nonzero theta-e23}}
%
%
 For $\theta^e_{23} = 0$ we have in the scheme we are
considering: $\theta_{23}\cong \pi/4 - 0.5\sin^2\theta_{13}$.
A nonzero value of $\theta^e_{23}$ allows for a significant
deviation of $\theta_{23}$ from $\pi/4$.  Such deviation
is not excluded by the current data on $\sin^2\theta_{23}$,
eq.~(\ref{th23values}): at $3\sigma$, values of $\sin^2\theta_{23}$
in the interval (0.37~--~0.64) are allowed, the best fit value
being $\sin^2\theta_{23} = 0.437~(0.455)$.
The exact sum rules for $\cos\delta$ and $\cos\phi$,
eqs.~(\ref{cosdthnu}), (\ref{s2th12cosdthnu}) and (\ref{s2th12cosphithnu}),
depend on $\theta_{23}$,
while the leading order sum rules,
eqs.~(\ref{AKth12th12nucosd}) and (\ref{th12th12nucosphi}),
are independent of $\theta_{23}$.
In this Section we are going to investigate how the dependence
on $\theta_{23}$ affects the predictions for  $\cos\delta$ and $\cos\phi$,
based on the exact sum rules.

 We note first that  from the exact sum rules
in eqs.~(\ref{s2th12cosdthnu}) and (\ref{s2th12cosphithnu}) we get
to leading order in $\sin\theta_{13}$:
\be
\sin^2 \theta_{12} = \sin^2 \theta^{\nu}_{12}  +
\frac{\sin2\theta_{12}}{\tan\theta_{23}} \sin\theta_{13} \cos \delta
+ O(\sin^2 \theta_{13})\,,
\label{th23LOcosd}
\ee
\be
\sin^2 \theta_{12} = \sin^2 \theta^{\nu}_{12}  +
\frac{\sin2\theta^{\nu}_{12}}{\tan\theta_{23}} \sin\theta_{13} \cos \phi
+ O(\sin^2 \theta_{13}) \,.
\label{th23LOcosphi}
\ee
%
It follows from eqs.~(\ref{th23hat}) and (\ref{s2th23})
that in the case of $|\sin\theta^e_{23}| \ll 1$ considered
in ref. \cite{Antusch:2005kw}, we have \cite{Petcov:2014laa}
$(\tan \theta_{23})^{-1} \cong 2\cos^2\theta_{23} = 1 + O(\sin\theta^e_{23})$.
Applying the approximation employed in ref. \cite{Antusch:2005kw},
in which terms of order of, or smaller than,
$\sin^2\theta_{13}$, $\sin^2\theta^e_{23}$ and
$\sin\theta_{13}\sin\theta^e_{23}$, in the sum rules of interest
are neglected, we have to set $(\tan \theta_{23})^{-1} = 1$
in eqs.~(\ref{th23LOcosd}) and (\ref{th23LOcosphi}).
This leads to eqs. (\ref{s2th12cosdthnu2})
and (\ref{s2th12cosphithnu2}) and, correspondingly,
to eqs.~(\ref{AKth12th12nucosd}) and (\ref{th12th12nucosphi}).

\begin{figure}[t!]
  \begin{center}
     \hspace{-1.6cm}
   \subfigure
 {\includegraphics[width=12cm]{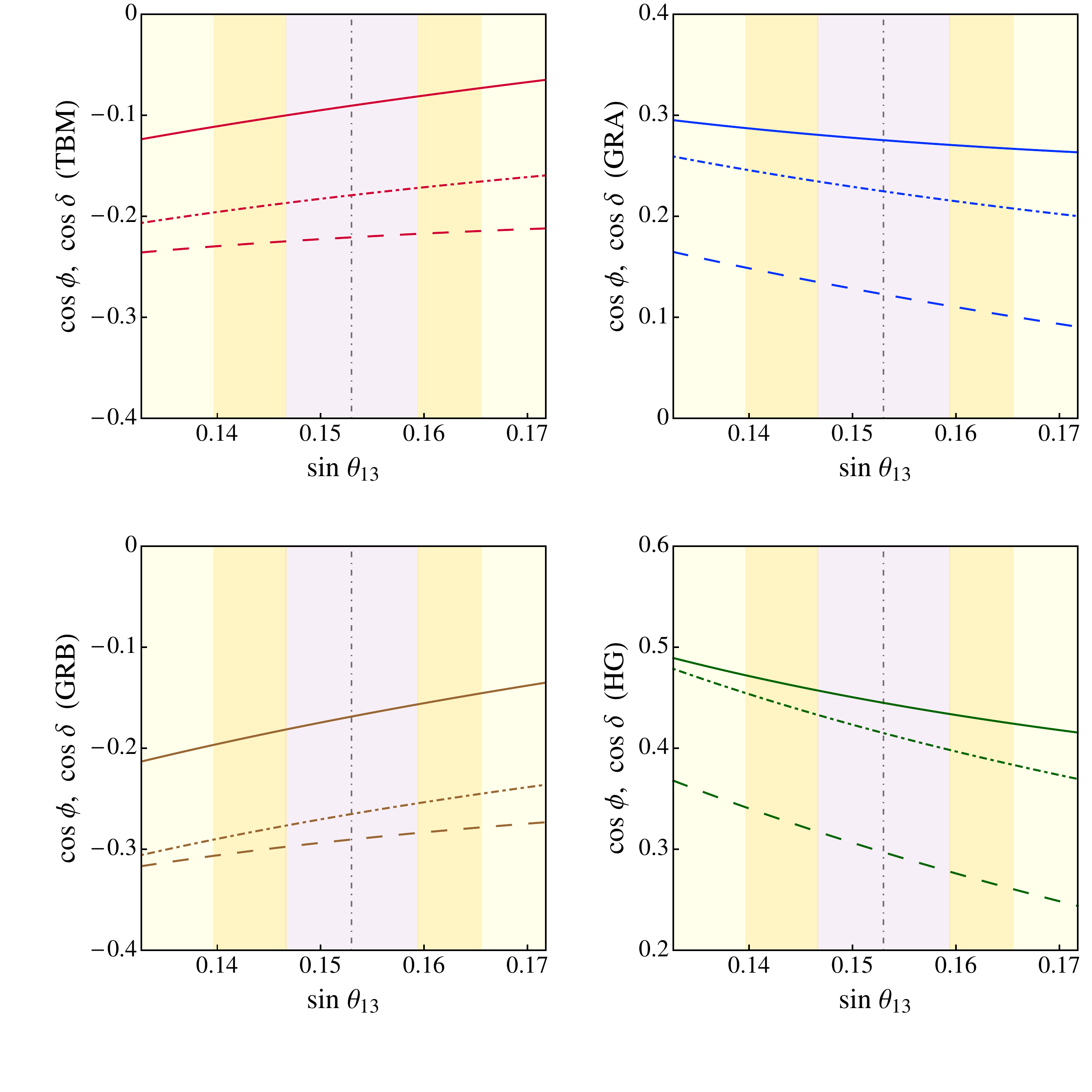}}
  \vspace{5mm}
     \end{center}
\vspace{-1.0cm} \it \caption{\label{Fig:1}
Predictions for $\cos\delta$ and $\cos \phi$
in the cases of the TBM (upper left panel),
GRA (upper right panel), GRB (lower left panel)
and HG (lower right panel)
forms of the matrix  $\tilde{U}_{\nu}$,
as functions of $\sin\theta_{13}$
and for the best fit values of
$\sin^2 \theta_{12} = 0.308$ and
 $\sin^2 \theta_{23} = 0.437$.
The solid lines (dashed lines) correspond
to $\cos\delta$ ($\cos \phi$)
determined from the exact sum rule
given in eq.~(\ref{s2th12cosdthnu}) (eq.~(\ref{s2th12cosphithnu})).
The dash-dotted line in each of the 4 panels
represents $(\cos \delta)_{\rm LO} = (\cos\phi)_{\rm LO}$
obtained from the leading order sum rule in
eq.~(\ref{sinth12cosd}) (eq.  (\ref{sth12cosphi10})).
The vertical dash-dotted line corresponds to the
best fit value of $\sin^2 \theta_{13} = 0.0234$;
the three coloured vertical bands indicate
the 1$\sigma$, 2$\sigma$ and 3$\sigma$ experimentally
allowed ranges of  $\sin\theta_{13}$
(see text for further details).
}
\end{figure}

  In Fig.~\ref{Fig:1} we show
the predictions for $\cos\delta$ and $\cos\phi$
in the cases of the TBM, GRA,
GRB and HG forms of the matrix $\tilde{U}_{\nu}$,
derived from the exact sum rules
in  eqs. (\ref{s2th12cosdthnu}) and (\ref{s2th12cosphithnu}),
$(\cos \delta)_{\rm E}$ (solid line)
and $(\cos \phi)_{\rm E}$ (dashed line), and
from the leading order sum rule in
eq.~(\ref{sinth12cosd}) (eq.~(\ref{sth12cosphi10})),
$(\cos \delta)_{\rm LO} = (\cos \phi)_{\rm LO}$ (dash-dotted line).
The results presented in  Fig.~\ref{Fig:1} are obtained
for the best fit values of $\sin^2 \theta_{12} = 0.308$ and
$\sin^2 \theta_{23} = 0.437$. The parameter $\sin^2 \theta_{13}$
is varied in its  3$\sigma$ allowed range,
eq.~(\ref{th13values}).
In Table~\ref{tab:4} we give the values of
$(\cos \delta)_{\rm E}$, $(\cos \delta)_{\rm LO}$,
$(\cos \phi)_{\rm E}$ and of their ratios,
corresponding to the best fit values
of  $\sin^2 \theta_{12}$, $\sin^2 \theta_{23}$
and  $\sin^2 \theta_{13}$.
We see from Table~\ref{tab:4} that for
the TBM, GRA, GRB and HG forms of $\tilde{U}_{\nu}$,
$\cos\delta$ determined from the exact sum rule
takes respectively the values
$(-0.091)$, $0.275$, $(-0.169)$ and $0.445$.
The values of $\cos \delta$, found using
the exact sum rule,
eq.~(\ref{s2th12cosdthnu}),
differ in the TBM, GRA, GRB and HG cases
from those calculated using the leading order
sum rule, eq.~(\ref{sinth12cosd}),
by the factors $0.506$, $1.22$, $0.636$ and $1.07$,
respectively. Thus, the largest difference
between the predictions of
the exact and the leading order sum rules occurs
for the TBM form of $\tilde{U}_{\nu}$.

\begin{figure}[h!]
  \begin{center}
     \hspace{-1.6cm}
   \subfigure
 {\includegraphics[width=12cm]{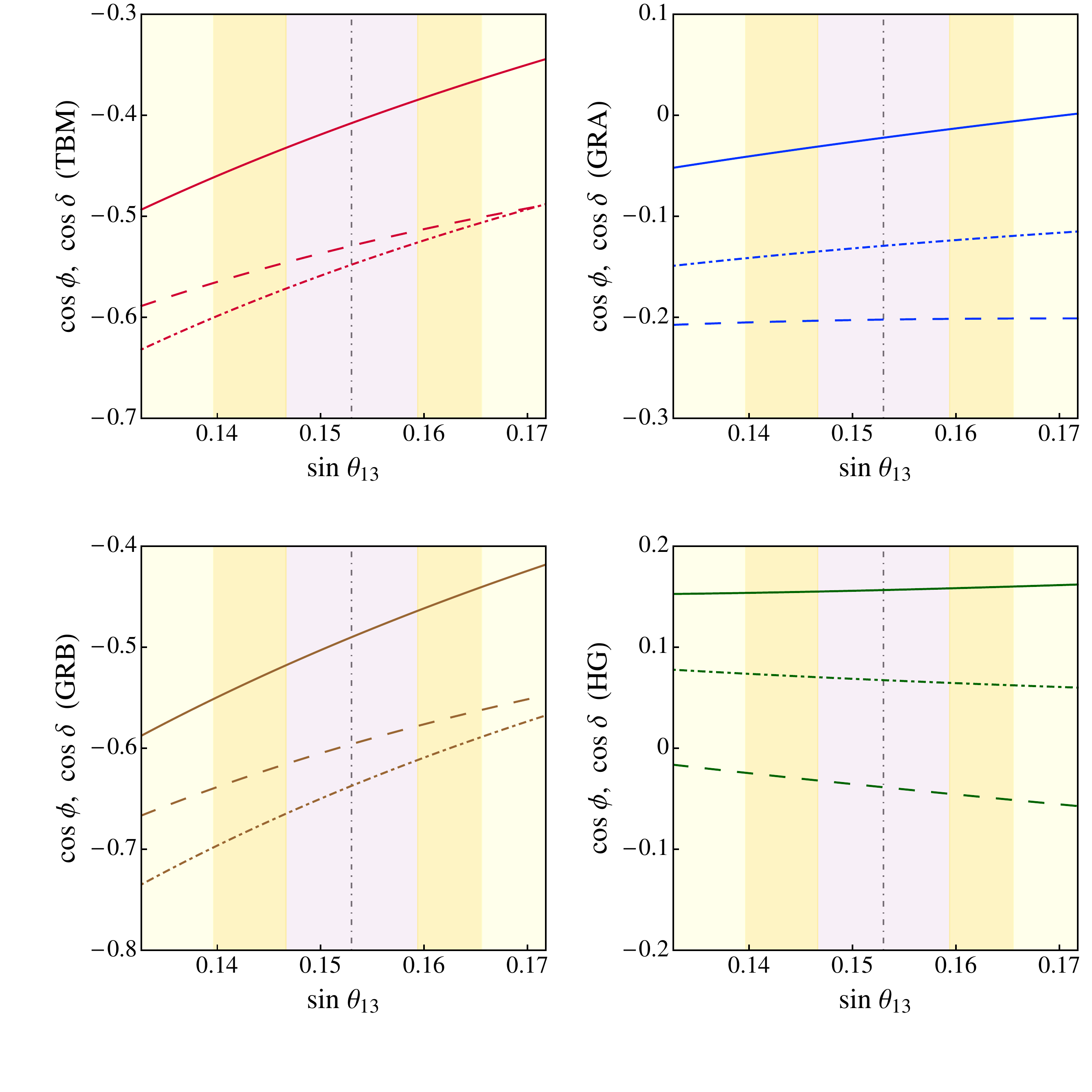}}
  \vspace{5mm}
     \end{center}
\vspace{-1.0cm} \it \caption{\label{Fig:2}
The same as in Fig. \ref{Fig:1}, but for
$\sin^2 \theta_{12} = 0.259$ (lower bound of the $3\sigma$ interval
in eq.~(\ref{th12values}))
and $\sin^2 \theta_{23} = 0.437$ (best fit value).
}
\end{figure}

\begin{figure}[h!]
  \begin{center}
     \hspace{-1.6cm}
   \subfigure
 {\includegraphics[width=12cm]{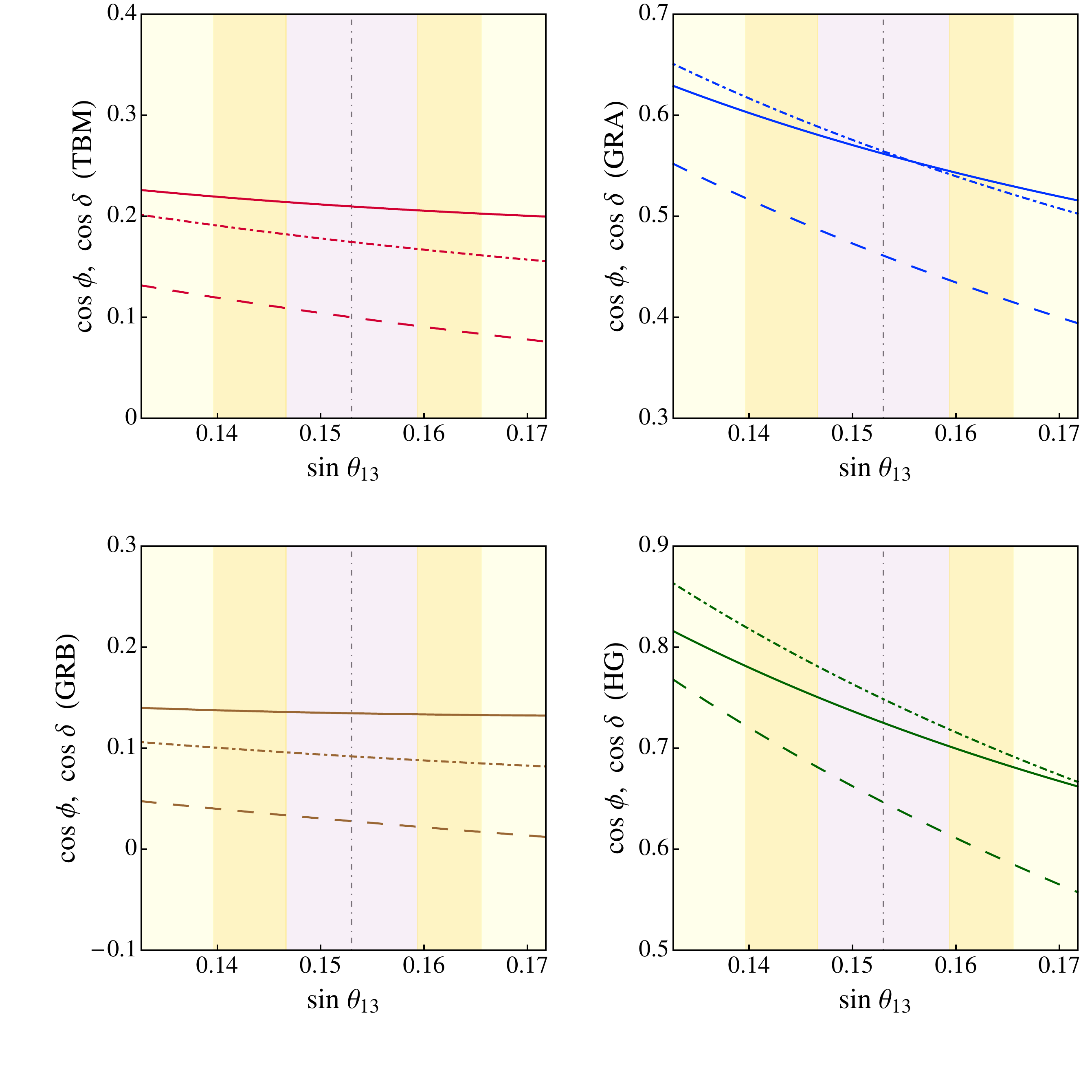}}
  \vspace{5mm}
     \end{center}
\vspace{-1.0cm} \it \caption{\label{Fig:3}
The same as in Fig. \ref{Fig:1}, but for
$\sin^2\theta_{12} = 0.359$  (upper bound of the $3\sigma$ interval
in eq.~(\ref{th12values}))
and $\sin^2\theta_{23} = 0.437$ (best fit value).
}
\end{figure}

\begin{figure}[h!]
  \begin{center}
     \hspace{-1.6cm}
   \subfigure
 {\includegraphics[width=12cm]{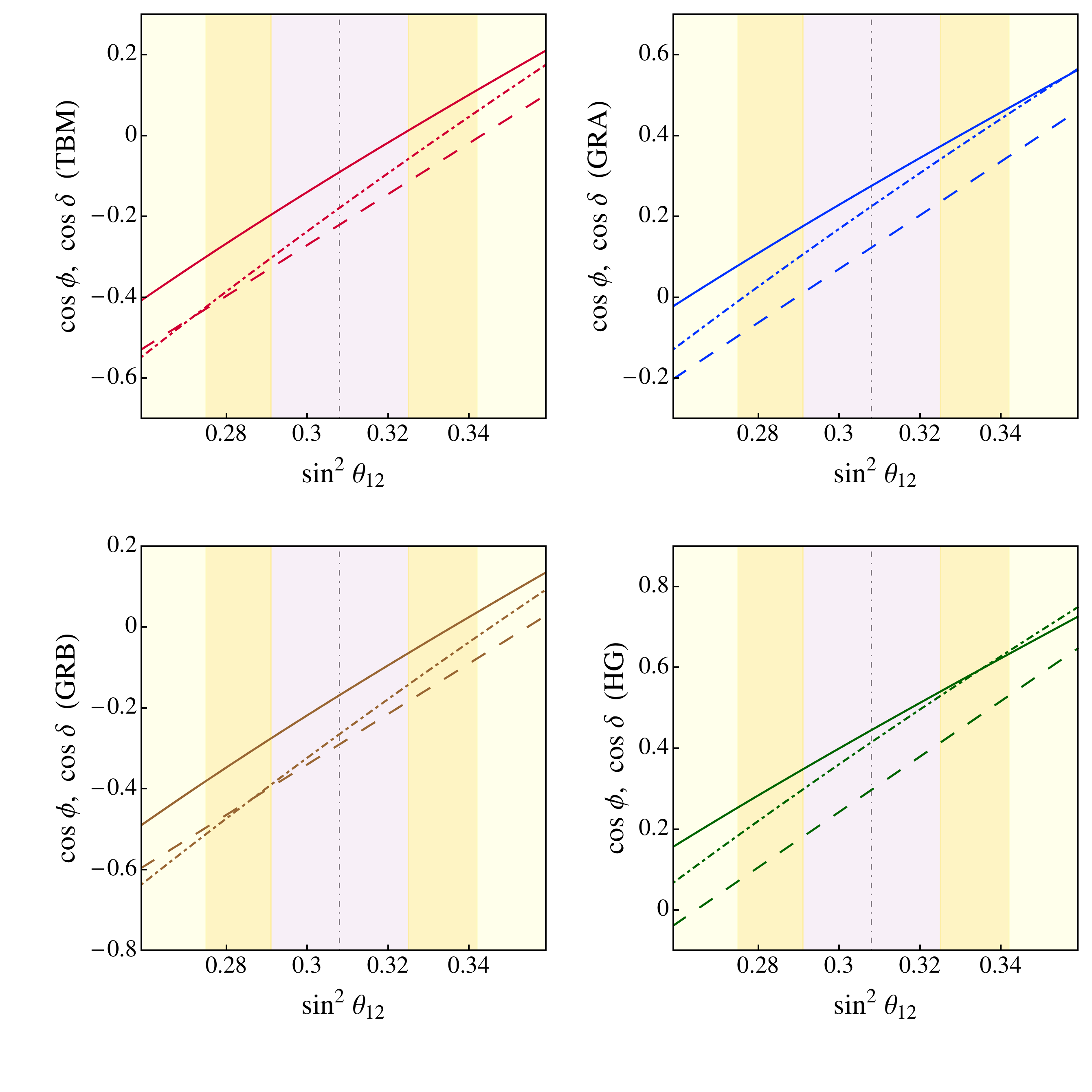}}
  \vspace{5mm}
     \end{center}
\vspace{-1.0cm} \it \caption{
\label{Fig:6}
The same as in Fig. \ref{Fig:1}, but for
$\sin^2 \theta_{13} = 0.0234$, $\sin^2 \theta_{23} = 0.437$ (best fit values)
and varying $\sin^2 \theta_{12}$ in the $3\sigma$ range.
The vertical dash-dotted line corresponds to
the best fit value of $\sin^2 \theta_{12} = 0.308$.
}
\end{figure}

  Since the predictions of the sum rules
depend on the value of $\theta_{12}$,
we show in Fig.~\ref{Fig:2} and Fig.~\ref{Fig:3}
also results for the values of  $\sin^2 \theta_{12}$,
corresponding to the lower
and the upper bounds of the $3 \sigma$ allowed
range of $\sin^2 \theta_{12}$, $\sin^2 \theta_{12} = 0.259$ and $0.359$,
keeping $\sin^2 \theta_{23}$ fixed to its best fit value.
The predictions for $(\cos \delta)_{\rm E}$,
$(\cos \phi)_{\rm E}$, $(\cos \delta)_{\rm LO} = (\cos \phi)_{\rm LO}$
and their ratios, obtained
for the best fit values of $\sin^2 \theta_{13} = 0.0234$ and
$\sin^2 \theta_{23} = 0.437$, and
for $\sin^2 \theta_{12} = 0.259$ ($\sin^2 \theta_{12} = 0.359$)
are given in Table~\ref{tab:5} (Table~\ref{tab:6}).
For  $\sin^2 \theta_{12} = 0.259$,
the exact sum rule predictions of $\cos\delta$
for the TBM, GRA, GRB and HG forms of $\tilde{U}_{\nu}$ read
(see Table~\ref{tab:5}):
$(\cos \delta)_{\rm E} = (-0.408)$, $(-0.022)$, $(-0.490)$ and $0.156$.
As in the case of negligible $\theta^e_{23}$
analysed in the preceding Section,
these values differ drastically (in general,
both in magnitude and sign)
from the exact sum rule values of $\cos\delta$
corresponding to the best fit value and the
$3\sigma$ upper bound of $\sin^2 \theta_{12} = 0.308$ and $0.359$.
The dependence of $(\cos \delta)_{\rm E}$,
$(\cos \delta)_{\rm LO}$ and $(\cos \phi)_{\rm E}$
on $\sin^2\theta_{12}$ under discussion
is shown graphically in Fig.~\ref{Fig:6}.

 Further, for  $\sin^2 \theta_{12} = 0.259$,
the ratio $(\cos \delta)_{\rm E}/(\cos \delta)_{\rm LO}$
in the TBM, GRA, GRB and HG
cases reads, respectively,
$0.744$, $0.172$, $0.769$ and $2.32$ (see Table~\ref{tab:5}).
Thus, the predictions for $\cos\delta$
of the exact and the leading order
sum rules differ by the factors
of $5.8$ and 2.3 in the GRA and HG cases.
For the upper bound of the $3 \sigma$ range
of $\sin^2 \theta_{12} = 0.359$,
the ratio  $(\cos \delta)_{\rm E}/(\cos \delta)_{\rm LO}$
takes the values $1.2$, $0.996$, $1.46$ and $0.969$
for the TBM, GRA, GRB and HG forms of $\tilde{U}_{\nu}$,
respectively (see Table~\ref{tab:6}).
For the GRA and HG symmetry forms
the leading order sum rule prediction for
$\cos\delta$ is very close to the exact sum rule prediction,
which can also be seen in Fig.~\ref{Fig:3}.

\begin{figure}[h!]
  \begin{center}
     \hspace{-1.6cm}
   \subfigure
 {\includegraphics[width=12cm]{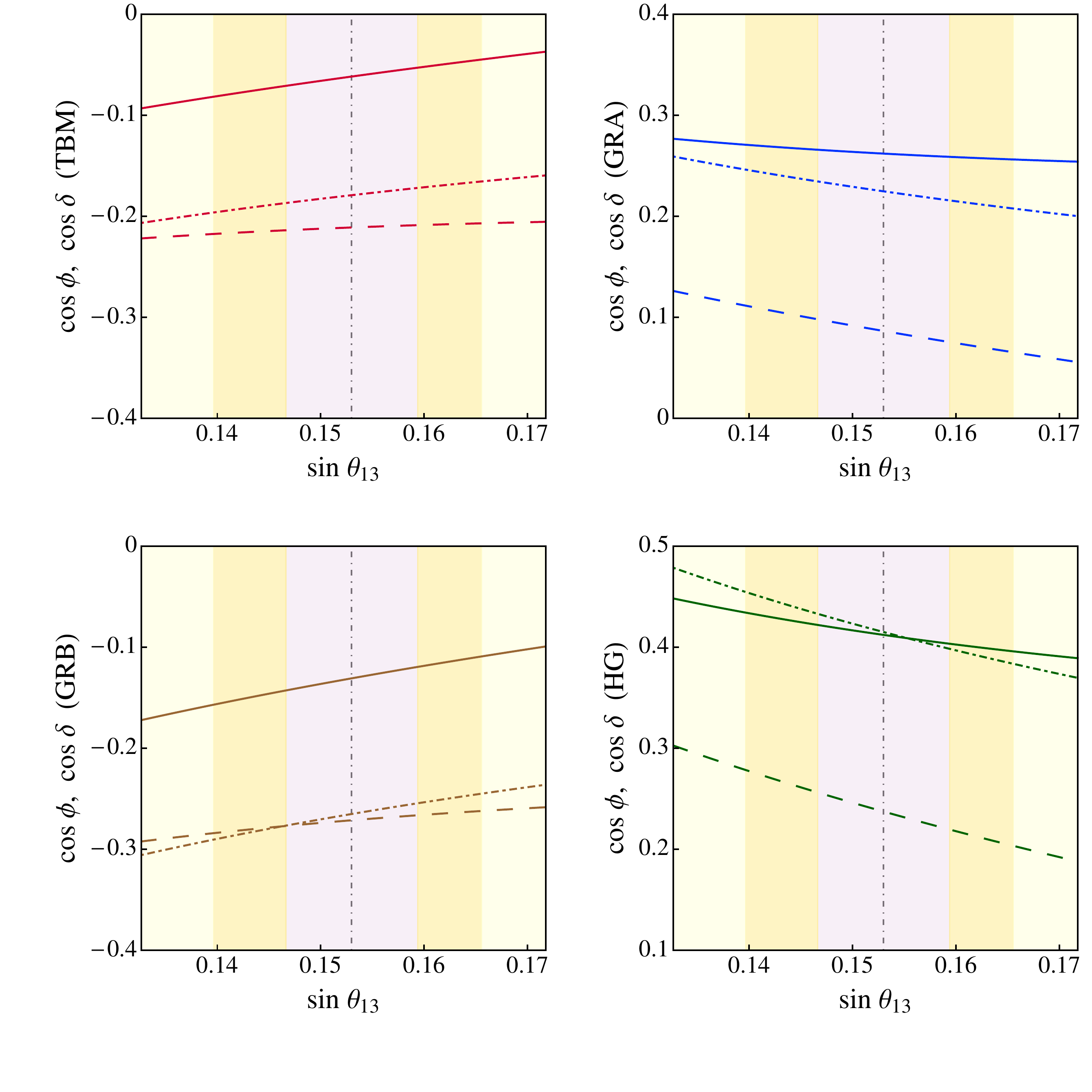}}
  \vspace{5mm}
     \end{center}
\vspace{-1.0cm} \it \caption{
\label{Fig:4}
The same as in Fig. \ref{Fig:1}, but for
$\sin^2 \theta_{12} = 0.308$ (best fit value) and
$\sin^2 \theta_{23} = 0.374$
(lower bound of the $3\sigma$ interval
in eq.~(\ref{th23values})).
}
\end{figure}

\begin{figure}[h!]
  \begin{center}
     \hspace{-1.6cm}
   \subfigure
 {\includegraphics[width=12cm]{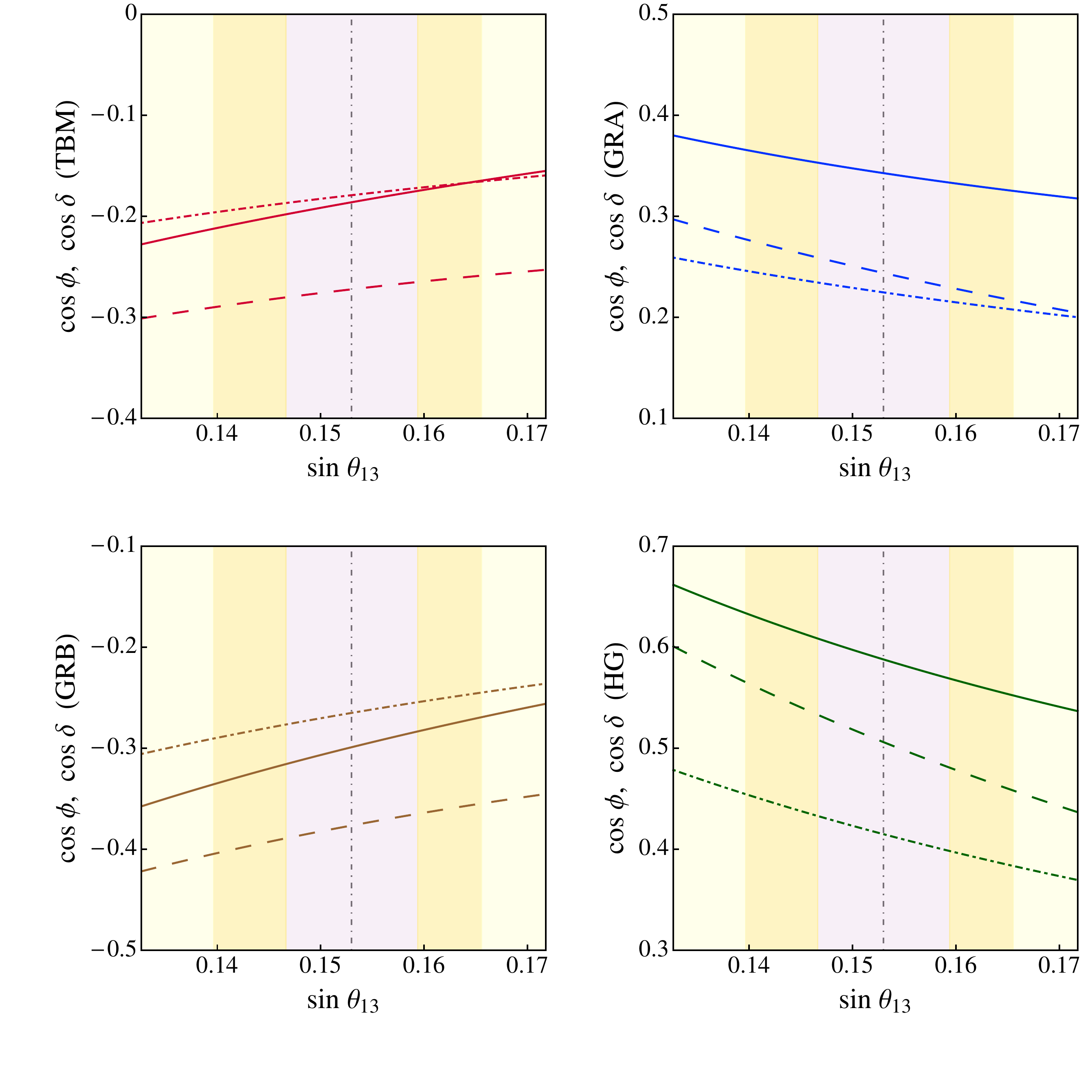}}
  \vspace{5mm}
     \end{center}
\vspace{-1.0cm} \it \caption{
\label{Fig:5}
The same as in Fig. \ref{Fig:1}, but for
$\sin^2 \theta_{12} = 0.308$ (best fit value) and $\sin^2 \theta_{23} = 0.626$
(upper bound of the $3\sigma$ interval
in eq.~(\ref{th23values})).
}
\end{figure}

 We will investigate next the dependence of the
predictions for $\cos\delta$ and $\cos\phi$
on the value of $\theta_{23}$ given the facts that
i) $\sin^2 \theta_{23}$ is determined experimentally with
a relatively large uncertainty, and
ii) in contrast to the leading order sum rule predictions
for $\cos\delta$ and $\cos\phi$,
the exact sum rule predictions depend on $\theta_{23}$.
In Figs.~\ref{Fig:4} and \ref{Fig:5}
we show the dependence of predictions for
$\cos\delta$ and $\cos\phi$ on  $\sin \theta_{13}$
for the best fit
value of $\sin^2 \theta_{12} = 0.308$
and the $3\sigma$ lower and upper bounds
of $\sin^2 \theta_{23} = 0.374$ and 0.626, respectively.
For $\sin^2 \theta_{23} = 0.374~(0.626)$ and
the best fit values of $\sin^2\theta_{13}$ and
$\sin^2 \theta_{12}$, the exact and the leading order
sum rule results $(\cos\delta)_{\rm E}$,
$(\cos\phi)_{\rm E}$, $(\cos\delta)_{\rm LO} =
(\cos\phi)_{\rm LO}$ and their ratios are given in
Tables \ref{tab:7} and \ref{tab:8}.
Comparing the values of  $(\cos\delta)_{\rm E}$
quoted in Tables \ref{tab:7} and \ref{tab:8}
with the values given in  Table \ref{tab:4} we note
that the exact sum rule predictions for $\cos\delta$
for  $\sin^2 \theta_{23} = 0.374$ (lower $3\sigma$ bound)
and $\sin^2 \theta_{23} = 0.437$ (best fit value)
do not differ significantly in the cases of the
TBM, GRA, GRB and HG forms of $\tilde{U}_{\nu}$
considered. However, the differences between
the predictions for  $\sin^2 \theta_{23} = 0.437$ and
$\sin^2 \theta_{23} = 0.626$ are rather large~---~by
factors of 2.05, 1.25, 1.77 and 1.32
in the TBM, GRA, GRB and HG cases, respectively.

In what concerns the difference between the
exact and leading order sum rules predictions
for $\cos\delta$, for the best fit values of
$\sin^2 \theta_{13}$ and  $\sin^2 \theta_{12}$,
and for  $\sin^2 \theta_{23} = 0.374$,
the ratio $(\cos \delta)_{\rm E}/(\cos \delta)_{\rm LO} =
0.345$, $1.17$, $0.494$ and $0.993$ for TBM, GRA, GRB and HG
forms of $\tilde{U}_{\nu}$. For $\sin^2 \theta_{23} = 0.626$
we have for the same ratio
$(\cos \delta)_{\rm E}/(\cos \delta)_{\rm LO} =
1.04$, $1.52$, $1.13$ and $1.42$.
Thus, for  $\sin^2 \theta_{23} = 0.374$ ($0.626$),
the leading order sum rule prediction for
$\cos\delta$ is rather precise in the HG (TBM) case.
For the other symmetry forms of $\tilde{U}_{\nu}$
the leading order sum rule prediction for
$\cos\delta$ is largely incorrect.
As can be seen from Figs \ref{Fig:1}~--~\ref{Fig:5}
and Tables \ref{tab:4}~--~\ref{tab:8},
we get similar results for $\cos\phi$.

In the case of the BM (LC) form of $\tilde{U}_{\nu}$,
physical values of $(\cos \delta)_{\rm E}$, $(\cos \phi)_{\rm E}$
and $(\cos \delta)_{\rm LO}$ can be obtained
for the best fit values of $\sin^2 \theta_{13}$
and $\sin^2 \theta_{23}$ if  $\sin^2 \theta_{12}$
has a relatively large value.
For, e.g., $\sin^2 \theta_{12} = 0.359$,
$\sin^2 \theta_{13} = 0.0234$ and $\sin^2 \theta_{23} = 0.437$
we find $(\cos \delta)_{\rm E} = -0.821$, $(\cos \delta)_{\rm LO} = -0.998$,
$(\cos \phi)_{\rm E} = -0.837$, and
 $(\cos \delta)_{\rm E} / (\cos \delta)_{\rm LO} = 0.823$.
%
\begin{table}[t!]
\centering
\begin{tabular}{ccccc}
 \toprule
  $(\sin^2 \theta_{12},\sin^2 \theta_{23})= (0.308,0.437)$ & TBM & GRA & GRB & HG \\
 \midrule
$(\cos \delta)_{\rm E}$ & $ -0.0906$ & $0.275$ & $-0.169$ & $0.445$ \\
$(\cos \delta)_{\rm LO}$ & $ -0.179$ & $0.225$ & $-0.265$ & $0.415$ \\
$(\cos\delta)_{\rm E}/(\cos\delta)_{\rm LO}$ &$0.506$ &$1.22$ &$0.636$ &$1.07$ \\
$(\cos \phi)_{\rm E}$ & $ -0.221$ & $0.123$ & $-0.290$ & $0.297$ \\
$(\cos\delta)_{\rm E}/(\cos\phi)_{\rm E}$ & $0.41$ & $2.24$ & $0.581$ & $1.50$ \\
$(\cos\phi)_{\rm E}/(\cos\phi)_{\rm LO}$ & $ 1.23$ & $0.547$ &$1.10$ &$0.716$\\
 \bottomrule
  \end{tabular}
\caption{
The predicted values of
$\cos\delta$ and $\cos \phi$,
obtained from the exact sum rules
in eqs.~(\ref{s2th12cosdthnu}) and (\ref{s2th12cosphithnu}),
$(\cos \delta)_{\rm E}$ and $(\cos \phi)_{\rm E}$,
and from the leading order sum rule in
eq.~(\ref{sinth12cosd}) (eq.  (\ref{sth12cosphi10})),
 $(\cos \delta)_{\rm LO} = (\cos \phi)_{\rm LO}$,
using the best fit values of
$\sin^2 \theta_{13} = 0.0234$,
$\sin^2 \theta_{12} = 0.308$ and
$\sin^2 \theta_{23} = 0.437$,
for the  TBM, GRA,  GRB  and HG forms of the matrix
$\tilde{U}_{\nu}$. The values of the ratios
$(\cos\delta)_{\rm E}/(\cos\delta)_{\rm LO}$,
$(\cos\delta)_{\rm E}/(\cos\phi)_{\rm E}$ and
$(\cos\phi)_{\rm E}/(\cos\phi)_{\rm LO}$ are also shown.
\label{tab:4}
}
\end{table}

%
\begin{table}[h!]
\centering
\begin{tabular}{ccccc}
 \toprule
  $(\sin^2 \theta_{12},\sin^2 \theta_{23})= (0.259,0.437)$ & TBM & GRA & GRB & HG \\
 \midrule
$(\cos \delta)_{\rm E}$ & $ -0.408$ & $-0.0223$ & $-0.490$ & $0.156$ \\
$(\cos \delta)_{\rm LO}$ & $ -0.548$ & $-0.129$ & $-0.637$ & $0.0673$ \\
$(\cos \delta)_{\rm E} / (\cos \delta)_{\rm LO}$  & $ 0.744$ & $0.172$ & $0.769$ & $2.32$ \\
$(\cos \phi)_{\rm E}$ & $ -0.529$ & $-0.202$ & $-0.596$ & $-0.0386$ \\
$(\cos \delta)_{\rm E} / (\cos \phi)_{\rm E}$  & $0.771$ & $0.110$ & $0.822$ & $-4.05$ \\
$(\cos\phi)_{\rm E}/(\cos\phi)_{\rm LO}$  & $ 0.966$ & $1.57$ & $0.935$ & $-0.573$ \\
 \bottomrule
  \end{tabular}
\caption{The same as in Table \ref{tab:4}, but
for $\sin^2 \theta_{13} = 0.0234$ (best fit value),
$\sin^2 \theta_{12} = 0.259$ (lower bound of the $3\sigma$ range)
and $\sin^2 \theta_{23} = 0.437$ (best fit value).
\label{tab:5}
}
\end{table}

\begin{table}[h!]
\centering
\begin{tabular}{ccccc}
 \toprule
  $(\sin^2 \theta_{12},\sin^2 \theta_{23})= (0.359,0.437)$ & TBM & GRA & GRB & HG \\
 \midrule
$(\cos \delta)_{\rm E}$ & $ 0.210$ & $0.562$ & $0.135$ & $0.725$ \\
$(\cos \delta)_{\rm LO}$ & $ 0.175$ & $0.564$ & $0.092$ & $0.749$ \\
$(\cos \delta)_{\rm E} / (\cos \delta)_{\rm LO}$  & $ 1.20$ & $0.996$ & $1.46$ & $0.969$ \\
$(\cos \phi)_{\rm E}$ & $ 0.100$ & $0.461$ & $0.0279$ & $0.647$ \\
$(\cos \delta)_{\rm E} / (\cos \phi)_{\rm E}$  & $2.09$ & $1.22$ & $4.83$ & $1.12$ \\
$(\cos \phi)_{\rm E} / (\cos \phi)_{\rm LO}$  & $ 0.573$ & $0.817$ & $0.303$ & $0.864$ \\
 \bottomrule
  \end{tabular}
\caption{
The same as in Table \ref{tab:4}, but
for $\sin^2 \theta_{13} = 0.0234$ (best fit value),
$\sin^2 \theta_{12} = 0.359$ (upper bound of the $3 \sigma$ range)
and $\sin^2 \theta_{23} = 0.437$ (best fit value).
\label{tab:6}
}
\end{table}

%

\begin{table}[h!]
\centering
\begin{tabular}{ccccc}
  \toprule
  $(\sin^2 \theta_{12},\sin^2 \theta_{23})= (0.308,0.374)$ & TBM & GRA & GRB & HG \\
 \midrule
$(\cos \delta)_{\rm E}$ & $ -0.0618$ & $0.262$ & $-0.131$ & $0.412$ \\
$(\cos \delta)_{\rm LO}$ & $ -0.179$ & $0.225$ & $-0.265$ & $0.415$ \\
$(\cos \delta)_{\rm E} / (\cos \delta)_{\rm LO}$  & $ 0.345$ & $1.17$ & $0.494$ & $0.993$ \\
$(\cos \phi)_{\rm E}$ & $ -0.211$ & $0.0866$ & $-0.271$ & $0.237$ \\
$(\cos \delta)_{\rm E} / (\cos \phi)_{\rm E}$  & $0.293$ & $3.03$ & $0.483$ & $1.74$ \\
$(\cos \phi)_{\rm E} / (\cos \phi)_{\rm LO}$  & $ 1.18$ & $0.385$ & $1.02$ & $0.572$ \\
 \bottomrule
  \end{tabular}
\caption{
The same as in Table \ref{tab:4}, but
for $\sin^2 \theta_{13} = 0.0234$ (best fit value),
$\sin^2 \theta_{12} = 0.308$ (best fit value)
and $\sin^2 \theta_{23} = 0.374$ (lower bound of the $3 \sigma$ range).
\label{tab:7}
}
\end{table}

%
\begin{table}[h!]
\centering
\begin{tabular}{ccccc}
 \toprule
  $(\sin^2 \theta_{12},\sin^2 \theta_{23})= (0.308,0.626)$ & TBM & GRA & GRB & HG \\
 \midrule
$(\cos \delta)_{\rm E}$ & $ -0.186$ & $0.343$ & $-0.299$ & $0.588$ \\
$(\cos \delta)_{\rm LO}$ & $ -0.179$ & $0.225$ & $-0.265$ & $0.415$ \\
$(\cos \delta)_{\rm E} / (\cos \delta)_{\rm LO}$  & $ 1.04$ & $1.52$ & $1.13$ & $1.42$ \\
$(\cos \phi)_{\rm E}$ & $ -0.272$ & $0.244$ & $-0.376$ & $0.506$ \\
$(\cos\delta)_{\rm E}/(\cos\phi)_{\rm E}$  & $0.684$ & $1.41$ & $0.794$ & $1.16$ \\
$(\cos \phi)_{\rm E} / (\cos\phi)_{\rm LO}$  & $ 1.52$ & $1.09$ & $1.42$ & $1.22$ \\
 \bottomrule
  \end{tabular}
\caption{
The same as in Table \ref{tab:4}, but
for $\sin^2 \theta_{13} = 0.0234$ (best fit value),
$\sin^2 \theta_{12} = 0.308$ (best fit value)
and $\sin^2 \theta_{23} = 0.626$ (upper bound of the $3 \sigma$ range).
\label{tab:8}
}
\end{table}

\newpage

\section{Statistical Analysis}
\label{sec:5}
 In the present Section we perform a statistical analysis
of the predictions for $\delta$, $\cos\delta$ and the rephasing
invariant $J_{\rm CP}$ which controls the magnitude
of CPV effects in neutrino oscillations \cite{PKSP3nu88},
in the cases of the TBM, BM (LC), GRA, GRB and HG symmetry
forms of the matrix $\tilde{U}_{\nu}$ (see eq.~(\ref{Unu1})).
In this analysis we use as input
the latest results on $\sin^2\theta_{12}$, $\sin^2\theta_{13}$,
$\sin^2\theta_{23}$ and $\delta$, obtained in the global analysis of the
neutrino oscillation data performed in \cite{Capozzi:2013csa}.
Our goal is to derive the allowed ranges
for $\delta$, $\cos\delta$ and $J_{\rm CP}$,
predicted on the basis of the current data on
the neutrino mixing parameters for each of the
symmetry forms of $\tilde{U}_{\nu}$ considered.
We recall that in the standard parametrisation of the PMNS matrix,
the $J_{\rm CP}$ factor reads (see, e.g., \cite{PDG2014}):
\be
J_{\rm CP} = {\rm Im} \left\{ U^*_{e1} U^*_{\mu 3} U_{e3} U_{\mu 1} \right\}
= \frac{1}{8} \sin \delta \sin 2\theta_{13} \sin 2\theta_{23}
\sin 2\theta_{12} \cos \theta_{13} \,.
\ee
%

 We construct $\chi^2$ for the schemes considered~---~TBM,
BM (LC), GRA, GRB and HG~---~as described in Appendix \ref{AppendixB}.
We will focus on the general case of non-vanishing $\theta^e_{23}$
in order to allow for possible sizeable deviations of
$\theta_{23}$ from the symmetry value $\pi / 4$.

\begin{figure}[h!]
  \begin{center}
     \hspace{-1.6cm}
   \subfigure
 {\includegraphics[width=14cm]{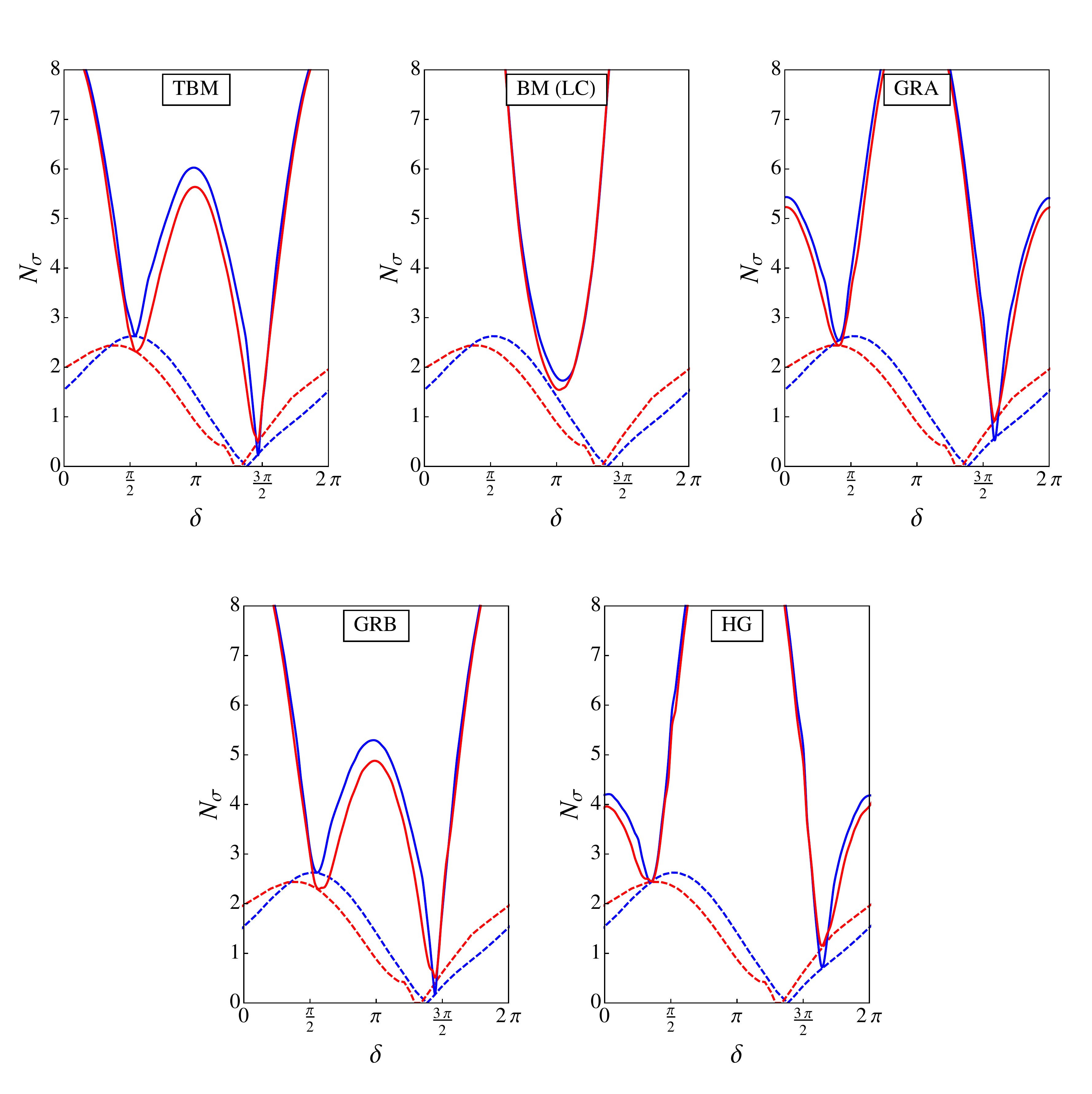}}
  \vspace{5mm}
     \end{center}
\vspace{-1.0cm} \caption{\label{Fig:chi2delta}
$N_{\sigma} \equiv \sqrt{\chi^2}$ as a function of $\delta$.
The dashed lines represent the results of the global fit
\cite{Capozzi:2013csa}, while the solid lines represent the
results we obtain for the TBM, BM (LC), GRA
(upper left, central, right panels),
GRB and HG (lower left and right panels)
symmetry forms of $\tilde{U}_{\nu}$. The
blue (red) lines are for NO (IO)
neutrino mass spectrum (see text for further details).
}
\end{figure}

 In the five panels in Fig.~\ref{Fig:chi2delta} we show
$N_{\sigma} \equiv \sqrt{\chi^2}$
as a function of $\delta$ for the five symmetry forms of
$\tilde{U}_{\nu}$ we have studied.
The dashed lines correspond to the results of the global fit
\cite{Capozzi:2013csa}. The solid lines represent the
results we obtain
by minimising the value of $\chi^2$
in $\sin^2\theta_{13}$ and $\sin^2\theta_{23}$
(or, equivalently,
in $\sin^2\theta^e_{12}$ and $\sin^2\hat \theta_{23}$)
for a fixed value of $\delta$~
\footnote{We note that in the scheme considered
by us, fixing the value of $\delta$ implies
that one of the three neutrino mixing angles
is expressed in terms of the other two.
We choose for convenience
this angle to be $\theta_{12}$.}.
The blue (red) lines correspond to NO (IO)
neutrino mass spectrum.
The value of $\chi^2$ at the minimum, $\chi^2_{\rm min}$,
which determines the best fit value of
$\delta$ predicted for each symmetry
form of $\tilde{U}_{\nu}$,
allows us to make conclusions about the
compatibility of a given symmetry form of
$\tilde{U}_{\nu}$ with the current global
neutrino oscillation data.

 It follows from the results shown in
Fig.~\ref{Fig:chi2delta} that
the BM (LC) symmetry form is disfavoured
by the data at approximately $1.8\sigma$,
all the other symmetry forms considered
being compatible with the data.
We note that for the TBM, GRA,
GRB and HG  symmetry forms, a value of $\delta$
in the vicinity of 3$\pi/2$ is preferred statistically.
For the TBM symmetry form this result was
first obtained in \cite{Marzocca:2013cr}
while for the GRA,
GRB and HG symmetry forms it was first
found in \cite{Petcov:2014laa}.
In contrast, in the case of the BM (LC) form
the best fit value is very close to $\pi$
\cite{Marzocca:2013cr,Petcov:2014laa}.
The somewhat larger value of $\chi^2$
at the second local minimum
in the vicinity of $\pi/2$ in the TBM, GRA,
GRB and HG cases, is a consequence of the
fact that the best fit value of $\delta$ obtained in the
global analysis of the current neutrino oscillation data
is close to $3\pi/2$ and that the value of $\delta = \pi/2$
is statistically disfavoured (approximately at $2.5\sigma$).
In the absence of any information on $\delta$,
the two minima would have exactly the
same value of $\chi^2$, because they correspond to the same
value of $\cos \delta$.  In the schemes considered, as we
have discussed,  $\cos \delta$ is determined by the values of $\theta_{12}$,
$\theta_{13}$ and  $\theta_{23}$.
The degeneracy in the sign of $\sin \delta$ can only be solved
by an experimental input on $\delta$.
In Table~\ref{table:1223} we give the best fit values of $\delta$
and the corresponding $3\sigma$ ranges
for the TBM, BM (LC), GRA, GRB and HG forms of $\tilde{U}_{\nu}$,
found by fixing $\sqrt{\chi^2-\chi^2_{\rm min}} = 3$.

\begin{figure}[h!]
  \begin{center}
   \subfigure
 {\includegraphics[width=8cm]{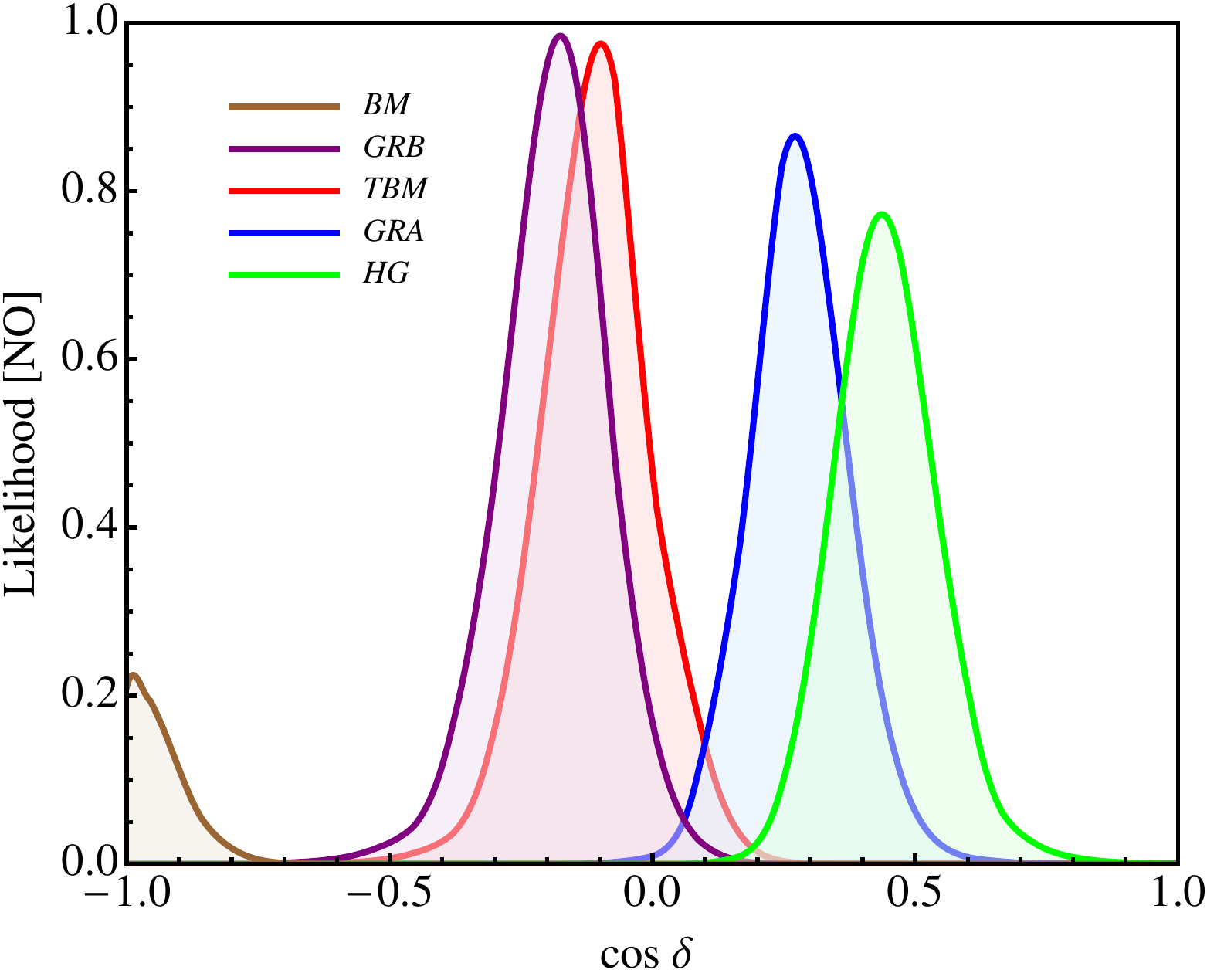}}
   \vspace{5mm}
     \end{center}
\vspace{-1.0cm} \caption{\label{Fig:cosdeltaNO}
The likelihood function 
versus $\cos \delta$ for NO
neutrino mass spectrum after marginalising over
$\sin^2\theta_{13}$ and $\sin^2\theta_{23}$
for the TBM, BM (LC), GRA, GRB and HG symmetry forms
of the mixing matrix $\tilde{U}_{\nu}$ (see text for further details).
}
\end{figure}

 In Fig.~\ref{Fig:cosdeltaNO} we show the
likelihood function versus $\cos \delta$ for NO
neutrino mass spectrum. The results shown are obtained
by marginalising over all the
other relevant parameters of the scheme considered
(see Appendix \ref{AppendixB} for details).
The dependence of the likelihood function on $\cos \delta$
in the case of IO neutrino mass spectrum
differs little from that shown
in Fig.~\ref{Fig:cosdeltaNO}.
Given the global fit results, the likelihood function, i.e.,
\be
L(\cos \delta) \propto \exp \left(- \frac{\chi^2 (\cos \delta)}{2} \right) \,,
\ee
%
represents the most probable value of $\cos \delta$ for
each of the considered symmetry forms of $\tilde U_{\nu}$.
The $n \sigma$ confidence level region corresponds to the
interval of values of $\cos \delta$ in which
$L(\cos \delta) \geq L(\chi^2 = \chi^2_{\rm min}) \cdot L(\chi^2 = n^2)$.

As can be observed from Fig.~\ref{Fig:cosdeltaNO},
a rather precise measurement of $\cos \delta$ would allow one
to distinguish between the different symmetry forms of $\tilde U_{\nu}$
considered by us. For the TBM and GRB forms
there is a significant overlap of the corresponding likelihood functions.
The same observation is valid for the GRA and HG forms.
However, the overlap of the likelihood functions of these
two groups of symmetry forms occurs only at $3\sigma$ level
in a very small interval of values of $\cos \delta$, as can also
be seen from Table~\ref{table:1223}.
This implies that in order to distinguish
between TBM/GRB, GRA/HG and BM symmetry forms a
not very demanding measurement (in terms of accuracy) of $\cos \delta$
might be sufficient. The value of the non-normalised likelihood
function at the maximum in Fig.~\ref{Fig:cosdeltaNO}
is equal to $\exp( - \chi^2_{\rm min}/2)$, which
allows us to make conclusions
about the compatibility of the symmetry schemes
with the current global data,
as has already been pointed out.

\begin{figure}[h!]
  \begin{center}
     \hspace{-0.9cm}
   \subfigure
  {\includegraphics[width=8cm]{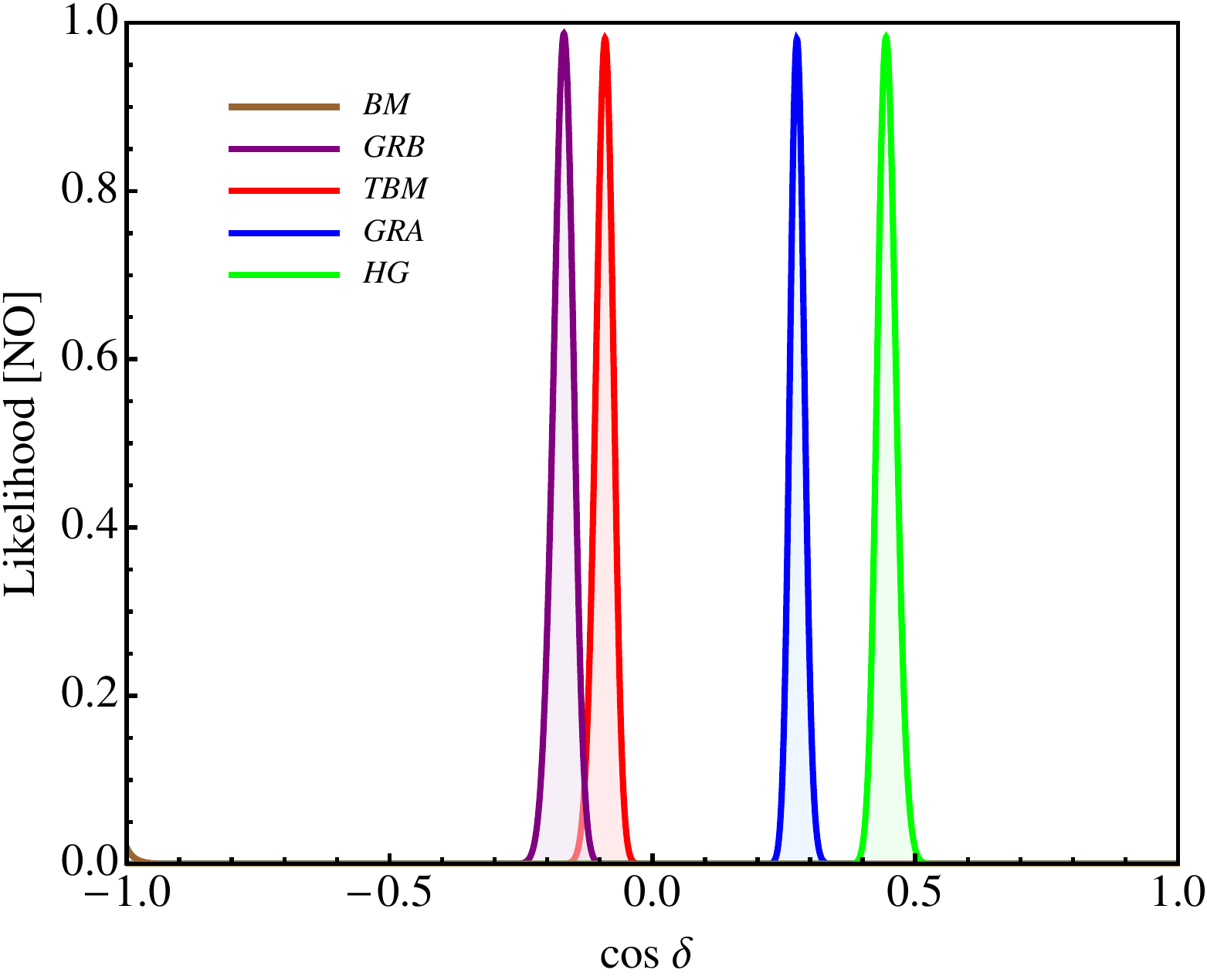}}
  {\includegraphics[width=8cm]{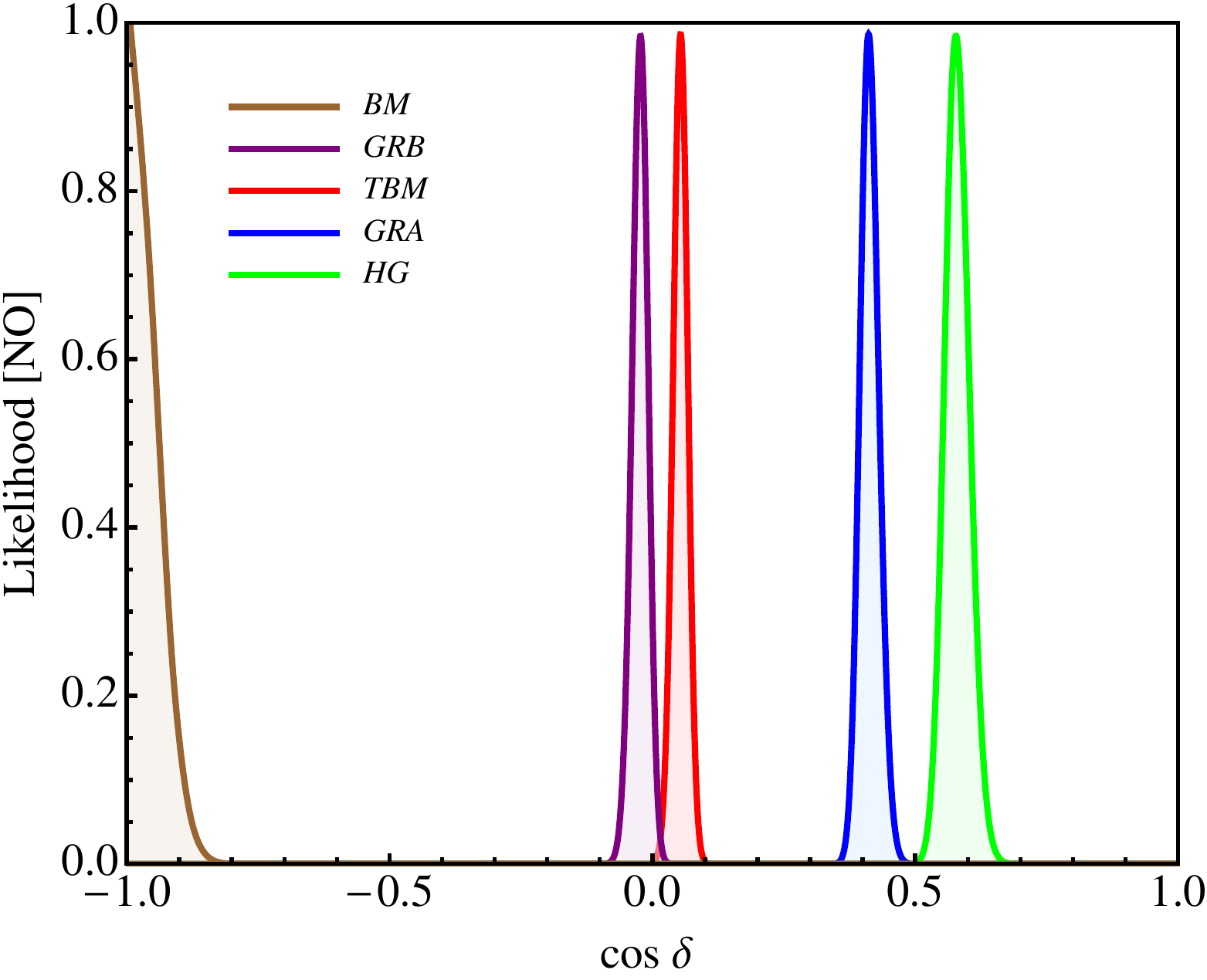}}
  \vspace{5mm}
     \end{center}
      
\vspace{-1.0cm} \caption{ \label{Fig:cosdeltaNO_fut}
The same as in Fig.~\ref{Fig:cosdeltaNO},
but using the prospective $1\sigma$ uncertainties in the 
determination of the neutrino
mixing angles within the Gaussian approximation (see text for further details).
In the left (right) panel $\sin^2 \theta_{12} = 0.308$ ($0.332$), the other mixing angles
being fixed to their NO best fit values.
}
\end{figure}

\begin{figure}[h!]
  \begin{center}
     \hspace{-0.9cm}
   \subfigure
  {\includegraphics[width=8cm]{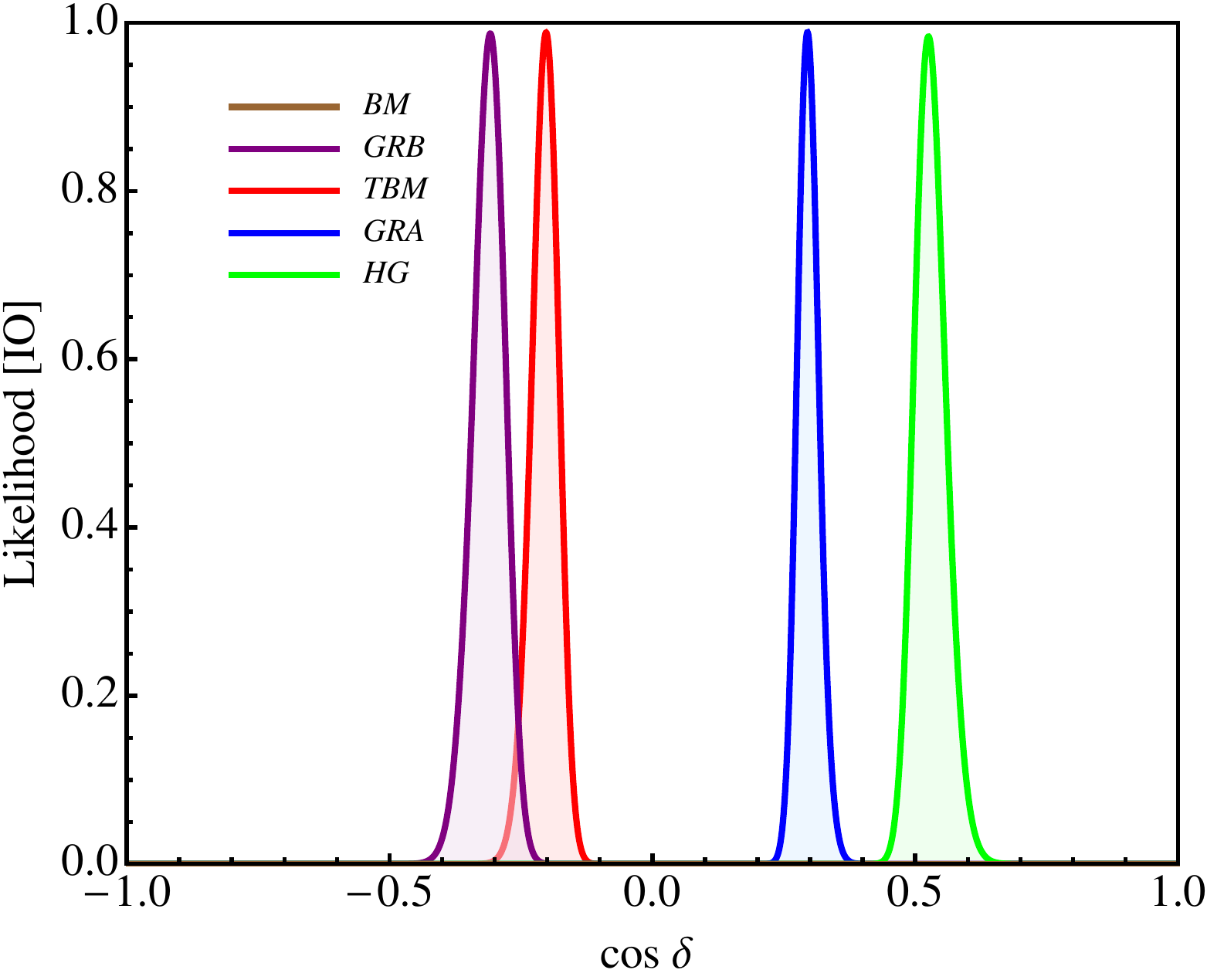}}
  \vspace{5mm}
     \end{center}
\vspace{-1.0cm} \caption{ \label{Fig:cosdeltaNO_GGIO}
The same as in Fig.~\ref{Fig:cosdeltaNO_fut},
but using the IO best fit values taken from \cite{Gonzalez-Garcia:2014bfa}.
}
\end{figure}

In the left panel of Fig.~\ref{Fig:cosdeltaNO_fut} we present the likelihood function
versus $\cos \delta$ within the Gaussian approximation
(see Appendix \ref{AppendixB} for details), using the current best fit values 
of the mixing angles for NO neutrino mass spectrum in 
eqs.~(\ref{th12values})~--~(\ref{th13values}) and the prospective 
$1\sigma$ uncertainties in the determination of 
$\sin^2 \theta_{12}$ (0.7\% from JUNO \cite{Wang:2014iod}),
$\sin^2 \theta_{13}$ (almost 3\% derived from an expected error on 
$\sin^2 2 \theta_{13}$ of 3\% from Daya Bay, see A.~de Gouvea {\it et al.} in \cite{LBLFuture13}) 
and $\sin^2 \theta_{23}$ (5\%~\footnote{This sensitivity can be achieved
in future neutrino facilities \cite{Coloma:2014kca}.} derived from the potential sensitivity of 
NOvA and T2K on $\sin^2 2 \theta_{23}$ of 2\%, 
see A.~de Gouvea {\it et al.} in \cite{LBLFuture13}).
The BM case is very sensitive to the best fit values of $\sin^2 \theta_{12}$
and $\sin^2 \theta_{23}$ and is disfavoured at more than $2\sigma$ for
the current best fit values quoted in eqs.~(\ref{th12values})~--~(\ref{th13values}). 
This case might turn out to be compatible with the data 
for larger (smaller) measured values of $\sin^2 \theta_{12}$ ($\sin^2 \theta_{23}$), 
as can be seen from the right 
panel of Fig.~\ref{Fig:cosdeltaNO_fut}, which was obtained
for $\sin^2 \theta_{12} = 0.332$.
With the increase of the value of $\sin^2 \theta_{23}$
the BM form becomes increasingly disfavoured, 
while the TBM/GRB (GRA/HG) predictions
for $\cos \delta$ are shifted somewhat~---~approximately by
0.1~---~to the left (right) with respect to those shown in the 
left panel of Fig.~\ref{Fig:cosdeltaNO_fut}. This shift is illustrated
in Fig.~\ref{Fig:cosdeltaNO_GGIO}, which is obtained
for $\sin^2 \theta_{23} = 0.579$, more precisely, for the best 
fit values found in \cite{Gonzalez-Garcia:2014bfa} and
corresponding to IO neutrino mass spectrum.
The measurement of $\sin^2 \theta_{12}$, $\sin^2 \theta_{13}$
and $\sin^2 \theta_{23}$ with the quoted precision will open up
the possibility to distinguish between the BM, TBM/GRB, GRA and HG forms of
$\tilde U_{\nu}$. Distinguishing between the TBM and GRB forms would
require relatively high precision measurement of $\cos \delta$.

\begin{figure}[h!]
  \begin{center}
     \hspace{-1.6cm}
   \subfigure
 {\includegraphics[width=14cm]{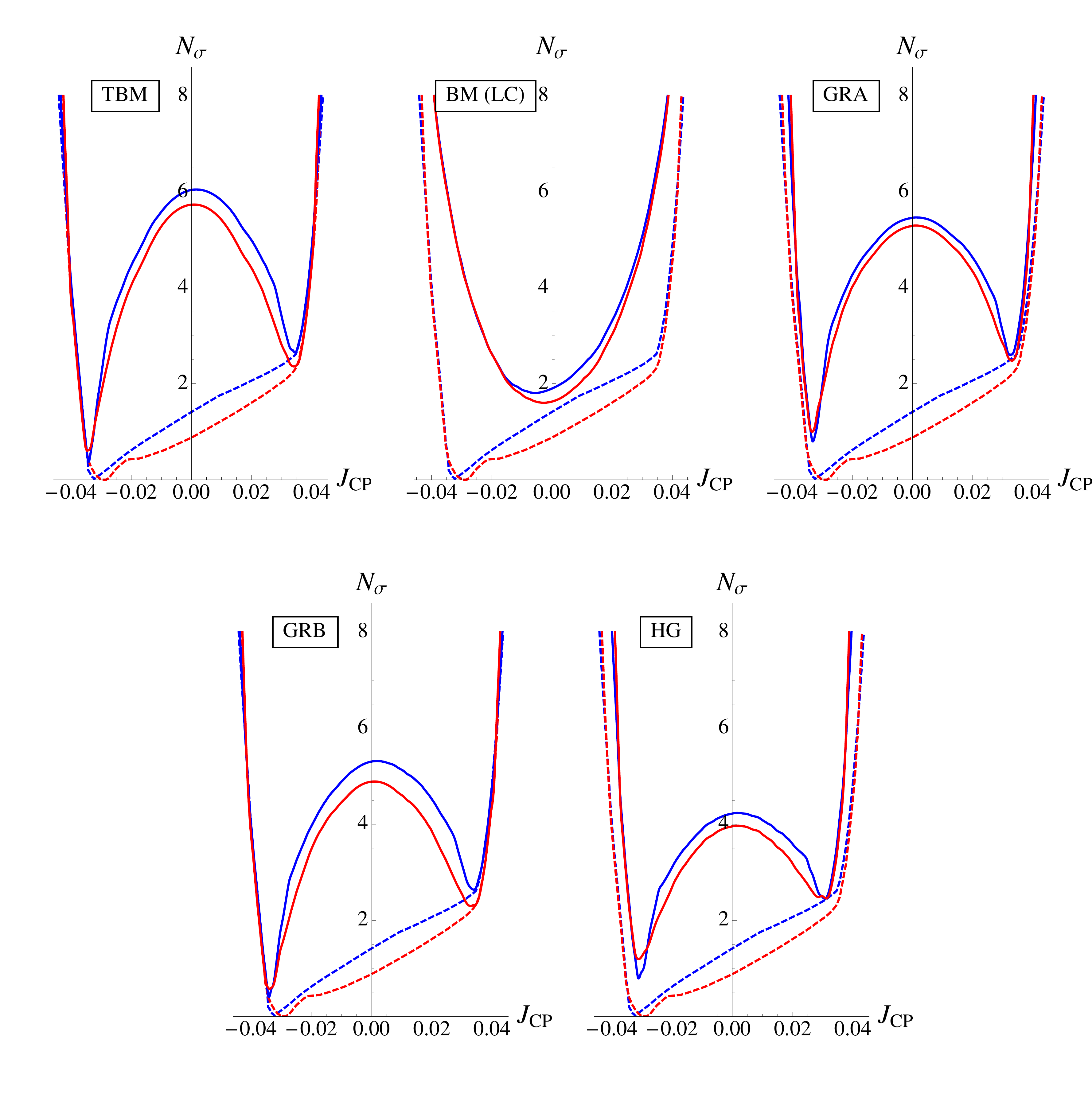}}
  \vspace{5mm}
     \end{center}
\vspace{-1.0cm} \caption{\label{Fig:chi2JCP}
$N_{\sigma} \equiv \sqrt{\chi^2}$ as a function of $J_{\rm CP}$.
The dashed lines represent the results of the global fit
\cite{Capozzi:2013csa},
 while the solid lines represent the
results we obtain for the TBM, BM (LC), GRA
(upper left, central, right panels),
GRB and HG (lower left and right panels)
neutrino mixing symmetry forms.
The blue (red) lines are for NO (IO)
neutrino mass spectrum (see text for further details).
}
\end{figure}

 We have performed also a  statistical analysis in order to derive
predictions for $J_{\rm CP}$.
In Fig.~\ref{Fig:chi2JCP} we present
$N_{\sigma} \equiv \sqrt{\chi^2}$ as a function of $J_{\rm CP}$
for NO and IO neutrino mass spectra.
Similarly to the case of $\delta$, we minimise
the value of $\chi^2$ for a fixed value of $J_{\rm CP}$
by varying $\sin^2\theta_{13}$ and $\sin^2\theta_{23}$
(or, equivalently, $\sin^2\theta^e_{12}$ and $\sin^2\hat \theta_{23}$).
The best fit value of $J_{\rm CP}$
and the corresponding $3\sigma$ range
for each of the considered symmetry forms
of $\tilde{U}_{\nu}$
are summarised in Table~\ref{table:1223}.
%
As  Fig.~\ref{Fig:chi2JCP} shows,
the CP-conserving value of $J_{\rm CP} = 0$ is
excluded  in the cases of the  TBM, GRA, GRB
and HG neutrino mixing symmetry forms, respectively,
at approximately $5\sigma$, $4\sigma$, $4\sigma$
and $3\sigma$ confidence levels
 with respect to the confidence level of the
corresponding best fit values
\footnote{The confidence levels under discussion
differ in the cases of NO and IO neutrino mass spectra,
but as Fig.~\ref{Fig:chi2JCP} indicates, in the cases
considered these differences are rather small
and we have not given them.}.
These results correspond to
those we have obtained for $\delta$,
more specifically to the confidence levels at which
the CP-conserving values of  $\delta = 0$, $\pi$, $2\pi$,
are excluded (see Fig.~\ref{Fig:chi2delta}).

In contrast, for the BM (LC) symmetry form,
the CP-conserving value of $\delta$,
namely, $\delta \cong \pi$, is preferred
and therefore the CP-violating effects in neutrino
oscillations are predicted to be suppressed.
At the best fit point we obtain a value of
$J_{\rm CP} = -0.005 \, (-0.002)$
for  NO (IO) neutrino mass spectrum,
which corresponds to the best fit value of
$\delta / \pi = 1.04 \, (1.02)$.
The allowed range of the $J_{\rm CP}$ factor in the
BM (LC) includes the CP-conserving value
 $J_{\rm CP} = 0$ at practically
any confidence level.
As can be seen from
Table~\ref{table:1223}, the $3\sigma$ allowed intervals of values
of $\delta$ and $J_{\rm CP}$ are rather narrow for
all the symmetry forms considered, except for the BM (LC) form.

 Finally, for completeness, we present
in Appendix \ref{AppendixC}
also results of a statistical analysis
of the predictions for
the values of  $\sin^2 \theta_{23}$ for the
TBM, BM (LC), GRA, GRB and HG neutrino mixing symmetry
forms considered. We recall that of the three neutrino mixing
parameters, $\sin^2 \theta_{12}$,  $\sin^2 \theta_{13}$ and
$\sin^2 \theta_{23}$, $\sin^2 \theta_{23}$ is determined
in the global analyses of the neutrino
oscillation data with the largest uncertainty.
\begin{table}[h!]
\centering
\begin{tabular}{ l l | c | c }
\toprule
Symmetry form	& 	& Best fit & 3$\sigma$ range  \\ \midrule
TBM	& $J_{\rm CP}$ (NO)   & $-0.034$ & $-0.038 \div -0.028 \oplus 0.031 \div 0.036$ \\
 & $J_{\rm CP}$ (IO)    & $-0.034$ & $-0.039 \div -0.025 \oplus 0.029 \div 0.037$ \\
        & $\delta / \pi$ (NO)  & $1.48$ & $0.49 \div 0.58 \oplus 1.34 \div 1.57$ \\
         & $\delta / \pi$ (IO)  & $1.48$ & $0.47 \div 0.65 \oplus 1.30 \div 1.57$ \\
         & $\cos \delta$ (NO) & $-0.07$ & $-0.47\div 0.21$\\
         & $\cos \delta$ (IO) & $-0.07$ & $-0.60 \div 0.23$\\
         \midrule

BM (LC)	& $J_{\rm CP}$ (NO)   & $-0.005$ & $-0.026 \div 0.021$ \\
& $J_{\rm CP}$ (IO)    & $-0.002$ & $-0.025 \div 0.023$ \\
         & $\delta / \pi$ (NO)  & $1.04$ & $0.80 \div 1.24$ \\
         & $\delta / \pi$ (IO)  &  $1.02$  & $0.79 \div 1.23$ \\
         & $\cos \delta$ (NO) & $-0.99$ & $-1.00\div -0.72$\\
         & $\cos \delta$ (IO)  & $-1.00$ & $-1.00 \div -0.72$\\
         \midrule

GRA    & $J_{\rm CP}$ (NO)   & $-0.033$ & $-0.037 \div -0.027 \oplus 0.030 \div 0.035$ \\
& $J_{\rm CP}$ (IO)    & $-0.033$ & $-0.037 \div -0.025 \oplus 0.028 \div 0.036$ \\
        & $\delta / \pi$ (NO)  & $1.58$ & $0.35 \div 0.46 \oplus 1.50 \div 1.70$ \\
         & $\delta / \pi$ (IO)  & $1.58$ & $0.31 \div 0.48 \oplus 1.47 \div 1.74$ \\
         & $\cos \delta$ (NO) & $0.25$ & $-0.08\div 0.69$\\
         & $\cos \delta$ (IO) & $0.25$ & $-0.08 \div 0.69$\\
         \midrule

GRB    & $J_{\rm CP}$ (NO)   & $-0.034$ & $-0.039 \div -0.026 \oplus 0.031 \div 0.036$ \\
& $J_{\rm CP}$ (IO)    & $-0.033$ & $-0.039 \div -0.022 \oplus 0.026 \div 0.037$ \\
        & $\delta / \pi$ (NO)  & $1.45$ & $0.51 \div 0.61 \oplus 1.31 \div 1.54$ \\
         & $\delta / \pi$ (IO)  & $1.45$ & $0.50 \div 0.70 \oplus 1.25 \div 1.54$ \\
         & $\cos \delta$ (NO) & $-0.15$ & $-0.57\div 0.13$\\
         & $\cos \delta$ (IO) & $-0.15$ & $-0.70 \div 0.13$\\
         \midrule

HG	& $J_{\rm CP}$ (NO)   & $-0.031$ & $-0.035 \div -0.020 \oplus 0.026 \div 0.034$ \\
 & $J_{\rm CP}$ (IO)    & $-0.031$ & $-0.036 \div -0.015 \oplus 0.019 \div 0.034$ \\
         & $\delta / \pi$ (NO)  & $1.66$ & $0.27 \div 0.41 \oplus 1.55 \div 1.80$ \\
         & $\delta / \pi$ (IO)  & $1.63$ & $0.19 \div 0.42 \oplus 1.55 \div 1.86$ \\
         & $\cos \delta$ (NO) & $0.47$ & $0.16\div0.80$\\
         & $\cos \delta$ (IO) & $0.40$ & $0.16\div0.80$\\

\bottomrule
\end{tabular}
\caption{\label{table:1223} Best fit values of $J_{\rm CP}$, $\delta$
and $\cos \delta$
and corresponding 3$\sigma$ ranges
(found fixing $\sqrt{\chi^2-\chi^2_{\rm min}} = 3$)
in our setup
using the data from \cite{Capozzi:2013csa}.
}
\end{table}

\section*{Summary and Conclusions}
%
%
 Using the fact that the neutrino mixing matrix
$U = U^\dagger_{e}U_{\nu}$, where $U_{e}$ and $U_{\nu}$
result from the diagonalisation of the charged lepton
and neutrino mass matrices, we have analysed the sum rules which
the Dirac phase $\delta$ present in $U$
satisfies when $U_{\nu}$ has a form dictated by,
or associated with, discrete symmetries and
$U_e$ has a ``minimal'' form (in
terms of angles and phases it contains)
that can provide the requisite
corrections to $U_{\nu}$, so that the
reactor, atmospheric and solar neutrino mixing angles
$\theta_{13}$, $\theta_{23}$ and  $\theta_{12}$
have values compatible with the current data.

 We have considered the following symmetry forms of $U_{\nu}$:
i) tri-bimaximal (TBM), ii) bimaximal (BM)
(or corresponding to the conservation of the
lepton charge $L' = L_e - L_\mu - L_{\tau}$ (LC)),
iii) golden ratio type A (GRA),
iv) golden ratio type B (GRB),
and v) hexagonal (HG).
For all these symmetry forms $U_\nu$ can be written as
$ U_\nu  = \Psi_1\,\tilde{U}_{\nu}\,Q_0 =
\Psi_1\,R_{23}(\theta^\nu_{23})
R_{12}(\theta^\nu_{12}) Q_0$,
%
where $R_{23}(\theta^\nu_{23})$ and $R_{12}(\theta^\nu_{12})$ are
orthogonal matrices describing rotations in the 2-3 and 1-2 planes,
respectively, and $\Psi_1$ and $Q_0$ are diagonal phase matrices
each containing two phases. The phases in the matrix $Q_0$
give contribution to the Majorana phases in the PMNS matrix.
The symmetry forms of $\tilde{U}_{\nu}$ of interest,
TBM, BM (LC), GRA, GRB and HG, are
characterised by the same value of the
angle $\theta^\nu_{23} = -\pi/4$,
but correspond to different fixed values of
the angle $\theta^\nu_{12}$ and
thus of $\sin^2\theta^\nu_{12}$, namely, to
i)  $\sin^2\theta^{\nu}_{12} = 1/3$ (TBM),
ii)  $\sin^2\theta^{\nu}_{12} = 1/2$ (BM (LC)),
iii)  $\sin^2\theta^{\nu}_{12} =  (2 + r)^{-1} \cong 0.276$ (GRA),
$r$ being the golden ratio, $r = (1 +\sqrt{5})/2$,
iv) $\sin^2\theta^{\nu}_{12} = (3 - r)/4 \cong 0.345$ (GRB), and
v) $\sin^2\theta^{\nu}_{12} = 1/4$ (HG).

The minimal form of $U_e$ of interest
that can provide the requisite corrections to $U_{\nu}$,
so that the neutrino mixing angles
$\theta_{13}$, $\theta_{23}$ and  $\theta_{12}$
have values compatible with the current data,
including a possible sizeable deviation of $\theta_{23}$ from $\pi/4$,
includes a product of two orthogonal matrices describing
rotations in the 2-3 and 1-2 planes \cite{Marzocca:2013cr},
$R_{23}(\theta^e_{23})$ and $R_{12}(\theta^e_{12})$,
$\theta^e_{23}$ and $\theta^e_{12}$ being two (real) angles.
This leads to the parametrisation of the PMNS matrix $U$
given in eq. (\ref{UUedagUnu}),
which can be recast in the form \cite{Marzocca:2013cr}:
$U =R_{12}(\theta^e_{12})\Phi(\phi)R_{23}(\hat\theta_{23})\,
R_{12}(\theta^{\nu}_{12})\,\hat{Q}$,
where $\Phi = {\rm diag} \left(1,\text{e}^{i\phi},1\right)$, $\phi$
being a CP violation phase, $\hat\theta_{23}$
is a function of $\theta^e_{23}$ (see eq. (\ref{th23hat})),
and $\hat{Q}$ is a diagonal phase matrix.
The phases in  $\hat{Q}$ give contributions
to the Majorana phases in the PMNS matrix.
The angle $\hat\theta_{23}$, however, can be expressed
in terms of the angles $\theta_{23}$ and $\theta_{13}$
of the PMNS matrix (eq.~(\ref{s2th23})) and the value
of  $\hat\theta_{23}$ is fixed by the values of
$\theta_{23}$ and $\theta_{13}$.

 In this scheme the four observables
$\theta_{12}$, $\theta_{23}$, $\theta_{13}$
and  the Dirac phase $\delta$ in the PMNS matrix
are functions of three parameters
$\theta^e_{12}$, $\hat\theta_{23}$ and $\phi$.
As a consequence, the Dirac phase $\delta$
can be expressed as a function of the
three PMNS angles $\theta_{12}$, $\theta_{23}$
and $\theta_{13}$, leading to a new ``sum rule''
relating $\delta$ and $\theta_{12}$, $\theta_{23}$
and $\theta_{13}$. This sum rule is exact
within the scheme considered.
Its explicit form depends on the symmetry
form of the matrix $\tilde{U}_{\nu}$, i.e.,
on the value of the angle $\theta^{\nu}_{12}$.
For arbitrary fixed value of $\theta^{\nu}_{12}$
the sum rule of interest
is given in eq.~(\ref{cosdthnu})
(or the equivalent eq.~(\ref{s2th12cosdthnu}))
\cite{Petcov:2014laa}. A similar exact sum rule
can be derived for the phase $\phi$
(eq.~(\ref{s2th12cosphithnu})) \cite{Petcov:2014laa}.

  A parametrisation of the PMNS matrix, similar
to that given in eq.~(\ref{UUedagUnu}),
has been effectively employed
in ref. \cite{Antusch:2005kw}. Treating $\sin\theta^e_{12}$
and  $\sin\theta^e_{23}$ as small parameters,
$|\sin\theta^e_{12}| \ll 1$, $|\sin\theta^e_{23}| \ll 1$, and
neglecting terms of order of, or smaller than,
$O((\theta^e_{12})^2)$, $O((\theta^e_{23})^2)$ and
$O(\theta^e_{12} \theta^e_{23})$,
the following ``leading order'' sum rule
was obtained in \cite{Antusch:2005kw}:
$ \theta_{12} \cong \theta^{\nu}_{12} + \theta_{13} \cos \delta$.
%
This sum rule, in the approximation used to
obtain it, is equivalent to the sum rule
$\sin\theta_{12} \cong
\sin\theta^{\nu}_{12}
 + \cos \theta^{\nu}_{12} \sin\theta_{13} \cos \delta$,
%
which was shown in ref. \cite{Petcov:2014laa}
to be the leading order approximation
of the exact sum rule given in eq.~(\ref{cosdthnu})
(or the equivalent eq.~(\ref{s2th12cosdthnu})).
In the present article we have investigated the predictions
for $\cos\delta$ in the cases of
TBM, BM (LC), GRA, GRB and HG symmetry forms of
the matrix $\tilde{U}_{\nu}$
using the exact and the leading order sum rules
for $\cos\delta$ discussed above and given in
eqs.~(\ref{s2th12cosdthnu}) and (\ref{sinth12cosd}).
It was shown in \cite{Petcov:2014laa}, in particular,
using the best fit values of the neutrino mixing parameters
$\sin^2\theta_{12}$, $\sin^2\theta_{23}$ and $\sin^2\theta_{13}$
and the exact sum rule results for $\cos\delta$ derived for the
TBM, GRA, GRB and HG forms of $\tilde{U}_{\nu}$,
that the leading order sum rule provides largely
imprecise predictions for $\cos\delta$.
Here we have performed a thorough study
of the exact and leading order sum rule
predictions for $\cos\delta$ in the
TBM, BM (LC), GRA, GRB and HG cases
taking into account the uncertainties
in the measured values of
$\sin^2\theta_{12}$, $\sin^2\theta_{23}$ and $\sin^2\theta_{13}$.
This allowed us, in particular, to assess
the accuracy of the predictions for $\cos\delta$
based on the leading order sum rules and
its dependence on the values
of the indicated neutrino mixing parameters
when the latter are varied in their respective
3$\sigma$ experimentally allowed ranges.
In contrast to the leading order sum rule,
the exact sum rule for $\cos\delta$ depends not only
on $\theta_{12}$ and $\theta_{13}$, but also
on $\theta_{23}$, and we have investigated
this dependence as well.

In the present study we have analysed both the cases of $\theta^e_{23} = 0$,
in which $\sin^2 \theta_{23} \cong 0.5(1 - \sin^2 \theta_{13})$, and of arbitrary $\theta^e_{23}$.
In the second case $\theta_{23}$ can deviate significantly from $\pi/4$.

  We confirm the result found in \cite{Petcov:2014laa}
that the exact sum rule  predictions for
$\cos\delta$ vary significantly
with the symmetry form of $\tilde{U}_{\nu}$.
This result implies that the measurement of $\cos\delta$ 
can allow us to distinguish between the different symmetry forms of
$\tilde{U}_{\nu}$ \cite{Petcov:2014laa}
provided $\sin^2\theta_{12}$, $\sin^2\theta_{13}$ and
$\sin^2\theta_{23}$ are known with a
sufficiently good precision.
Even determining the sign of $\cos\delta$
will be sufficient to eliminate some of
the possible symmetry forms of $\tilde{U}_{\nu}$.

 We find also that the exact sum rule predictions for
$\cos\delta$ exhibit strong dependence on the value of
$\sin^2\theta_{12}$ when the latter is varied in
its $3\sigma$ experimentally allowed range
(0.259\,--\,0.359) (Tables \ref{tab:1}~--~\ref{tab:6}).
The predictions for $\cos\delta$ change significantly
not only in magnitude, but in the cases
of TBM, GRA and GRB forms of $\tilde U_{\nu}$
also the sign of $\cos\delta$ can change.
These significant changes take place both
for $\theta^e_{23} = 0$ and $\theta^e_{23} \neq 0$.

 We have investigated the dependence of
the exact sum rule predictions for $\cos\delta$
in the cases of the symmetry forms of $\tilde{U}_{\nu}$
considered on the value of $\sin^2\theta_{23}$
varying the latter in the respective $3\sigma$ allowed interval
$0.374\leq \sin^2\theta_{23}\leq 0.626$
(Figs.~\ref{Fig:4} and \ref{Fig:5},
and Tables \ref{tab:7} and \ref{tab:8}).
The results we get for $\sin^2\theta_{23} =0.374$ and
$\sin^2\theta_{23} =0.437$, setting
$\sin^2\theta_{12}$ and $\sin^2\theta_{13}$
to their best fit values, do not differ significantly.
However, the differences between the predictions
for $\cos\delta$ obtained
for  $\sin^2 \theta_{23} = 0.437$ and
for $\sin^2 \theta_{23} = 0.626$ are relatively large (they
differ by the factors of 2.05, 1.25, 1.77 and 1.32
in the TBM, GRA, GRB and HG cases, respectively).

 In all cases considered, having the exact sum rule results
for $\cos\delta$, we could investigate  the
precision of the leading order sum rule predictions for
$\cos\delta$. We found that the leading order sum rule predictions for
$\cos\delta$ are, in general, imprecise and in many cases
are largely incorrect, the only exception being the case of the BM (LC)
form of $\tilde{U}_{\nu}$ \cite{Petcov:2014laa}.

 We have performed a similar analysis of the predictions for the
cosine of the phase $\phi$. The phase $\phi$ is related to,
but does not coincide with, the Dirac phase $\delta$.
The parameter $\cos\phi$ obeys a leading order sum rule
which is almost identical to the leading order sum rule
satisfied by $\cos\delta$. This leads to the confusing
identification of $\phi$  with $\delta$:
the exact sum rules satisfied by $\cos\phi$ and
$\cos\delta$ differ significantly. Correspondingly,
the predicted values of $\cos\phi$ and $\cos\delta$
in the cases of the  TBM, GRA, GRB and HG
symmetry forms of $\tilde{U}_{\nu}$ considered by us
also differ significantly (see Figs.
\ref{Fig:a}~--~\ref{Fig:5}  and
Tables  \ref{tab:1}~--~\ref{tab:8}).
This conclusion is not valid for the
BM (LC) form: for this form the
exact sum rule predictions for
$\cos\phi$ and $\cos\delta$ are
rather similar. The phase $\phi$ appears
in a large class of models
of neutrino mixing and neutrino mass generation and
serves as a  ``source'' for the Dirac phase $\delta$
in these models.

Finally, we have performed a statistical analysis
of the predictions for $\delta$, $\cos\delta$ and the rephasing
invariant $J_{\rm CP}$ which controls the magnitude
of CPV effects in neutrino oscillations \cite{PKSP3nu88},
in the cases of the TBM, BM (LC), GRA, GRB and HG symmetry
forms of the matrix $\tilde{U}_{\nu}$ considered.
In this analysis we have used as input
the latest results on $\sin^2\theta_{12}$, $\sin^2\theta_{13}$,
$\sin^2\theta_{23}$ and $\delta$, obtained in the global analysis of the
neutrino oscillation data performed in \cite{Capozzi:2013csa}.
Our goal was to derive the allowed ranges
for $\delta$, $\cos\delta$ and $J_{\rm CP}$,
predicted on the basis of the current data on
the neutrino mixing parameters for each of
the symmetry forms of $\tilde{U}_{\nu}$ considered.
The results of this analysis are
shown in Figs.~\ref{Fig:chi2delta}, \ref{Fig:cosdeltaNO}
and \ref{Fig:chi2JCP}, and are summarised in Table
\ref{table:1223}, in which we give
the predicted best fit values and $3\sigma$ ranges
of $J_{\rm CP}$, $\delta$ and $\cos\delta$
for each of the symmetry forms
of $\tilde{U}_{\nu}$ considered.
We have shown, in particular, that
the CP-conserving value of $J_{\rm CP} = 0$ is
excluded  in the cases of the  TBM, GRA, GRB
and HG neutrino mixing symmetry forms, respectively,
at approximately $5\sigma$, $4\sigma$, $4\sigma$
and $3\sigma$ confidence levels
 with respect to the confidence level of the
corresponding best fit values (Fig.~\ref{Fig:chi2JCP}).
 These results reflect the predictions
we have obtained for $\delta$,
more specifically, the confidence levels at which
the CP-conserving values of  $\delta = 0$, $\pi$, $2\pi$,
are excluded in the discussed cases
(see Fig.~\ref{Fig:chi2delta}).
We have found also that the $3\sigma$
allowed intervals of values of
$\delta$ and $J_{\rm CP}$ are rather narrow for
all the symmetry forms considered,
except for the BM (LC) form
(Table \ref{table:1223}).
More specifically, for the TBM, GRA, GRB
and HG symmetry forms we have obtained
at $3\sigma$: $0.020 \leq |J_{\rm CP}| \leq 0.039$.
For the best fit values of $J_{\rm CP}$ we have found,
respectively:
$J_{\rm CP} = (-0.034)$, $(-0.033)$, $(-0.034)$, and  $(-0.031)$.
Our results indicate that distinguishing between
the TBM, GRA, GRB and HG symmetry forms of the neutrino mixing
would require extremely high precision measurement of
the $J_{\rm CP}$ factor.

Using the likelihood method, we have derived also
the ranges of the predicted values of $\cos \delta$ for the different 
forms of $\tilde U_{\nu}$ considered, using the prospective 
$1\sigma$ uncertainties in the determination of
$\sin^2 \theta_{12}$, $\sin^2 \theta_{13}$ and $\sin^2 \theta_{23}$
respectively in JUNO, Daya Bay and accelerator and atmospheric
neutrino experiments (Fig.~\ref{Fig:cosdeltaNO_fut}).
In this analysis the current best fit values of 
$\sin^2 \theta_{12}$, $\sin^2 \theta_{13}$ and $\sin^2 \theta_{23}$
have been utilised (left panel of Fig.~\ref{Fig:cosdeltaNO_fut}).
The results thus obtained show that i) the measurement of 
the sign of $\cos \delta$ will allow to distinguish between the TBM/GRB, BM
and GRA/HG forms of $\tilde U_{\nu}$, ii) for a best fit value
of $\cos \delta = -1 \, (-0.1)$ distinguishing at $3\sigma$ between 
the BM (TBM/GRB) and the other forms of $\tilde U_{\nu}$
would be possible if $\cos \delta$ is measured with $1\sigma$
uncertainty of $0.3 \, (0.1)$.

The predictions for $\delta$, $\cos\delta$ and $J_{\rm CP}$
in the case of the BM (LC) symmetry form of $\tilde{U}_{\nu}$,
as the results of the statistical analysis performed by us showed,
differ significantly from those found for the
TBM, GRA, GRB and HG forms:
the best fit value of $\delta \cong \pi$, and, correspondingly,
of $J_{\rm CP} \cong 0$. For the $3\sigma$ range of
 $J_{\rm CP}$ we have obtained in the case of
NO (IO) neutrino mass spectrum:
$-0.026~(-0.025)\leq J_{\rm CP} \leq 0.021~(0.023)$, i.e.,
it includes a sub-interval of values centred on zero, which does not
overlap with the $3\sigma$ allowed intervals of values
of  $J_{\rm CP}$ in  the TBM, GRA, GRB and HG cases.

  The results obtained in the present study, in particular,
reinforce the conclusion reached in ref. \cite{Petcov:2014laa}
that the experimental measurement of the cosine
of the Dirac phase $\delta$ of the PMNS neutrino mixing matrix
can provide unique information about the possible
discrete symmetry origin of the observed pattern of
neutrino mixing.

\vspace{0.2cm}
{\bf Acknowledgements.} This work was supported in part 
by the European Union FP7
ITN INVISIBLES (Marie Curie Actions, PITN-GA-2011-289442-INVISIBLES),
by the INFN program on Theoretical Astroparticle Physics (TASP),
by the research grant  2012CPPYP7 ({\sl  Theoretical Astroparticle Physics})
under the program  PRIN 2012 funded by the Italian Ministry of Education, University and Research (MIUR)
and by the World Premier International Research Center
Initiative (WPI Initiative, MEXT), Japan (STP).

\vspace{-0.2cm}
\appendix
\section{Relations Between Phases in Two Parametrisations}
\label{App:A}

In this section we present the relations between
the phases of the two different parametrisations
of the PMNS matrix employed in
\cite{Antusch:2005kw} and \cite{Petcov:2014laa}.
In the parametrisation used in \cite{Antusch:2005kw}
the PMNS matrix after setting $\theta^e_{13} = \theta^{\nu}_{13} = 0$ reads:
\be
U_{\rm PMNS} =
U^{e_L \dagger}_{12} \, U^{e_L \dagger}_{23} \, U^{\nu_L}_{23} \, U^{\nu_L}_{12} \,,
\ee
%
where the subscripts 12 and 23 stand
for the rotation plane, e.g., the matrix $U^{e_L }_{12}$ being defined as
\be
U^{e_L }_{12} =
\begin{pmatrix}
\cos \theta^e_{12} & \sin \theta^e_{12} \, e^{- i\delta^e_{12}}& 0\\
-\sin \theta^e_{12} \, e^{i\delta^e_{12}}& \cos \theta^e_{12} & 0\\
0 & 0 & 1 \\
\end{pmatrix} \,,
\ee
%
and the others analogously. We can factorise
the phases in the charged lepton and the neutrino
sectors in the following way:
\begin{align}
U^{e_L \dagger}_{12} \, U^{e_L \dagger}_{23} = &
\begin{pmatrix}
1 & 0 & 0\\
0 & e^{i(\delta^e_{12} + \pi)} & 0\\
0 & 0 & e^{i(\delta^e_{12} + \delta^e_{23})} \\
\end{pmatrix}
\begin{pmatrix}
\cos \theta^e_{12} & \sin \theta^e_{12} & 0\\
-\sin \theta^e_{12} & \cos \theta^e_{12} & 0\\
0 & 0 & 1 \end{pmatrix} \nonumber \\
\times &  \begin{pmatrix}
1 & 0 & 0\\
0 & \cos \theta^e_{23} & \sin \theta^e_{23} \\
0 & -\sin \theta^e_{23}  & \cos \theta^e_{23} \\
\end{pmatrix}
\begin{pmatrix}
1 & 0 & 0\\
0 & e^{-i(\delta^e_{12} + \pi)} & 0\\
0 & 0 & e^{-i(\delta^e_{12} + \delta^e_{23})} \\
\end{pmatrix} \,,
\label{Eq:par1}
\end{align}
\begin{align}
U^{\nu_L}_{23} \, U^{\nu_L}_{12} = &
\begin{pmatrix}
1  & 0 & 0\\
0 & e^{i\delta^{\nu}_{12}} & 0\\
0 & 0 & e^{i(\delta^{\nu}_{23} + \delta^{\nu}_{12})} \\
\end{pmatrix}
\begin{pmatrix}
1 & 0 & 0\\
0 & \cos \theta^\nu_{23} & \sin \theta^\nu_{23} \\
0 & -\sin \theta^\nu_{23}  & \cos \theta^\nu_{23} \\
\end{pmatrix} \nonumber \\
\times &  \begin{pmatrix}
\cos \theta^\nu_{12} & \sin \theta^\nu_{12} & 0\\
-\sin \theta^\nu_{12} & \cos \theta^\nu_{12} & 0\\
0 & 0 & 1 \end{pmatrix}
\begin{pmatrix}
1  & 0 & 0\\
0 & e^{-i\delta^{\nu}_{12}} & 0\\
0 & 0 & e^{-i(\delta^{\nu}_{23} + \delta^{\nu}_{12})} \\
\end{pmatrix} \,.
\label{Eq:par2}
\end{align}
%
Combining eqs.~(\ref{Eq:par1}) and (\ref{Eq:par2}) and
comparing with the parametrisation of the PMNS matrix
employed in \cite{Petcov:2014laa} and given
in eqs.~(\ref{UUedagUnu}) and (\ref{PsieQ0}), we find the
following relations:
\begin{align}
& \psi = \delta^e_{12} - \delta^{\nu}_{12}  +\pi \,, \quad \omega = \delta^e_{23} + \delta^e_{12} - \delta^{\nu}_{23} - \delta^{\nu}_{12} \,, \\
& \xi_{21} = -2 \delta^{\nu}_{12} \,, \quad \xi_{31} = -2 (\delta^{\nu}_{12} + \delta^{\nu}_{23}) \,.
\end{align}
%

%
\section{Statistical Details}
\label{AppendixB}
%

\begin{figure}[t]
    \hspace{-1.5cm}
   \subfigure
 {\includegraphics[width=17cm]{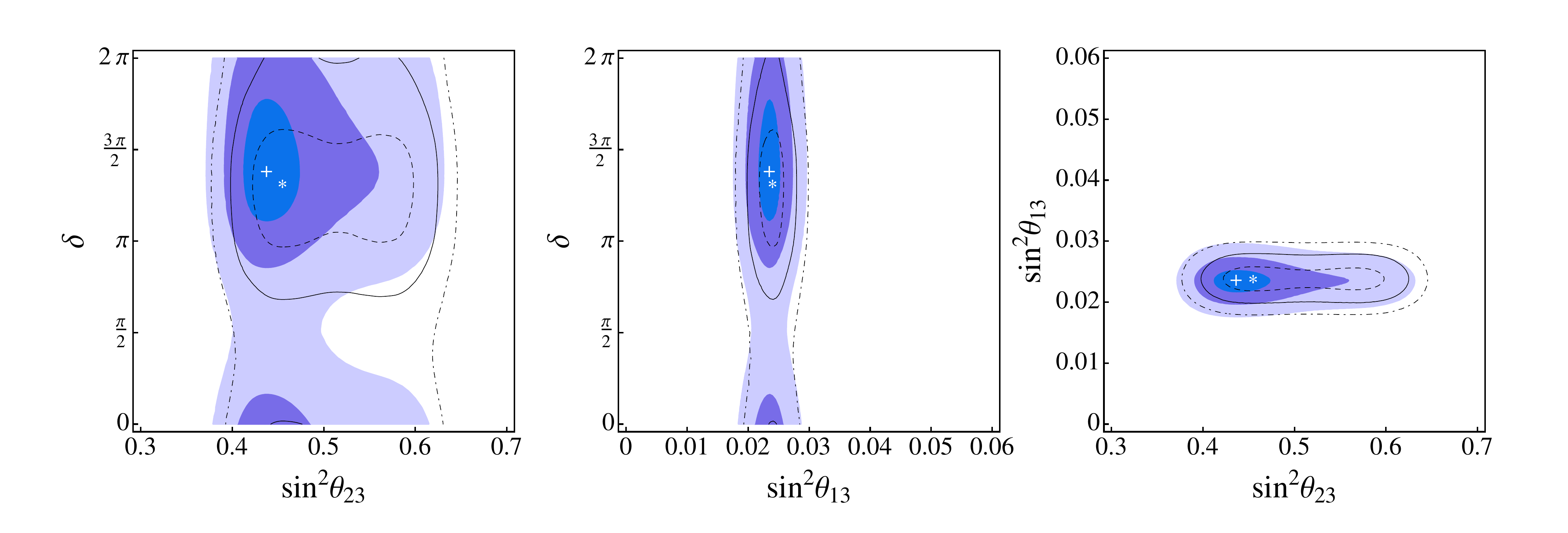}}
\vspace{-1.0cm} \caption{\label{Fig:St1}
Confidence regions at $1\sigma$, $2\sigma$ and $3\sigma$ for 1 degree of freedom
in the planes $(\sin^2 \theta_{23} , \delta)$, $(\sin^2 \theta_{13} , \delta)$ and $(\sin^2 \theta_{23} , \sin^2\theta_{13})$
in the blue (dashed lines), purple (solid lines) and light-purple (dash-dotted lines) for NO (IO)
neutrino mass spectrum, respectively,
obtained using eq.~(\ref{eq:chi2}). The best fit points are indicated with a cross (NO) and an asterisk (IO).
}
\end{figure}

\begin{figure}[t]
    \hspace{-1.5cm}
   \subfigure
 {\includegraphics[width=17cm]{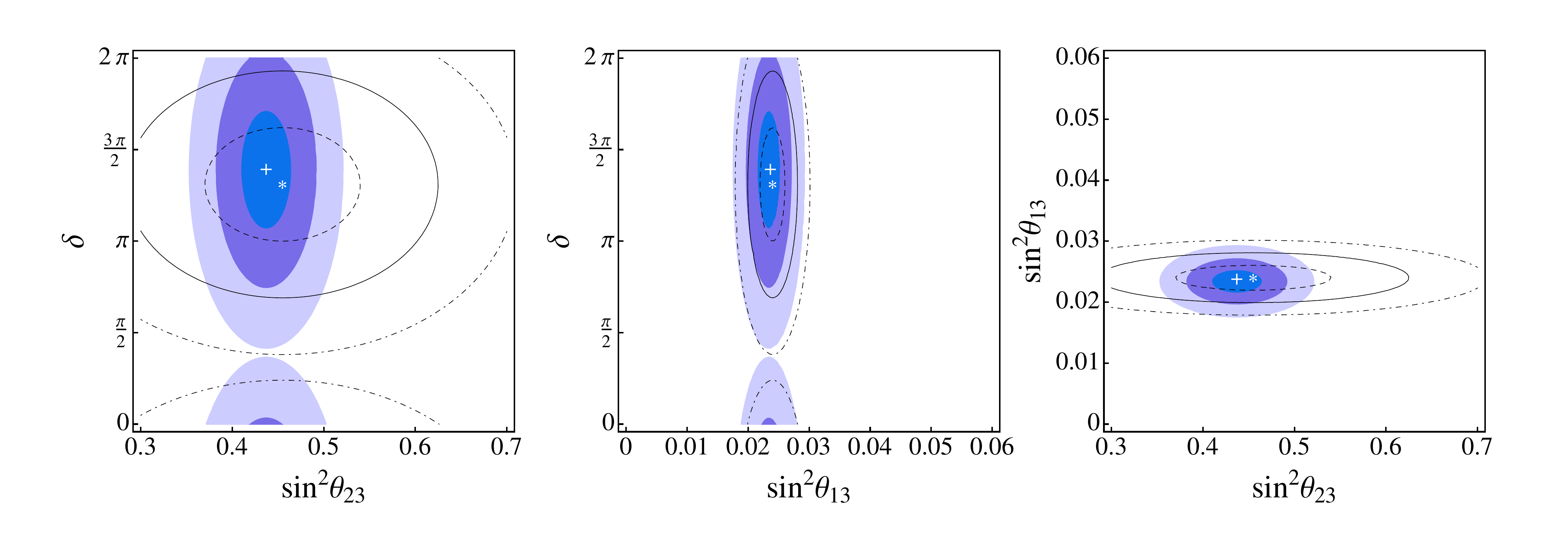}}
\vspace{-1.0cm} \caption{\label{Fig:St2}
The same as in Fig.~\ref{Fig:St1}, but using eq.~(\ref{eq:chi2G}).
}
\end{figure}

In order to perform a statistical analysis of the models considered we construct
the $\chi^2$ function in the following way:
\be
\chi^2(\sin^2 \theta_{12}, \sin^2 \theta_{13}, \sin^2 \theta_{23}, \delta) = \chi_1^2(\sin^2 \theta_{12})
+ \chi_2^2(\sin^2 \theta_{13}) + \chi_3^2(\sin^2 \theta_{23}) + \chi_4^2(\delta) \,,
\label{eq:chi2}
\ee
in which we have neglected the correlations among the oscillation parameters, since
the functions $\chi^2_i$ have been extracted from the 1-dimensional projections in \cite{Capozzi:2013csa}.
In order to quantify the accuracy of our approximation we show in Fig.~\ref{Fig:St1} the
confidence regions at $1\sigma$, $2\sigma$ and $3\sigma$ for 1 degree of freedom
in the planes $(\sin^2 \theta_{23} , \delta)$, $(\sin^2 \theta_{13} , \delta)$ and $(\sin^2 \theta_{23} , \sin^2\theta_{13})$
in blue (dashed lines), purple (solid lines) and light-purple (dash-dotted lines) for NO (IO)
neutrino mass spectrum, respectively, obtained using eq.~(\ref{eq:chi2}). The parameters not shown in the plot have been
marginalised.
It should be noted that what is also used in the literature is the Gaussian approximation,
in which $\chi^2$ can be simplified
using the best fit values and the $1\sigma$ uncertainties as follows:
\be
\chi^2_{\rm G} = \sum_i \dfrac{(x_i - \overline x_i)^2}{\sigma^2_{x_i}} \,.
\label{eq:chi2G}
\ee
Here $x_i = \{\sin^2 \theta_{12},\sin^2 \theta_{13},\sin^2 \theta_{23},\delta\}$,
$\overline x_i$ and $\sigma_{x_i}$ being the best fit values and the $1\sigma$ uncertainties
\footnote{In the case of asymmetric errors we take the mean value of the two errors.} taken from \cite{Capozzi:2013csa}.
We present in Fig.~\ref{Fig:St2} the results of a similar two-dimensional analysis
for the confidence level regions in the planes shown in Fig.~\ref{Fig:St1}, but using the
approximation for $\chi^2$ given in eq.~(\ref{eq:chi2G}).
It follows from these figures that the Gaussian approximation does not allow to reproduce the
confidence regions of \cite{Capozzi:2013csa} with sufficiently good accuracy. For this reason in our analysis we
use the more accurate procedure defined through eq.~(\ref{eq:chi2}).
In both the figures the best fit points are indicated with a cross and an asterisk for NO and
IO spectra, respectively.
%
%
%
%
%
%
%

Each symmetry scheme considered in our analysis, which we label with an index $m$, depends on a set of parameters $y^m_j$,
which are related to the standard oscillation parameters through expressions of the form
$x_i=x^m_i(y^m_j)$. In order to produce the 1-dimensional figures we minimise
\be
\chi^2\left(x^m_i(y^m_j)\right) = \sum_{i=1}^4 \chi_i^2\left(x^m_i(y^m_j)\right)
\label{eq:chi2m}
\ee
for a fixed value of the corresponding observable $\alpha$, i.e.,
\be
\chi^2(\alpha) = \min \left[\chi^2\left(x^m_i(y^m_j)\right) \big|_{\alpha = {\rm const}} \right] \,,
\label{eq:chi2m}
\ee
with $\alpha = \{\delta, J_{\rm CP}, \sin^2 \theta_{23}\}$.
The likelihood function for $\cos \delta$ has been computed
by taking
\be
L(\cos \delta) \propto \exp \left( - \frac{\chi^2(\cos \delta)}{2} \right) \,,
\ee
which was used to produce the likelihood function for the different symmetry forms in Fig.~\ref{Fig:cosdeltaNO}.
It is worth noticing that in the case of flat priors on the mixing parameters, the
posterior probability density function reduces to the likelihood function.
Although we did not use the Gaussian approximation for obtaining Figs.~\ref{Fig:chi2delta}, 
\ref{Fig:cosdeltaNO}, \ref{Fig:chi2JCP} and \ref{Fig:chi2th23}, we employed it to obtain 
Figs.~\ref{Fig:cosdeltaNO_fut} and \ref{Fig:cosdeltaNO_GGIO}.

\vspace{-0.2cm}
\section{Results for the Atmospheric Angle}
\label{AppendixC}

\begin{figure}[h!]
  \begin{center}
     \hspace{-1.6cm}
   \subfigure
 {\includegraphics[width=14cm]{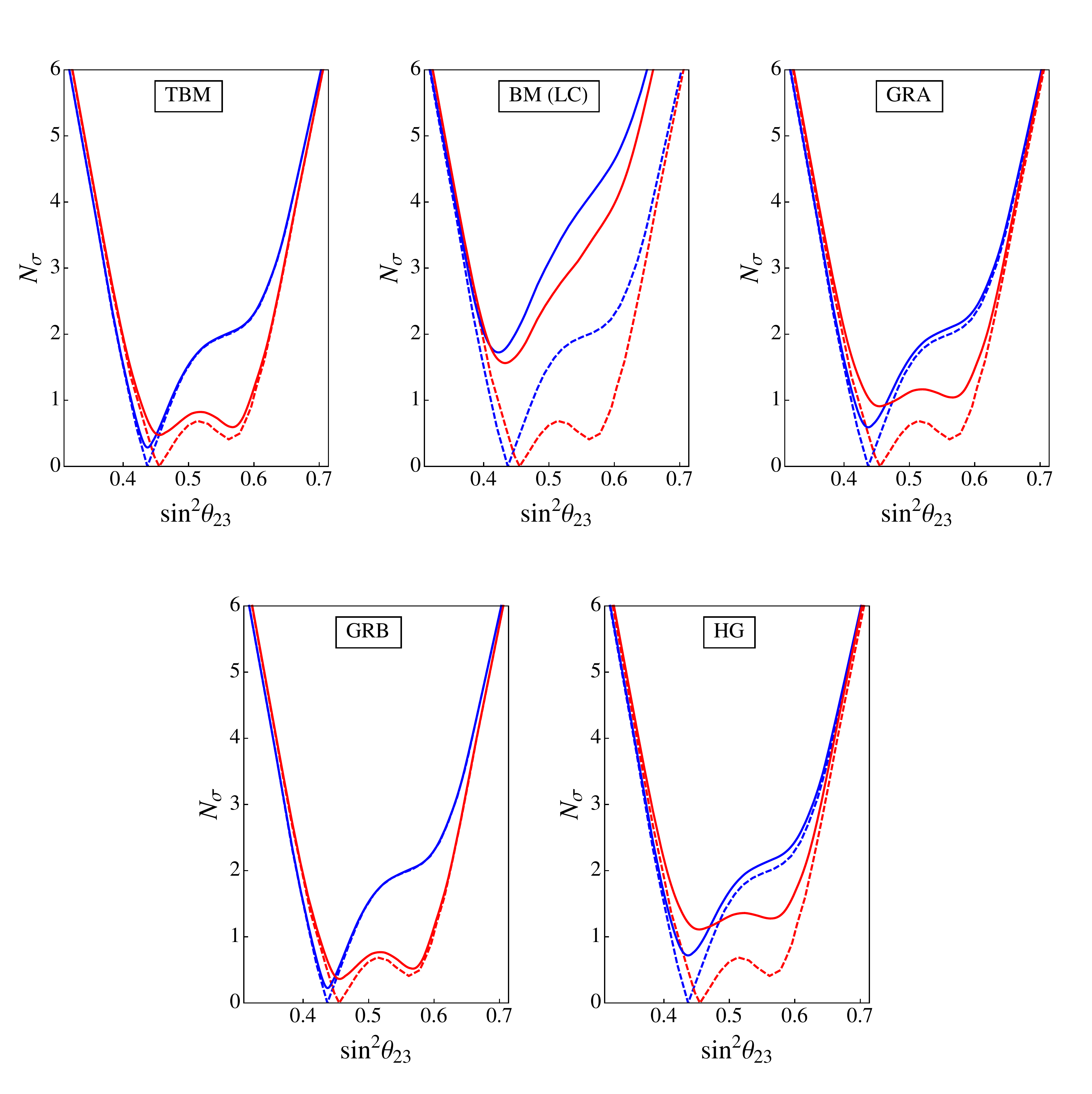}}
  \vspace{5mm}
     \end{center}
\vspace{-1.0cm} \caption{\label{Fig:chi2th23}
$N_{\sigma} \equiv \sqrt{\chi^2}$ as a function of $\sin^2 \theta_{23}$.
The dashed lines represent the results of the global fit
\cite{Capozzi:2013csa}, while the solid ones represent the
results we obtain for the TBM, BM (LC), GRA
(upper left, central, right panels), GRB and HG
(lower left and right panels) neutrino mixing symmetry forms.
The blue (red) lines are for NO (IO) neutrino mass spectrum.
}
\end{figure}

For completeness in Fig.~\ref{Fig:chi2th23} we give $N_{\sigma} \equiv \sqrt{\chi^2}$ as a function of $\sin^2 \theta_{23}$.
The best fit values and the $3\sigma$ regions are summarised in Table~\ref{table:1223_th23}.
%
\begin{table}[h!]
\centering
\begin{tabular}{ l l | c | c }
\toprule
Symmetry form	& 	& Best fit & 3$\sigma$ range  \\ \midrule
TBM	
	& $\sin^2 \theta_{23}$ (NO)  & $0.44$ & $0.37 \div 0.63$ \\
	& $\sin^2 \theta_{23}$ (IO)  & $0.46$ & $0.38 \div 0.65$ \\
         \midrule

BM (LC)	
	& $\sin^2 \theta_{23}$ (NO)  & $0.42$ & $0.37 \div 0.52$ \\
	& $\sin^2 \theta_{23}$ (IO)  & $0.42$ & $0.37 \div 0.56$ \\
         \midrule

GRA
	& $\sin^2 \theta_{23}$ (NO)  & $0.44$ & $0.37 \div 0.63$ \\
        & $\sin^2 \theta_{23}$ (IO)  & $0.46$ & $0.38 \div 0.65$ \\
         \midrule

GRB
	& $\sin^2 \theta_{23}$ (NO)  & $0.44$ & $0.37 \div 0.63$ \\
	& $\sin^2 \theta_{23}$ (IO)  & $0.46$ & $0.38 \div 0.65$ \\
         \midrule

HG
	& $\sin^2 \theta_{23}$ (NO)  & $0.44$ & $0.37 \div 0.63$ \\
	& $\sin^2 \theta_{23}$ (IO)  & $0.46$ & $0.38 \div 0.64$ \\

\bottomrule
\end{tabular}
\caption{\label{table:1223_th23} Best fit values of $\sin^2 \theta_{23}$
and corresponding 3$\sigma$ ranges
(found fixing $\sqrt{\chi^2-\chi^2_{\rm min}} = 3$)
in our setup
using the data from \cite{Capozzi:2013csa}.
}
\end{table}



\begin{thebibliography}{99}

\bibitem{PDG2014}
K.~Nakamura and S.~T. Petcov,
in  K.~A.~Olive {\it et al.} (Particle Data Group),
Chin.\ Phys.\ C {\bf 38} (2014) 090001.

\bibitem{LBLFuture13}
S.~K.~Agarwalla {\it et al.}, JHEP {\bf 1405} (2014) 094; 
C.~Adams {\it et al.}, arXiv:1307.7335 [hep-ex];
A.~de Gouvea {\it et al.}, arXiv:1310.4340 [hep-ex].

\bibitem{Cabibbo:1977nk}
  N.~Cabibbo,
  Phys.\ Lett.\ B {\bf 72} (1978) 333.

\bibitem{BHP80} S.~M.~Bilenky, J.~Hosek and S.~T.~Petcov,
              Phys.\ Lett. B {\bf 94} (1980) 495.


\bibitem{BiPet87} S.~M.~Bilenky and S.~T.~Petcov,
   Rev. Mod. Phys.  {\bf 59} (1987) 671.


\bibitem{BPP1} S.~M.~Bilenky, S.~Pascoli and S.~T.~Petcov,
      Phys.\ Rev.\ D {\bf 64} (2001) 053010;
S.~T.~Petcov,
 Phys.\ Scripta T {\bf 121} (2005) 94.
%
\bibitem{WRodej10}
W. Rodejohann,
 Int. J. Mod. Phys. E {\bf 20} (2011) 1833.

\bibitem{Lang87} P. Langacker {\it et al.},
 Nucl. Phys. B {\bf 282} (1987) 589.

\bibitem{Pascoli:2006ie}
  S.~Pascoli, S.~T.~Petcov and A.~Riotto,
  Phys.\ Rev.\ D {\bf 75} (2007) 083511.

\bibitem{Pascoli:2006ci}
  S.~Pascoli, S.~T.~Petcov and A.~Riotto,
  Nucl.\ Phys.\ B {\bf 774} (2007) 1.

\bibitem{Capozzi:2013csa}
  F.~Capozzi {\it et al.},
  Phys.\ Rev.\ D {\bf 89} (2014) 093018.

\bibitem{Gonzalez-Garcia:2014bfa}
  M.~C.~Gonzalez-Garcia, M.~Maltoni and T.~Schwetz,
  JHEP {\bf 1411} (2014) 052.

\bibitem{Marzocca:2013cr}
  D.~Marzocca, S.~T.~Petcov, A.~Romanino and M.~C.~Sevilla,
  JHEP {\bf 1305} (2013) 073.


\bibitem{Petcov:2014laa}
  S.~T.~Petcov,
  Nucl.\ Phys.\ B {\bf 892} (2015) 400.


\bibitem{Antusch:2005kw}
  S.~Antusch and S.~F.~King,
  Phys.\ Lett.\ B {\bf 631} (2005) 42.

 \bibitem{King:2005bj}
  S.~F.~King,
   JHEP {\bf 0508} (2005) 105.

\bibitem{King:2014nza}
  S.~F.~King {\it et al.},
  New J.\ Phys.\  {\bf 16} (2014) 045018.

\bibitem{King:2013eh}
  S.~F.~King and C.~Luhn,
  Rept.\ Prog.\ Phys.\  {\bf 76} (2013) 056201.
  
  
  \bibitem{TBM}
  P.~F.~Harrison, D.~H.~Perkins and W.~G.~Scott,
  Phys.\ Lett.\ B {\bf 530} (2002) 167;
   P.~F.~Harrison and W.~G.~Scott,
  Phys.\ Lett.\ B {\bf 535} (2002) 163; 
  Z.~Z.~Xing,
  Phys.\ Lett.\ B {\bf 533} (2002) 85;
  X.~G.~He and A.~Zee,
  Phys.\ Lett.\ B {\bf 560} (2003) 87;
see also
L.~Wolfenstein,
  Phys.\ Rev.\ D {\bf 18} (1978) 958.

\bibitem{SPPD82} S.~T.~Petcov, Phys.\ Lett.\ B {\bf 110} (1982) 245.

\bibitem{BM}
F.~Vissani,
  hep-ph/9708483;
V.~D.~Barger, S.~Pakvasa, T.~J.~Weiler and K.~Whisnant,
Phys.\ Lett.\ B {\bf 437} (1998) 107;
A.~J.~Baltz, A.~S.~Goldhaber and M.~Goldhaber,
Phys.\ Rev.\ Lett.\  {\bf 81} (1998) 5730.

\bibitem{Everett:2008et}
  L.~L.~Everett and A.~J.~Stuart,
  Phys.\ Rev.\ D {\bf 79} (2009) 085005.

\bibitem{GRAM}
  Y.~Kajiyama, M.~Raidal and A.~Strumia,
  Phys.\ Rev.\ D {\bf 76} (2007) 117301.

\bibitem{GRBM}
  W.~Rodejohann,
  Phys.\ Lett.\ B {\bf 671} (2009) 267;
  A.~Adulpravitchai, A.~Blum and W.~Rodejohann,
  New J.\ Phys.\  {\bf 11} (2009) 063026.

\bibitem{HGM}
  C.~H.~Albright, A.~Dueck and W.~Rodejohann,
  Eur.\ Phys.\ J.\ C {\bf 70} (2010) 1099.

\bibitem{Kim:2010zub}
  J.~E.~Kim and M.~S.~Seo,
  JHEP {\bf 1102} (2011) 097.


\bibitem{Girardi:2013sza}
  I.~Girardi, A.~Meroni, S.~T.~Petcov and M.~Spinrath,
  JHEP {\bf 1402} (2014) 050.


\bibitem{Chen:2009gf}
  M.~C.~Chen and K.~T.~Mahanthappa,
  Phys.\ Lett.\ B {\bf 681} (2009) 444;
  M.~C.~Chen, J.~Huang, K.~T.~Mahanthappa and A.~M.~Wijangco,
  JHEP {\bf 1310} (2013) 112.






\bibitem{FPR04}
P.~H.~Frampton, S.~T.~Petcov and W.~Rodejohann,
Nucl. Phys. B {\bf 687} (2004) 31.



\bibitem{Marzocca:2011dh}
 D.~Marzocca, S.~T.~Petcov, A.~Romanino, M.~Spinrath,
JHEP {\bf 11}  (2011) 009.

 \bibitem{Antusch:2011qg}
  S.~Antusch and V.~Maurer,
  Phys.\ Rev.\ D {\bf 84} (2011) 117301;
A.~Meroni, S.~T.~Petcov and M.~Spinrath,
 Phys. Rev. D {\bf 86} (2012) 113003;
  S.~Antusch, C.~Gross, V.~Maurer and C.~Sluka,
  Nucl.\ Phys.\ B {\bf 866} (2013) 255.




\bibitem{CarlMCC} C.~H.~Albright and M.~C.~Chen,
Phys. \ Rev. D {\bf 74} (2006) 113006.


\bibitem{Chao:2011sp}
  W.~Chao and Y.~j.~Zheng,
  JHEP {\bf 1302} (2013) 044;


\bibitem{GTani02}
C.~Giunti and M.~Tanimoto,
Phys. Rev. D {\bf 66} (2002) 053013,
 and Phys.\ Rev.\ D {\bf 66} (2002) 113006.


\bibitem{Romanino:2004ww}
  A.~Romanino,
  Phys. Rev.  D {\bf 70} (2004) 013003.

\bibitem{Shimizu:2014ria}
  Y.~Shimizu and M.~Tanimoto,
  arXiv:1405.1521 [hep-ph].
  
  
\bibitem{Hall:2013yha}
  L.~J.~Hall and G.~G.~Ross,
  JHEP {\bf 1311} (2013) 091;
  Z.~Liu and Y.~L.~Wu,
  Phys.\ Lett.\ B {\bf 733} (2014) 226;
  S.~K.~Garg and S.~Gupta,
  JHEP {\bf 1310} (2013) 128.



\bibitem{Gehrlein:2014wda}
  J.~Gehrlein, J.~P.~Oppermann, D.~Schäfer and M.~Spinrath,
  Nucl.\ Phys.\ B {\bf 890} (2015) 539.
  
  

\bibitem{King:2013vna}
   S.~F.~King, T.~Neder and A.~J.~Stuart,
  Phys.\ Lett.\ B {\bf 726} (2013) 312.

\bibitem{Hagedorn:2014wha}
  C.~Hagedorn, A.~Meroni and E.~Molinaro,
  Nucl.\ Phys.\ B {\bf 891} (2015) 499.

\bibitem{Luhn:2013lkn}
  C.~Luhn,
  Nucl.\ Phys.\ B {\bf 875} (2013) 80.

\bibitem{Altarelli:2012bn}
  G.~Altarelli {\it et al.}, 
  JHEP {\bf 1208} (2012) 021;
  G.~Altarelli, F.~Feruglio and L.~Merlo,
  Fortsch.\ Phys.\  {\bf 61} (2013) 507;
  F.~Bazzocchi and L.~Merlo,
  Fortsch.\ Phys.\  {\bf 61} (2013) 571.




  \bibitem{RGE} S.~Antusch {\it et al.},
{Nucl.\ Phys.} {\bf B674} (2003) 401;
J.~A.~Casas {\it et al.},
{Nucl.\ Phys.} {\bf B573} (2000) 652;
P.~H.~Chankowski and Z.~Pluciennik,
{Phys.\ Lett.} {\bf B316} (1993) 312;
K.~S.~Babu, C.~N.~Leung and J.~Pantaleone,
{Phys.\ Lett.} {\bf B319} (1993) 191.

\bibitem{Petcov:2005yh}
  S.~T.~Petcov, T.~Shindou and Y.~Takanishi,
  Nucl.\ Phys.\ B {\bf 738} (2006) 219.
  
  
\bibitem{Alta} G.~Altarelli, F.~Feruglio and I.~Masina,
  Nucl. Phys. B {\bf 689} (2004) 157;
 I.~Masina,
  Phys.\ Lett.\  B {\bf 633} (2006) 134.


\bibitem{HPR07}   K.~A.~Hochmuth, S.~T.~Petcov and W.~Rodejohann,
Phys. Lett B {\bf 654} (2007) 177.
  


\bibitem{PKSP3nu88}
P.~I.~Krastev and S.~T.~Petcov,
Phys.\ Lett.\  B {\bf 205} (1988) 84.


\bibitem{Wang:2014iod}
  Y.~Wang,
  PoS Neutel {\bf 2013} (2013) 030.
  
\bibitem{Coloma:2014kca}
  P.~Coloma, H.~Minakata and S.~J.~Parke,
  Phys.\ Rev.\ D {\bf 90} (2014) 9,  093003.

  
 

\end{thebibliography}
\end{document}